\documentclass{article}
\pdfoutput=1
\PassOptionsToPackage{numbers,sort&compress}{natbib}

\usepackage[final]{neurips_data_2021}
\usepackage[utf8]{inputenc} 
\usepackage[T1]{fontenc}    
\usepackage{hyperref}       
\usepackage{url}            
\usepackage{booktabs}       
\usepackage{amsfonts}       
\usepackage{nicefrac}       
\usepackage{microtype}      
\usepackage{xcolor}         
\usepackage{adjustbox}
\usepackage{multirow}

\usepackage{booktabs}
\usepackage{adjustbox}
\usepackage{multirow}
\usepackage{subcaption}
\usepackage{comment}
\usepackage{wrapfig}
\usepackage{pdfpages}

\usepackage{enumitem}

\title{Intelligent Sight and Sound: \\A Chronic Cancer Pain Dataset}

\author{
  Catherine Ordun \textsuperscript{\rm 1, \rm 2}, 
  Alexandra N. Cha \textsuperscript{\rm 1}, 
  Edward Raff \textsuperscript{\rm 1, \rm 2},\\ 
  \textbf{
  Byron Gaskin \textsuperscript{\rm 1}, 
  Alex Hanson \textsuperscript{\rm 1}, 
  Mason Rule \textsuperscript{\rm 3},
  Sanjay Purushotham \textsuperscript{\rm 2}, 
  James L. Gulley \textsuperscript{\rm 3}
  }\\
  \textsuperscript{\rm 1} Booz Allen Hamilton \\
  \textsuperscript{\rm 2}University of Maryland, Baltimore County \\
  \textsuperscript{\rm 3}Center for Cancer Research, National Cancer Institute, National Institutes of Health
}

\everypar{\looseness=-1}
\begin{document}

\maketitle

\begin{abstract}
Cancer patients experience high rates of chronic pain throughout the treatment process. Assessing pain for this patient population is a vital component of psychological and functional well-being, as it can cause a rapid deterioration of quality of life. Existing work in facial pain detection often have deficiencies in labeling or methodology that prevent them from being clinically relevant. 
This paper introduces the first chronic cancer pain dataset, collected as part of the Intelligent Sight and Sound (ISS) clinical trial, guided by clinicians to help ensure that model findings yield clinically relevant results. The data collected to date consists of 29 patients, 509 smartphone videos, 189,999 frames, and self-reported affective and activity pain scores adopted from the Brief Pain Inventory (BPI). Using static images and multi-modal data to predict self-reported pain levels, early models show significant gaps between current methods available to predict pain today, with room for improvement. Due to the especially sensitive nature of the inherent Personally Identifiable Information (PII) of facial images, the dataset will be released under the guidance and control of the National Institutes of Health (NIH). 
\end{abstract}

\section{Introduction}
The prevalence of chronic pain in cancer patients is high, with an estimated prevalence of 59\% in those undergoing anticancer treatment, 64\% of whom have advanced stage disease and 33\% who continue to experience pain following completion of curative treatment \cite{van2007prevalence}. Despite advances in pain management, prompt assessment and management of cancer pain remains a challenge and a large proportion of patients continue to experience moderate to severe pain.  
Sub-optimal pain management can block patient recovery and improvement, making the already difficult cancer experience, worse, for both patient and family \cite{sun2007overcoming, cleary2000cancer}. Manual clinical assessment requires accounting for a landscape of complex emotions and beliefs that clinicians must regularly take into account when assessing cancer patient pain - physical, psychological, social, and spiritual elements combined with severe distress for future outlook \cite{lotsch2018machine, sun2007overcoming}. For example, patients undergoing chemotherapy are more likely to believe that ``good patients" do not complain about pain which they believe can be distracting to clinicians and become non-communicative \cite{sun2007overcoming}. Further, few patients are actually screened for pain at each clinical visit \cite{cleary2000cancer}, and pain is under-reported in patient populations such as nursing home patients \cite{cleary2000cancer}. Due to the variety of complex conditions affecting cancer pain, experts recommend repeated, regular pain assessment, which can be difficult and impractical for manual assessment by clinicians.  

Currently, no facial pain datasets exists for chronic cancer pain and little research overall has been conducted into machine learning for the identification and evaluation of chronic pain. For example, in a review by ~\cite{lotsch2018machine}, only seven out of 52 machine learning papers evaluated pain in a non-acute context such as chronic fatigue, fibromyalgia, chronic pancreatitis, migraines, and genetic pain. Existing facial pain research focuses on acute, musculoskeletal pain such as chronic lower back pain \cite{hu2018using} and shoulder pain \cite{prkachin2008structure, lucey2011painful} or simulated pain induced by heat or electrical stimuli \cite{walter2013biovid, haque2018deep} where painful expressions are obvious through grimaces and eye raises. Such datasets are manually labeled by trained observers with Facial Action Coding Units \cite{ekmanw1978facial}, making the labeling procedure prohibitively expensive and impractical for clinical use. Further, external pain labels may be biased towards an outside observer's impression of a patient's pain, not the patient themselves. Research also shows that typical pain facial expressions that correlate with physical pain are less frequently observed among chronic pain cancer patients who exhibit subdued and placid expressions \cite{wilkie1995facial}. 

Given the limitations of existing facial expression pain data, the U.S. National Institutes of Health (NIH) National Cancer Institute (NCI) initiated \emph{``A Feasibility Study Investigating the Use of Machine Learning to Analyze Facial Imaging, Voice and Spoken Language for the Capture and Classification of Cancer Pain}" \cite{NIH}, or "Intelligent Sight and Sound" (ISS).  Details of the protocol are available publicly at \url{https://clinicaltrials.gov/ct2/show/NCT04442425}. This is an observational, non-interventional clinical study that aims to address the following problem statement \cite{pomeraniec2020intelligent, NIH}: ``\textit{To determine if a new observational based pain prediction algorithm can be produced that is accurate to standard, patient-reported pain measures and is generalizable for a diverse set of individuals, across sexes and skin types.}'' The study has two objectives: 1) investigate facial image data, and 2) analyze text and audio, as modalities for predicting self-reported chronic cancer pain. 

The study is ongoing and aims to recruit 112 patients. We report the initial dataset, which is less than a quarter of the final data consisting of 29 patients.  Data include multimodal extracts from video submitted in a spontaneous, home setting, and in a few cases of in-clinic capture at the NIH. It includes visual spectrum (RGB) video frames, facial images resulting from face detection models, facial landmarks from Active Appearance Models (AAMs) \cite{cootes2001active, kartynnik2019real}, audio files, Mel spectrograms, audio features, and self-reported pain scores adopted from the Brief Pain Inventory (BPI) \cite{cleeland1991pain, cleeland1991brief, kunz2017problems}.We will present details of the study design, data distribution, and storage procedures to ensure patient privacy. We also provide initial baseline results for pain classification using simple, traditional, machine learning models and neural networks. 

\vspace{-1mm}
\section{Related Works}
Automatically detecting pain from facial expressions has been extensively published following methods of facial emotion recognition (FER). The majority of these works have focused on acute or musculoskeletal physical pain \cite{lotsch2018machine, lucey2010automatically, zhou2016recurrent, neshov2015pain, khan2013pain, lo2015using, bellantonio2016spatio, hassan2019automatic, ashraf2009painful, bartlett2006automatic}. Primary pain datasets based upon facial imaging include UNBC-McMaster Shoulder Pain Expression Archive \cite{prkachin2008structure, lucey2011painful}, the Biopotential and Video Heat Pain (BioVid) Database using controlled, simulated heat to induce pain \cite{walter2013biovid}, Multimodal Intensity Pain (MIntPAIN) database using pain resulting from electrical stimulation \cite{haque2018deep}, the Experimentally Induced Thermal and Electrical (X-ITE) Pain Database \cite{gruss2019multi, werner2014automatic}, and the EmoPain for chronic, musculoskeletal pain \cite{aung2015automatic}. These datasets traditionally contain video sequences since video enables continuous clinical monitoring of pain response \cite{kunz2017problems}. These datasets also contain extensive offline annotations of pain ratings by external observers, and sometimes include additional modalities such as thermal and depth data. Additional video facial expression pain datasets exist that focus on different patient populations, but primarily focus on physical pain. These include multimodal behavioral and physiological data for neonatal pain \cite{zamzmi2016approach, zamzmi2018neonatal} and the University of Regina (UofR) Pain in Severe Dementia dataset \cite{rezaei2020unobtrusive, asgarian2019limitations}. A summary of the pain datasets is provided in Table \ref{related}.

\begin{table}[htbp]
\centering
\caption{\small{\textbf{Related Pain Datasets.}}}
\begin{adjustbox}{width=0.99\textwidth}
\begin{tabular}{@{}lllllll@{}}
\toprule
\textbf{Dataset} & \textbf{Stimulus} & \textbf{Subjects} & \textbf{Frames} & \textbf{Sequences} & \textbf{Seq. Duration} & \textbf{Modality} \\ \midrule
UNBC-McMaster   \cite{prkachin2008structure, lucey2011painful} & Shoulder pain & 25 & 48,398 & 200 & 10 - 30 sec., per & Unimodal \\
BioVid \cite{walter2013biovid} & Heat stimulus & 90 & 8700 & 87 & 5.5 sec. & Multimoda \\
MIntPAIN \cite{haque2018deep} & Electronic stimulus & 20 & 187,939 & 9366 & 1 - 10 sec. & Multimodal \\
EmoPain   \cite{aung2015automatic} & Chronic lower back pain & 22 & 44,820 & 35 & 3 sec. & Multimodal \\
Neonatal Pain, USF   \cite{salekin2021multimodal, zamzmi2018neonatal, salekin2019multi} & Heel lancing & 31 & 3026 & 200 & 9 sec. & Multimodal \\
UofR   \cite{rezaei2020unobtrusive} & Physical, painful movements & 102 & 162,629 & 95 & Unknown & Multimodal \\
X-ITE   \cite{othman2021automatic, gruss2019multi, werner2014automatic} & Heat and electronic stimuli & 134 & 26,454 &  N/A & 7 sec. & Multimodal \\
\toprule
\textbf{ISS (Dec. 2020 - Jul. 2021)} & \textbf{Chronic cancer pain} & \textbf{29} & \textbf{189,999} & \textbf{509} & \textbf{3.52 - 135.59} & \textbf{Multimodal} \\ \bottomrule
\end{tabular}
\end{adjustbox}
\label{related}
\end{table}

\section{ISS Dataset}
The ISS protocol is a single site study with a goal of enrolling a total of 112 patients (90 adult and 22 pediatric) who are actively receiving treatment for advanced malignancies and/or tumors at the NIH Clinical Center or treated with standard of care in the community. The study is overseen by the NIH Institutional Review Board (IRB), and the protocol was also reviewed by NCI's Center for Cancer Research (CCR) Scientific Review Committee. New patient enrollment was paused during the Covid-19 pandemic due to initially unknown risks, but has resumed with vaccine availability and clinician guidance.  

\subsection{Sample and Study Design}
To obtain as representative a sample as possible within the constraints of a feasibility study with an overall small sample size, the sample consists of twelve cohort groups of seven patients each. Patients represent a breadth of age, sex, skin tone (as a proxy for ethnicity), and pain experience. The current ISS dataset consists of 29 adult patients ages 18 years and over who have consented to participate in the study; no pediatric patients (ages 12-17) have been enrolled yet.

\begin{wraptable}[19]{r}{0.75\textwidth}
\vspace{-12pt}
\caption{\small{\textbf{ISS: Twelve Patient Cohorts.}}}
\label{recruitment}
\centering
\adjustbox{max width=0.70\textwidth}{%
\begin{tabular}{@{}lllllll@{}}
\toprule
\textbf{Number} & \textbf{Pain Target} & \textbf{Skin Types} & \textbf{Sex} & \textbf{Pain Class} & \textbf{Goal} & \textbf{Current} \\ \midrule
1A & 0 & I - III & Male & None & 7 & 7 \\
1B & 0 & I - III & Female & None & 7 & 2 \\
1C & 0 & IV - VI & Male & None & 7 & 4 \\
1D & 0 & IV - VI & Female & None & 7 & 0 \\
2A & 1-3 & I - III & Male & Low & 7 & 1 \\
2B & 1-3 & I - III & Female & Low & 7 & 1 \\
2C & 1-3 & IV - VI & Male & Low & 7 & 3 \\
2D & 1-3 & IV - VI & Female & Low & 7 & 0 \\
3A & 4-6 & I - III & Male & Moderate & 7 & 2 \\
3B & 4-6 & I - III & Female & Moderate & 7 & 2 \\
3C & 4-6 & IV - VI & Male & Moderate & 7 & 0 \\
3D & 4-6 & IV - VI & Female & Moderate & 7 & 0 \\
4A & 7-10 & I - III & Male & Severe & 7 & 1 \\
4B & 7-10 & I - III & Female & Severe & 7 & 2 \\
4C & 7-10 & IV - VI & Male & Severe & 7 & 2 \\
4D & 7-10 & IV - VI & Female & Severe & 7 & 3 \\
\toprule
\textbf{} & \textbf{} & \textbf{} & \textbf{} & \textbf{} & \textbf{112} & \textbf{30} \\ \bottomrule
\end{tabular}}
\end{wraptable}

The goal is to evenly split the sample by i) sex (Male or Female),  ii) Fitzpatrick Skin Type \cite{goldsmith2012fitzpatrick}, a self-reported, visual method of skin tone classification, where patients are asked to type themselves into one of two groups: ``light" skin tones in types I-III or ``dark" skin tones in types IV-VI, and iii) a self-reported ``worst" pain score reported on a 0 – 10 Numerical Rating Scale (NRS) \cite{haefeli2006pain}. The self-reported pain score is referred to as the ``Pain Target" and are grouped into levels 0, 1-3, 4-6, and 7-10. 

It represents the worst pain the patient has experienced in the past thirty days prior to the start of the study. It is fixed throughout the patient's enrollment and does not change. As a result, there is no variance for the ``Pain Target" score. The ``Pain Target" is the classification target which is later used for our baseline tasks. The twelve different cohorts are shown in Table \ref{recruitment}, along with the goal of seven patients to be enrolled per cohort, and the current distribution of patients enrolled. 

\begin{wraptable}{r}{8.5cm}
\centering
\caption{\small{\textbf{ISS Study Design Overview.} * Note that due to the global Covid-19 pandemic, the majority of patient videos were submitted in the remote setting.}}
\begin{adjustbox}{width=0.6\textwidth}
\begin{tabular}{@{}ll@{}}
\toprule
Study Duration & 3 months \\ \midrule
\# Remote Submission & 3 / week; Max. 1 /day \\ \midrule
\# In-Clinic Submissions & 1 - 4 /week* \\ \midrule
Survey Tool & Smartphone (iPhone or Android) or computer (camera/mic.) \\ \midrule
Time / Submission & Average total time: 3 min \\  \midrule
 Time / Question & Q 1-9: 1 min, Q10: 15 sec, Q11: 15 sec. -   3 min. \\ \midrule
Self-Reported Pain & Q 1-9: Questions with Likert-scale responses \\ \midrule
Voice/Video: Prompt & Q10: Read and record one of 3 randomized nursery rhymes. \\  \midrule
Voice/Video: Narrative & Q11: Record respond to "Describe how you feel right now." \\  \midrule
Compensation: & Min. 3 / week, they earn \$15. \\ \bottomrule
\end{tabular}
\end{adjustbox}
\label{studydesign}
\end{wraptable}

Note that data from one patient (0009) in Cohort 2B was unusable. As a result our analysis reports across 29 patients. Clinical inclusion criteria include individuals with a diagnosis of a cancer or tumor who are under active treatment for this condition at NIH/NCI. Patients must also have access to a smart phone or computer with camera, microphone, and internet access. Several clinical exclusion criteria apply. Excluded are patients with active central nervous system (CNS) metastases, with the exception of those who have completed curative intent radiotherapy or surgery and have been asymptomatic for three months prior to consent, patients with Parkinson's disease, and any psychiatric condition that would prohibit the understanding or rendering of informed consent. Additional exclusion criteria those who are non-English speaking or have known current alcohol or drug abuse. Each patient is enrolled for a three-month period and are financially incentivized to complete three check-ins per week remotely and up to four in-clinic check-ins. The study design is summarized in Table \ref{studydesign}. Patients engage using an electronic questionnaire and through video recording using a custom developed mobile or web application, using an Android, iPhone, or computer with camera and microphone.

\subsection{Patient Protocol}
\begin{figure}[htbp]
    \centering
    \includegraphics[width=0.9\textwidth]{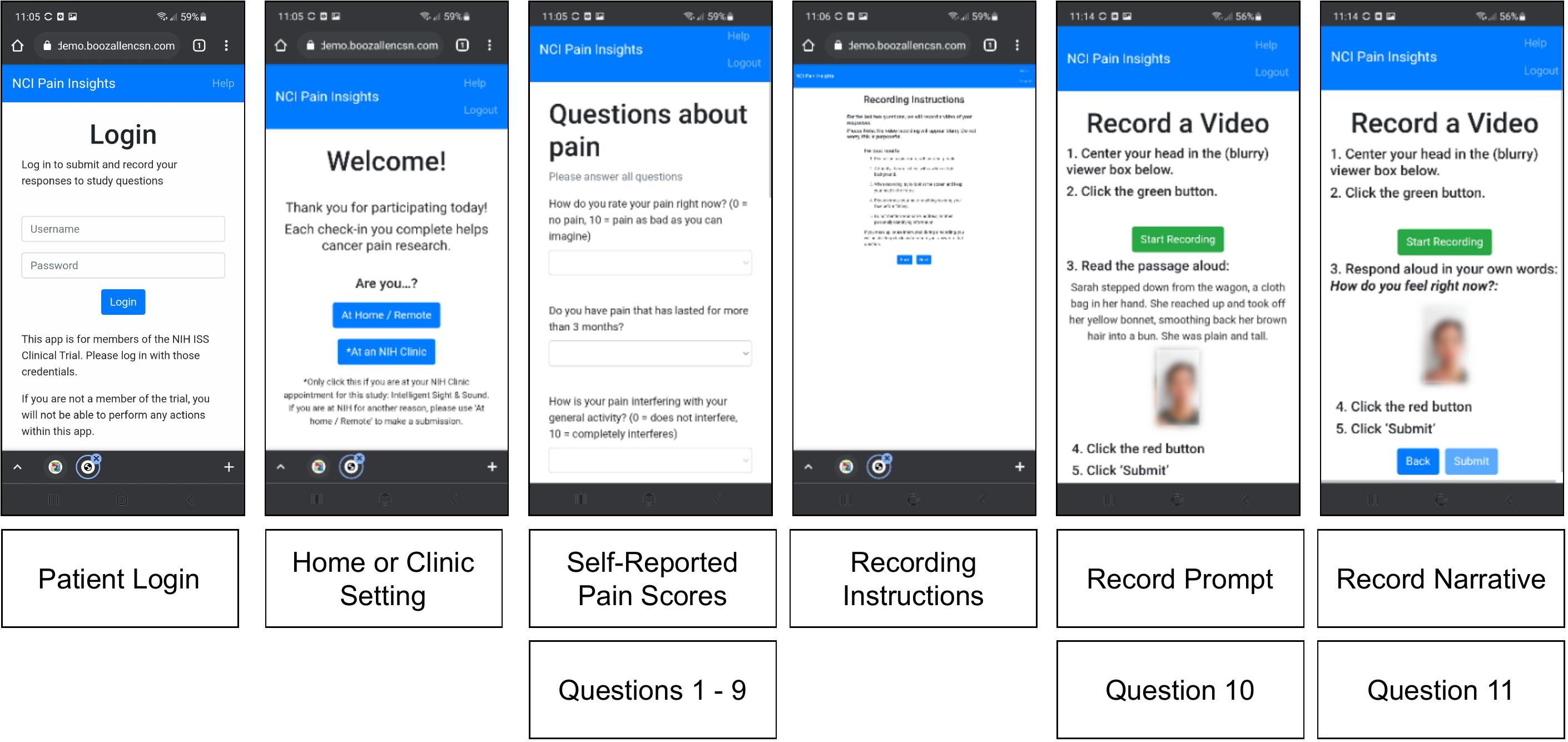}
    \caption{\small\textbf{Submitting a Video through the ISS Mobile Application.} }
    \label{dataflow}
\end{figure}

Figure \ref{dataflow} provides a series of screenshots showing the patient at-home or in-clinic check-in using the ISS application. For each approximately 3-minute check-in, patients respond to a nine element questionnaire based on the Brief Pain Inventory (BPI, licensed from MD Anderson) \cite{cleeland1991pain, cleeland1991brief, kunz2017problems} and two prompts to record videos of themselves.

\subsubsection{Questions 1 - 9: Self-Reported Pain Scores}
In addition to the self-reported ``Pain Target" which was assigned to each patient upon enrollment shown in Table \ref{recruitment}, there are nine additional self-reported pain scores. We capture these self-reported pain scores based on the cancer pain literature which indicates that cancer patients experience complex emotions and beliefs that can influence their perception of pain and as a result, clinical treatment \cite{cleary2000cancer, sun2007overcoming}. These nine pain scores are submitted at the time of video submission and change at each submission. They are distinct and unrelated to the ``Pain Target" which is used for the baseline classification tasks. There is no formula that relates the nine self-reported scores among themselves or to the ``Pain Target". In the below, Question 1 captures current pain intensity scored on an 11-point Likert Scale (0 No Pain - 10 Worst Possible Pain) followed by Question 2 which, when answered affirmatively, indicates the presence of chronic pain. Questions 3-9 utilize an 11-point Likert Scale (0 Does not Interfere - 10 Completely Interferes) to measure the interference of pain in an individual's activity (3, 5, 6) and an individual's affect or mood (4, 7, 8, 9).  
\begin{enumerate}[wide=0pt]
\itemsep -0.2em 
\item How do you rate your pain right now? (0 No Pain – 10 Worst Possible Pain on Likert-scale). 
\item Do you have pain, related to your cancer, that has lasted for more than 3 months? (Yes/No)
\item How is your pain interfering with your General Activity? 
\item How is your pain interfering with your Mood?  
\item How is your pain interfering with your Walking Ability?
\item How is your pain interfering with your Normal Work (both work outside the home and housework)?   
\item How is your pain interfering with your Relationships with other people? 
\item How is your pain interfering with your Sleep? 
\item How is your pain interfering with your Enjoyment of life?
\end{enumerate}

\subsubsection{Questions 10 and 11: Prompt and Narrative}
Following the questionnaire, Question 10 is a prompt to record a video where the patient reads a 10-15 second passage of text at a grade 3 reading level selected at random from three different passages. The use of this sort of prompt is common practice in mood induction or conditioning trials where a neutral, non-emotion inducing prompt is used as a control versus a potentially, emotionally charged response related to the experimental condition \cite{apolinario2018improving, fink2020interpretation, livesay1994emg}. The neutral passage options are: 

\begin{itemize}[wide=0pt]
\itemsep 0em 
    \item ``Sarah stepped down from the wagon, a cloth bag in her hand. She reached up and took off her yellow bonnet, smoothing back her brown hair into a bun. She was plain and tall.”  From Sarah Plain and Tall by Patricia MacLachlan \cite{maclachlan1997sarah}
    \item ``And then the dog came running around the corner. He was a big dog. And ugly. And he looked like he was having a real good time. His tongue was hanging out and he was wagging his tail.” From Because of Winn Dixie by Kate DiCamillo \cite{dicamillo2009because}
    \item ``You have brains in your head. You have feet in your shoes. You can steer yourself any direction you choose. You’re on your own. And you know what you know. And YOU are the one who’ll decide where to go.” From Oh the Places You Will Go by Dr. Seuss \cite{geisel1990oh}
\end{itemize}

Finally, in Question 11, the patient records a video responding to the prompt \emph{"Please describe how you feel right now.”} Narratives include discussion of medical conditions, mood, daily activities, current beliefs and attitudes about their pain. The allowable video length can range from 15 seconds to 3 minutes, with recording instructions  shown prior to each video prompt. For “at-home” check-ins, patients are instructed to complete the submission alone, in a quiet and brightly lit room, preferably with a white wall or background. In addition, patients are asked not to reveal personal information such as their name or address. In Figure \ref{dataflow}, the application screens for Questions 10 and 11 include a live video image to help the patient keep their face centered in the frame, but the application blurs the video. The blur effect is to prevent the patient from manipulating their facial expression and minimize self-conscious alteration of their appearance, allowing them to focus on their responses.

\subsection{Data Description}
A high level summary of the ISS dataset is provided in Table \ref{summary}. The ISS dataset is comprised of 29 patients submitting videos in a spontaneous, non-posed, home setting through a smartphone or computer. Patients are adults over the age of 18 y.o. and consist of the following demographics: 20 Male, 9 Female, 17 Skin Type I-III, 12 Skin Type IV - VI. All patients were enrolled between December 2020 and July 2021.  There are 189,999 total video frames. After facial detection, we extracted 173,011 facial images. After landmark detection on the facial images, the dataset was reduced by 2.86\% to 168,063 facial images with landmarks, since landmarks could not be detected for some faces. We show the ratio of data imbalance across four pain levels using the total frames in Table \ref{summary} where the "None" label is the majority class.  The dataset also contains self-reported pain scores from Questions 1 - 9, described in detail in the Study Design section, along with sex and skin type labels assigned upon enrollment. Additional descriptive analysis is provided in Supplementary Materials.

\begin{table}[htbp]
\centering
\caption{\small{\textbf{ISS Data Summary.}}}
\begin{adjustbox}{width=0.99\textwidth}
\begin{tabular}{@{}ll|ll|llll|llll@{}}
\toprule
\multicolumn{4}{c}{\textbf{ISS   Data Summary}} & \multicolumn{8}{c}{\textbf{Ratios of Total Frames   by Pain Levels}} \\ \toprule
\textbf{Total Patients} & \textbf{29} & \multicolumn{2}{c}{\textbf{20 M, 9 F, 17 Skin Type I-III, 12 Skin Type IV-VI}} & \textbf{4 Pain Levels} & \textbf{Frames} & \textbf{Ratio} & \textbf{No. Patients} & \textbf{2 Pain Levels} & \textbf{Frames} & \textbf{Ratio} & \textbf{No. Patients} \\
\midrule
\textbf{Total Videos} & 509 & \textbf{Avg. Videos per Patient} & 17.55 & None & 100984 & 1.00 & 13 & No Pain & 100984 & 1.00 & 13 \\
\textbf{Total Frames} & 189,999 & \textbf{Avg. Frames per Patient} & 6551 & Low & 11784 & 8.57 & 4 & Pain & 89015 & 1.13 & 16 \\
\textbf{Total Duration} & 316 min. & \textbf{Avg. Duration per Patient} & 655 sec. & Mod. & 25999 & 3.88 & 4 &  &  &  &  \\
\textbf{Avg. Duration per Video} & 37.32 sec. & \textbf{Range of Duration per Video} & 3.52 - 135.79 sec. & Severe & 51232 & 1.97 & 8 &  &  &  & \\ \bottomrule
\label{summary}
\end{tabular}
\end{adjustbox}
\end{table}

 \begin{figure}[htbp]
    \centering
    \includegraphics[width=1.0\textwidth]{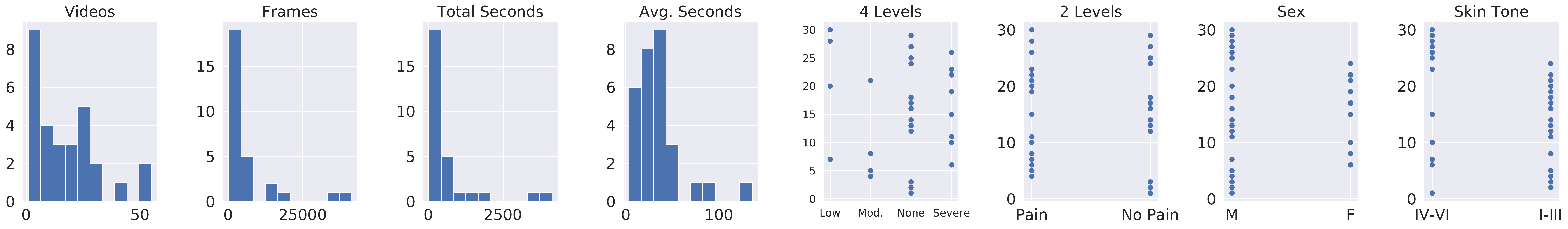}
    \caption{\small\textbf{Distribution of ISS Data.} Histograms for the total videos, frames, seconds, and average seconds per video, for the ISS dataset are in the four left-most plots. The four plots on the right illustrate the distribution of patients (y axis) by the four pain levels, when combined into two pain levels, by sex, and by skin type.}
    \label{hist}
\end{figure}

\begin{figure}[htbp]
    \centering
    \includegraphics[width=0.7\textwidth]{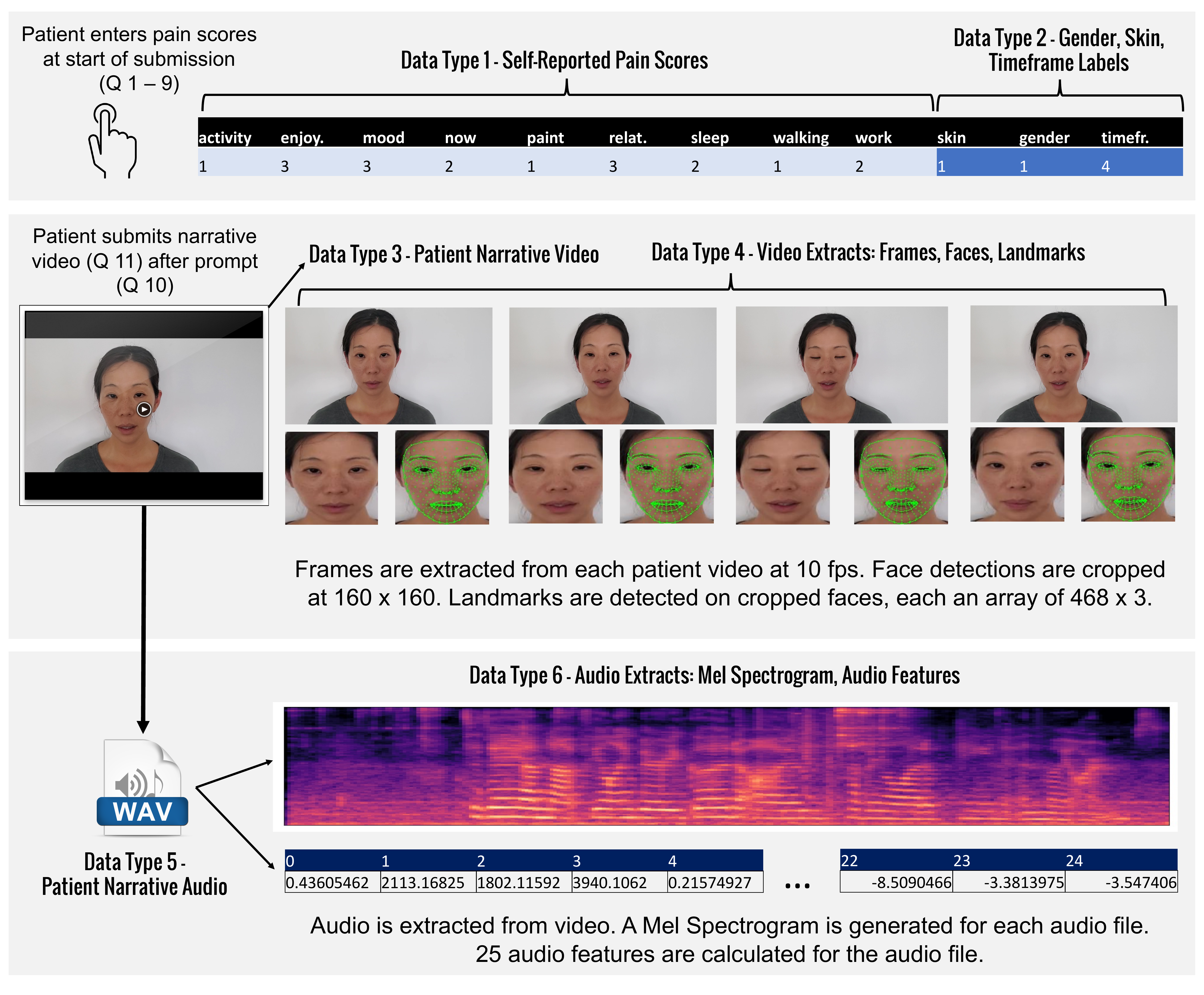}
    \caption{\small\textbf{ISS Data Types.} Facial images shown are \emph{not} actual patients from the ISS dataset due to privacy restrictions. The ISS Dataset currently consists of six types of data: 1) Nine Self-Reported Pain Scores, 2) Labels for Sex, Skin Type, and Timeframe, 3) Patient Narrative Video, 4) Video Extracts: Frames, Faces, Landmarks, and 5) Patient Audio, and 6) Audio Extracts: Mel Spectrogram, Audio Features.}
    \label{datadescription}
\end{figure}

A notional depiction of ISS data types is shown in Figure \ref{datadescription} to provide context for the data types. Due to the sensitivity of Personal Identifiable Information (PII) in the clinical study protocol, we are unable to display actual facial images from the dataset at this time. 

\subsubsection{Data Extraction}
We use the patient narrative (Question 11) video files (\texttt{.mp4}) and extract frames at 10 frames-per-second. We decide to use the narrative versus the prompt since it may contain greater signals of pain and emotion, compared to the neutral baseline recording. An audio \texttt{.wav} file of the patient narrative is simultaneously extracted using the ffmpeg library. We use the PyTorch FaceNet library that implements a fast and CUDA-enabled version of the Multi-task Cascaded Convolutional Networks (MTCNN) algorithm \cite{zhang2016joint} using an InceptionResnetV1 model pre-trained on VGGFace2 for face detection and cropping faces from frames. All patient faces were recorded in a frontally aligned position so no realignment was implemented. Similar to ~\cite{ashraf2009painful}, we extract features using AAMs. Specifically, we use the Google MediaPipe \cite{lugaresi2019mediapipe} Face Mesh AAM model based on 3D Morphable Models \cite{kartynnik2019real} to detect facial landmarks where each face returns an array of 468 points for three coordinates. From the audio \texttt{.wav}  file, we use the Librosa \cite{brian_mcfee_2021_4792298} library to generate a Mel Spectrogram (n\_fft=2048, hop\_length=512, n\_mels=128), and apply signal processing to capture audio features about the \texttt{.wav}  file to include Mel-frequency cepstral coefficients (MFCCs), chromogram, spectral centroid, spectral bandwidth, roll-off frequency, and zero crossing rate, leading to 25 audio features. We further break up the original video into 4-second chunks leading to 40 frames per video chunk, extracting its respective \texttt{.wav}  file, spectrogram, and audio features.

\subsubsection{No External Labels}
In contrast to existing acute pain datasets \cite{lucey2011painful, walter2013biovid, haque2018deep}, the ISS dataset lacks external offline labeling traditionally completed using the Facial Action Coding System (FACS) \cite{ekmanw1978facial}. Per the ISS problem statement, the goal is to predict patient (self)-reported pain, as opposed to observations made by non-patients via offline pain coders.  There are three reasons for not externally encoding ISS video frames using FACS. First, researchers agree that FACS is expensive due to the need for a trained coder to annotate each video frame, making the process time-consuming and clinically infeasible \cite{chen2018automated, kunz2017problems}. Second, ethicists and psychologists argue that there is limited evidence that facial expressions are reliably and specifically mapped to emotion production \cite{barrett2019emotional, crawford2021atlas}. Emotion production is not necessarily tied to a single set of facial expressions, but relies on the context of the situation and human culture \cite{barrett2019emotional}. Third, cancer patients with chronic pain may not display the typical set of facial action units (AUs) commonly associated with acute pain. For example ~\cite{wilkie1995facial} collected video data from 43 outpatient lung cancer patients obtained in a spontaneous home setting \cite{wilkie1995facial}. They found that the cancer patients were more subdued in expression, and displayed fewer AUs such as grimaces or clenched teeth, commonly found in facial pain images. As a result, AU labels associated with pain such as brow lowering (AU4), orbital tightening (AU6, AU7), levator contraction (AU9, AU10), and eye closure (AU43) may not be applicable to chronic cancer facial pain detection \cite{prkachin1992consistency}. However, when the ISS dataset is released, there are no prohibitions on researchers attempting to annotate using FACS.

\subsection{Data Storage and Access}
A secure cloud-based environment receives mobile and web-based submissions of patients’ video, audio, and survey (nine self-reported pain scores) data. No PII such as names or date of birth is stored, with the exception of face, voice, and sex information. The environment is AWS GovCloud FedRAMP Moderate, with Federal Information Security Management Act (FISMA) moderate Authority-to-Operate (ATO) credentials.

The ISS dataset consists of cancer patients discussing their medical conditions. The very nature of the images and videos make the data Protected Health Information (PHI) due to the NIH/NCI not being classified as a "covered entity". Extreme care must be exercised to ensure patient privacy and rights are not violated. As a result, we plan to ensure proper patient protections by placing the collected data in restricted access repositories under the stewardship of the NIH. Members of the scientific community will be able to request access to the data and code which may be granted on a per-case basis. This requirement is necessary to ensure legal requirements are met, avoid public spillage of PII data, and ensure patient trust that their data is used within the scope of the intended scientific use. In return researchers receive access to a dataset with numerous modalities and potential clinical relevance of results. 

\section{Baselines}
We conduct seven baseline experiments for a classification task to predict each patient's self-reported ``Pain Target" level assigned at the start of their enrollment shown in Table \ref{recruitment}. These levels are fixed upon enrollment for cohort assignment and remain unchanged throughout the study. As a result, one patient represents a single pain level throughout the study. All experiments are static models, which return predictions on a frame-by-frame level. Given how we are in the initial phase of the ISS study, we train models using facial images, landmarks, and the additional nine self-reported pain scores for emotion and activity. However, we do provide baseline results on 4-second chunks of audio via spectrograms and audio features. These are meant to be representative of common approaches to similar work, and establish the careful curation results in a task more difficult than prior literature with simpler labeling or collection. More details on all results are in the Supplementary Materials.

\textbf{Training Details} All experiments are trained using 10-fold cross validation where three test patients are withheld in the test set for nine splits and two patients set aside for the tenth split. There is no overlap between training and test sets for each split. Please refer to the Supplementary Materials Appendix Section F.1. Table 10 that shows the ``10-fold-CV details - Test Patients per Split." For neural networks in Experiments 1 and 3 - 7, we use a batch size of 16, Adam optimizer with 1e-4 learning rate, and cross entropy as the loss function, training for 10 epochs, for all experiments.  The batch size of 16 was selected empirically based on cross validation accuracy, after running several experiments varying batch size from 4, 8, 16, 32, and 64. We selected Adam optimizer since it has been used in recent facial pain detection studies such as ~ \cite{bellantonio2016spatio, rodriguez2017deep}. We fine-tune ResNet50 as the convolutional neural network (CNN) backbone for all multimodal experiments, which is pretrained on ImageNet. We use PyTorch for model training and train on four NVIDIA Tesla T4 GPUs. Experiments 2 and 3 are trained using the Scikit-Learn library for the Random Forest Classifier, using 100 estimators, gini criterion, min\_samples\_split=2, and min\_samples\_leaf=1.

\textbf{Experiment 1: Pain Prediction using Static Face Images}
The first set of experiments only uses static, facial images. We fine-tune ResNet-50 \cite{he2016deep} pretrained on ImageNet \cite{deng2009imagenet} to predict four and two levels of pain. Four levels are ``None" (Self-Reported Pain Level 0), ``Low" (1-3), ``Moderate" (4-6), or ``Severe" (7-10), and two levels combine ``Low", ``Moderate", and ``Severe" pain levels into a single ``Pain" class. Training binary classifiers for``No Pain"/``Pain" prediction is similar to many existing facial pain detection works \cite{lucey2011painful, khan2013pain}. We found that the binary classifier leads to better test patient accuracy scores, and continued Experiments 2 through 7 using only two pain levels.

\textbf{Experiments 2, 3: Pain Prediction using Static Landmarks or Pain} In these experiments, we use only one modality to train two separate models and use traditional machine learning models, specifically the Random Forest algorithm \cite{breiman2001random}. Experiment 2 uses the landmark arrays detected for each facial image and Experiment 3 uses the nine self-reported pain scores explained in Section 3.2.1 that represent how pain interferes with the patient's emotions and activity, plus labels for sex, skin type, and timeframe. The timeframe label is categorical and is extracted from the video submission timestamp representing what time of day (early AM, late PM, etc.) the video was submitted. For both Experiments 2 and 3, we train a Random Forest Classifier. Note, that the target ``Pain Target" is not in the set of the nine self-reported pain scores, which are distinct and separate.

\textbf{Experiments 4 - 6: Pain Prediction using Static Multimodal Data} 
We train three multimodal networks using an early, joint fusion strategy as proposed by \cite{huang2020fusion}. For Experiment 4 (``Fusion 1"), we concatenate the fully connected outputs of ResNet50 with raw landmarks. The feature vector is then inputted to a feedforward neural network for binary pain prediction. Experiment 5 (``Fusion 2") concatenates the fully connected outputs of ResNet50 with raw landmarks, in addition to the nine pain scores, skin, sex, and timeframe labels. Similarly, the feature vector is inputted to the same feedforward network architecture for binary pain prediction.  Experiment 6 (``Fusion 3") concatenates three vectors: the feature map from \texttt{layer-4-conv2D-1}, the landmark features outputted from a landmark-specific feedforward network, and the nine pain scores, sex, skin, and timeframe features outputted from a pain-specific feedforward network. The resulting feature vector is inputted to a CNN for binary pain prediction.

\textbf{Experiment 7: Audio Models}
Experiment 7 is a binary pain prediction model that uses the Mel spectrogram image and 25 audio features from 4-second chunks of audio extracted from each patient video. A feature vector resulting from the concatenation of the spectrogram feature map from \texttt{layer-4-conv2D-1} and audio features learned by a feedforward network, are inputted to the same CNN architecture as used in Experiment 6.  Diagrams for all experimental architectures are provided in the Supplementary Materials.

\section{Results}
\textbf{Accuracy Calculation} The accuracy of each model is evaluated for each test patient using the tenth model checkpoint. Using the checkpoint, we evaluate each test patient individually. We only evaluate test patients using their respective, assigned split per 10-fold cross-validation (See Supplementary Materials Section F.1. Table 10 ``10-fold-CV details - Test Patients per Split" for details). For example, test patients 0002, 0029, and 0021 are only evaluated using the trained model from Split 1, not Split 2 which would have included these three patients in its training set. We evaluate each test patient using a batch size of 1, predicting the target pain score for each patient image. We then calculate accuracy for the test patient in question as simply $accuracy\_score(y\_true, y\_pred)$ where $y\_true$ is the set of true ``Pain Target" labels and $y\_pred$ is the set of predicted ``Pain Target" labels. 

\begin{figure}[ht!]
     \centering
     \begin{subfigure}[b]{0.24\textwidth}
         \centering
         \includegraphics[width=\textwidth]{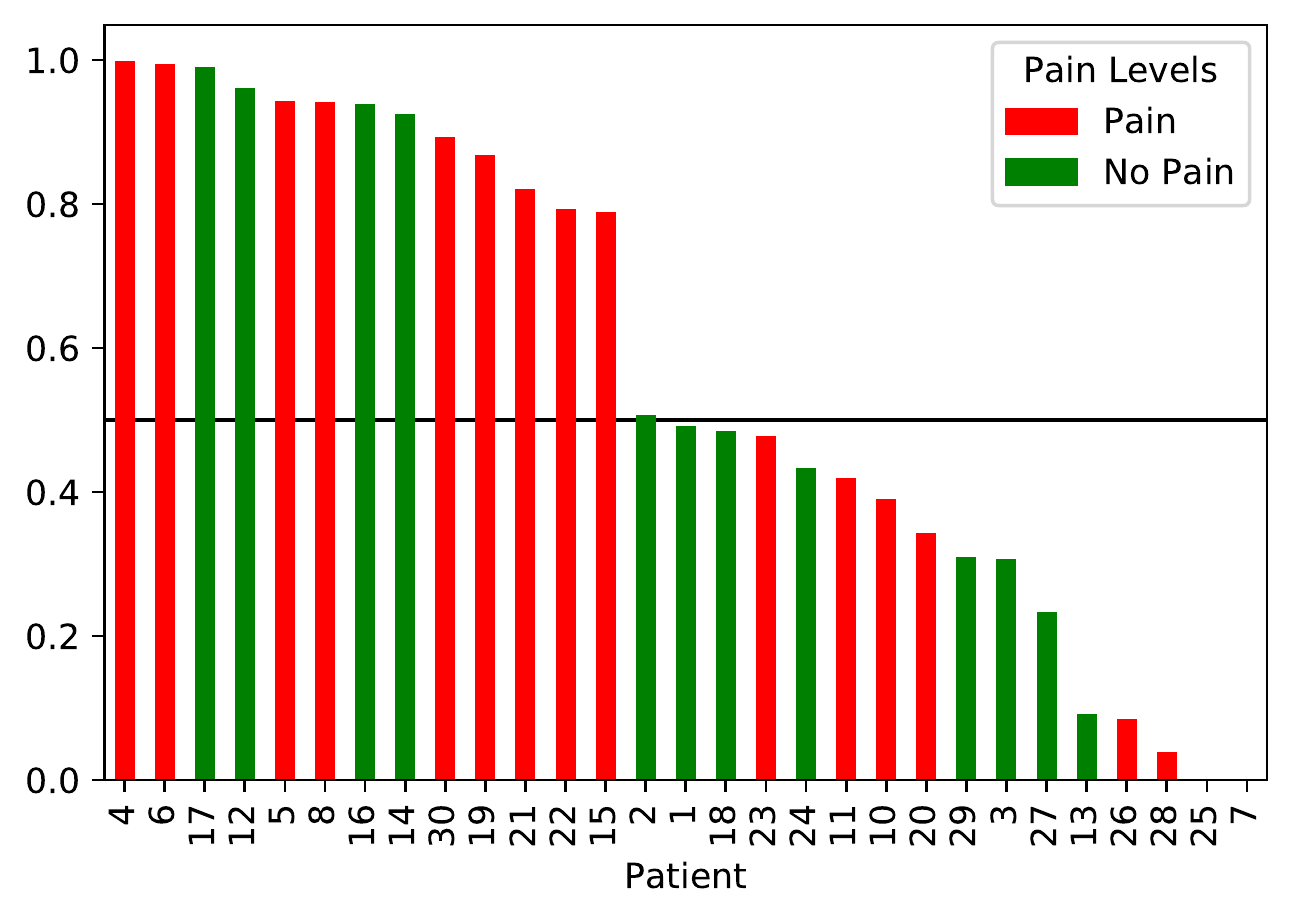}
         \caption{Faces, 48.3\%}
         \label{a}
     \end{subfigure}
    \begin{subfigure}[b]{0.24\textwidth}
        \centering
        \includegraphics[width=\textwidth]{images/LM_2.pdf}
        \caption{LM, 37.9\%}
        \label{b}
     \end{subfigure}
    \begin{subfigure}[b]{0.24\textwidth}
        \centering
        \includegraphics[width=\textwidth]{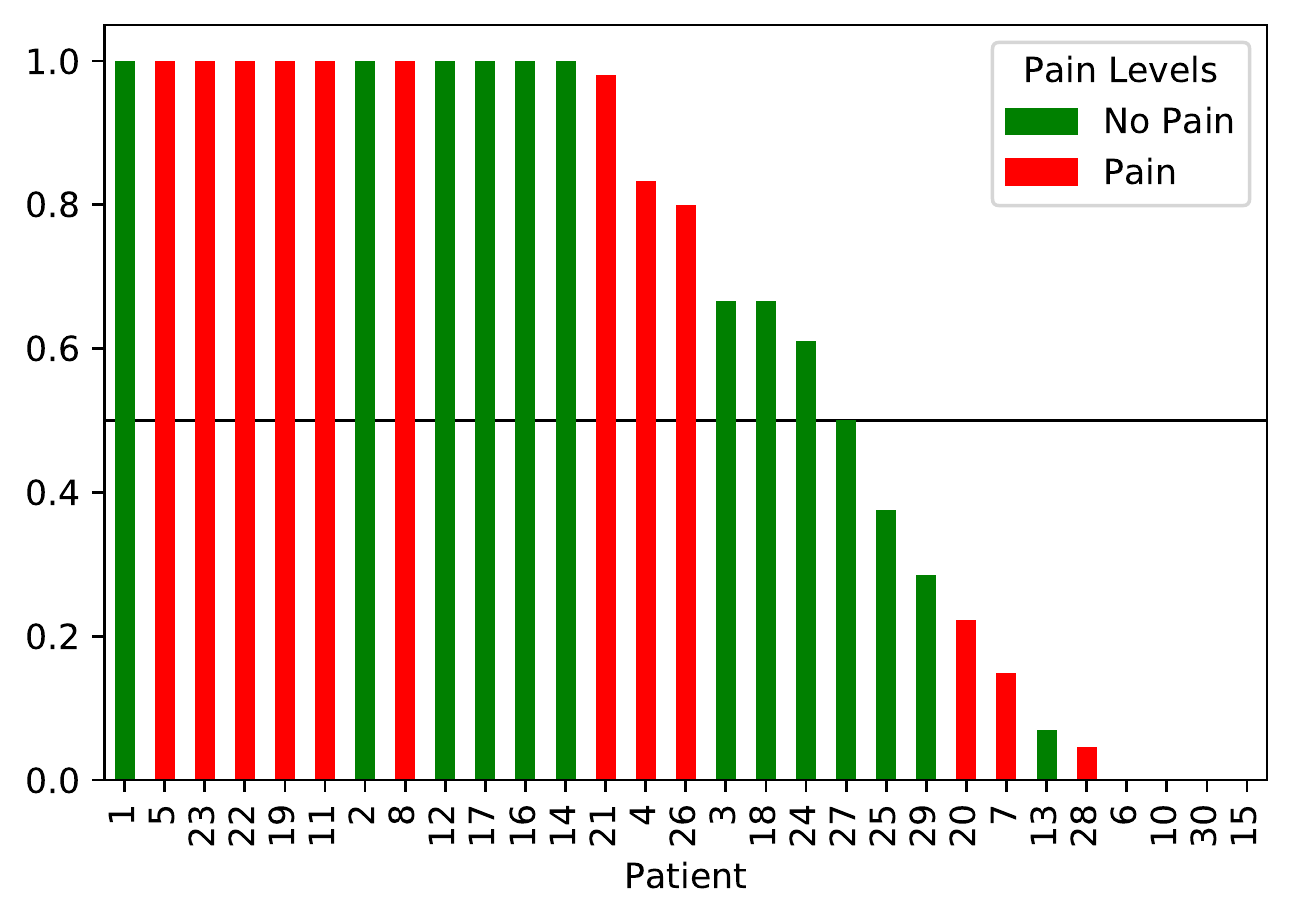}
        \caption{Pain, 65.5\%}
        \label{c}
     \end{subfigure}
    \begin{subfigure}[b]{0.24\textwidth}
        \centering
        \includegraphics[width=\textwidth]{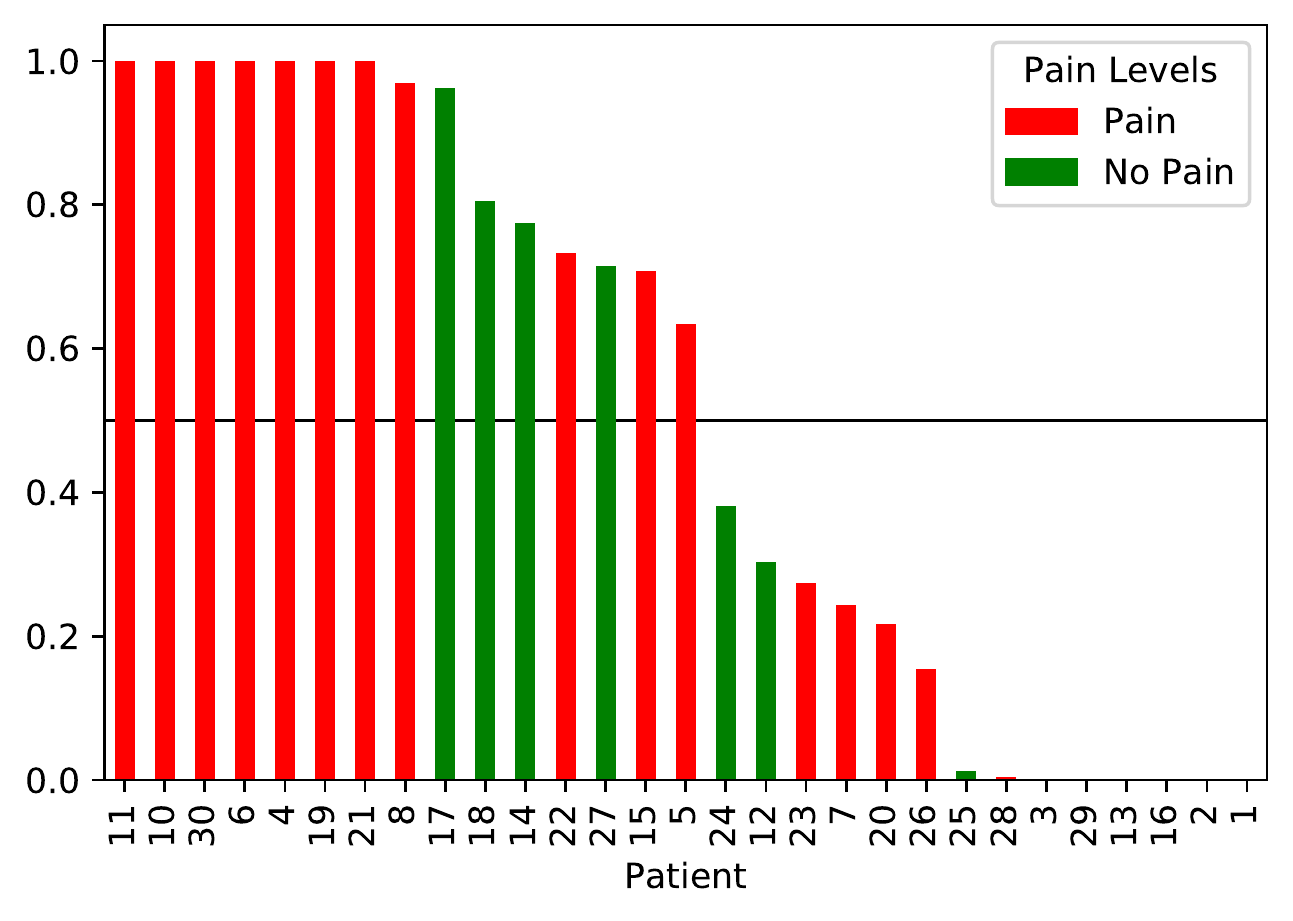}
        \caption{F1, 51.7\%}
        \label{d}
     \end{subfigure}
    \caption{
    \small{
    \textbf{Accuracy Scores per Test Patient by Model: Faces, Landmarks, Pain, and Fusion 1.} We show the resulting scores per test patient for the binary pain classifiers. Horizontal bar indicates 50\% accuracy. Percentages in sub-captions indicates the number of patients exceeding 50\% test accuracy. Notation: Faces=ResNet50-2-static; LM=Random Forest LM (landmarks); Pain=Random Forest Pain; F1=Fusion1. Best viewed in color and zoomed in.}}
    \label{faceplot1}
\end{figure}

\begin{figure}[ht!]
     \centering
     \begin{subfigure}[b]{0.24\textwidth}
         \centering
         \includegraphics[width=\textwidth]{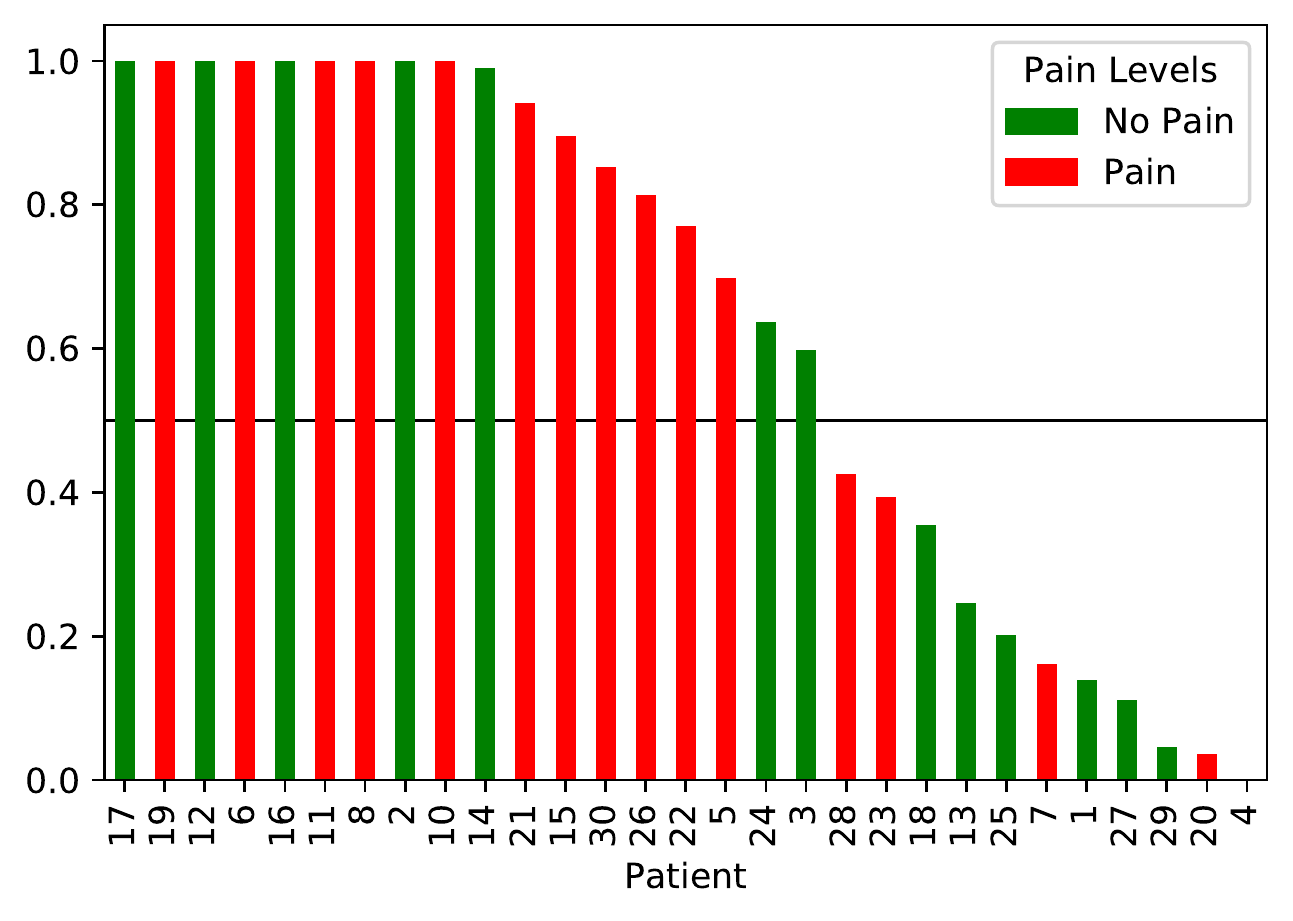}
         \caption{F2, 62.1\%}
         \label{A}
     \end{subfigure}
    \begin{subfigure}[b]{0.24\textwidth}
        \centering
        \includegraphics[width=\textwidth]{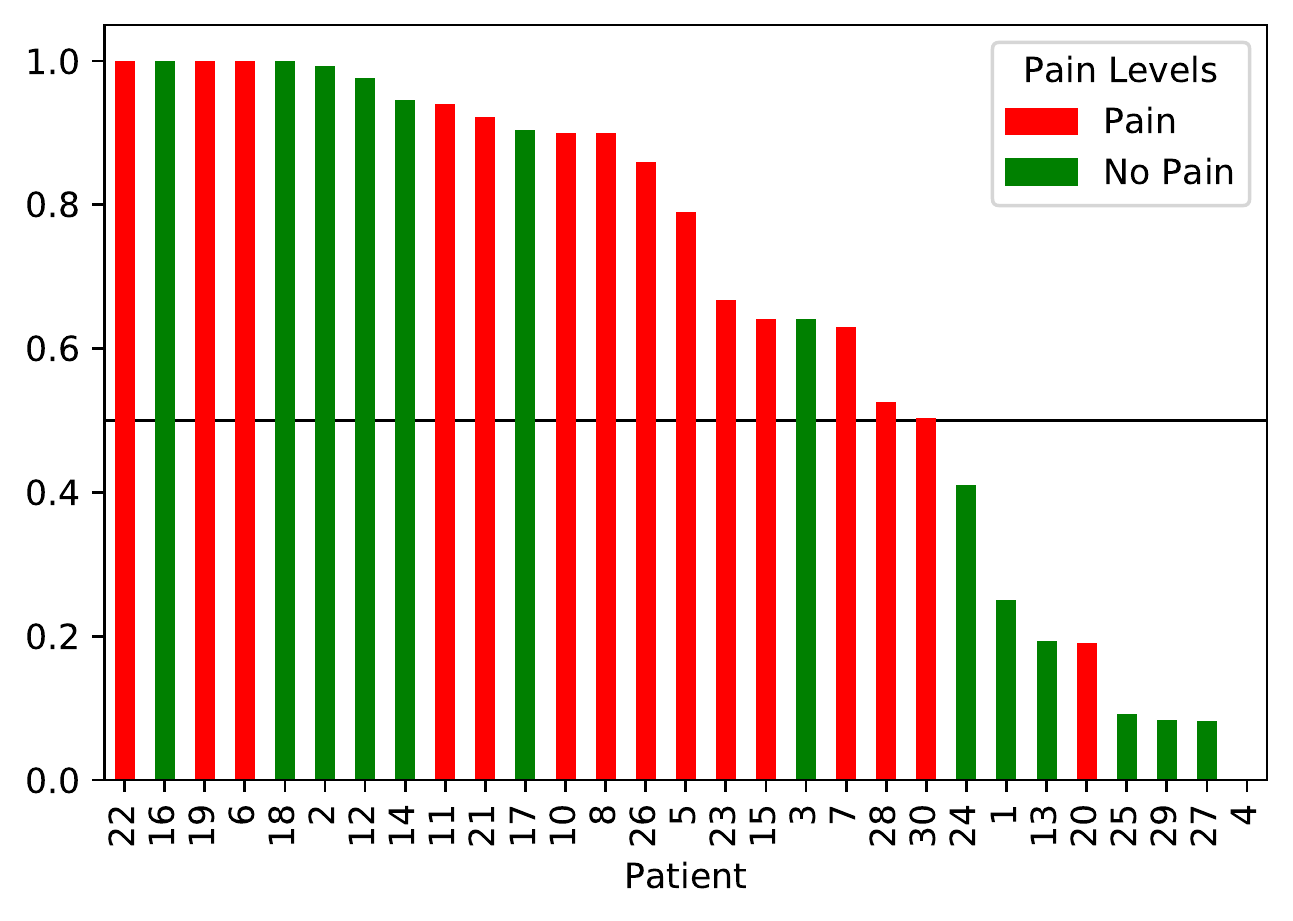}
        \caption{F3, 72.4\%}
        \label{B}
     \end{subfigure}
    \begin{subfigure}[b]{0.24\textwidth}
        \centering
        \includegraphics[width=\textwidth]{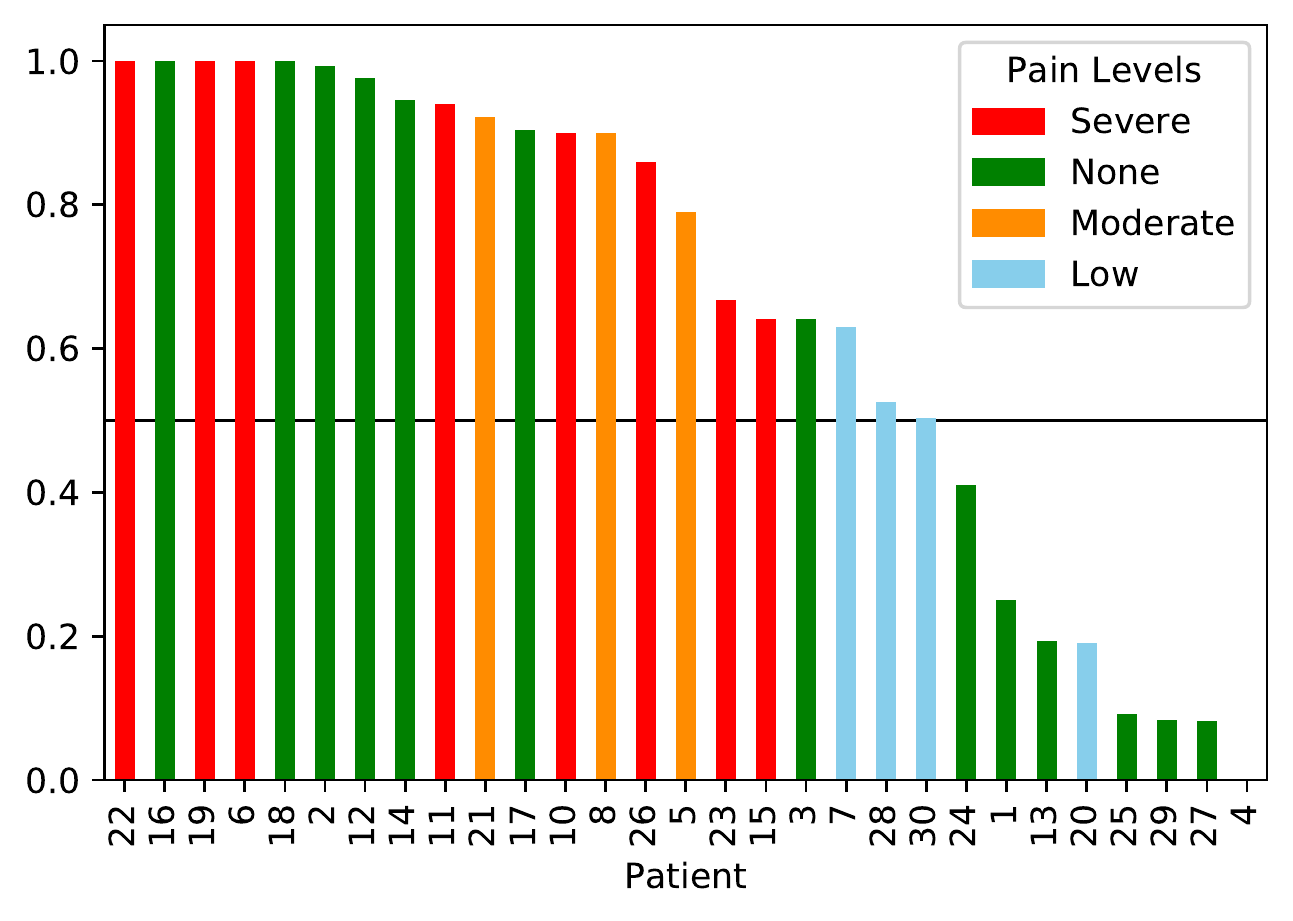}
        \caption{F3 - 4 Labels}
        \label{mmn6_4}
     \end{subfigure}
    \begin{subfigure}[b]{0.24\textwidth}
        \centering
        \includegraphics[width=\textwidth]{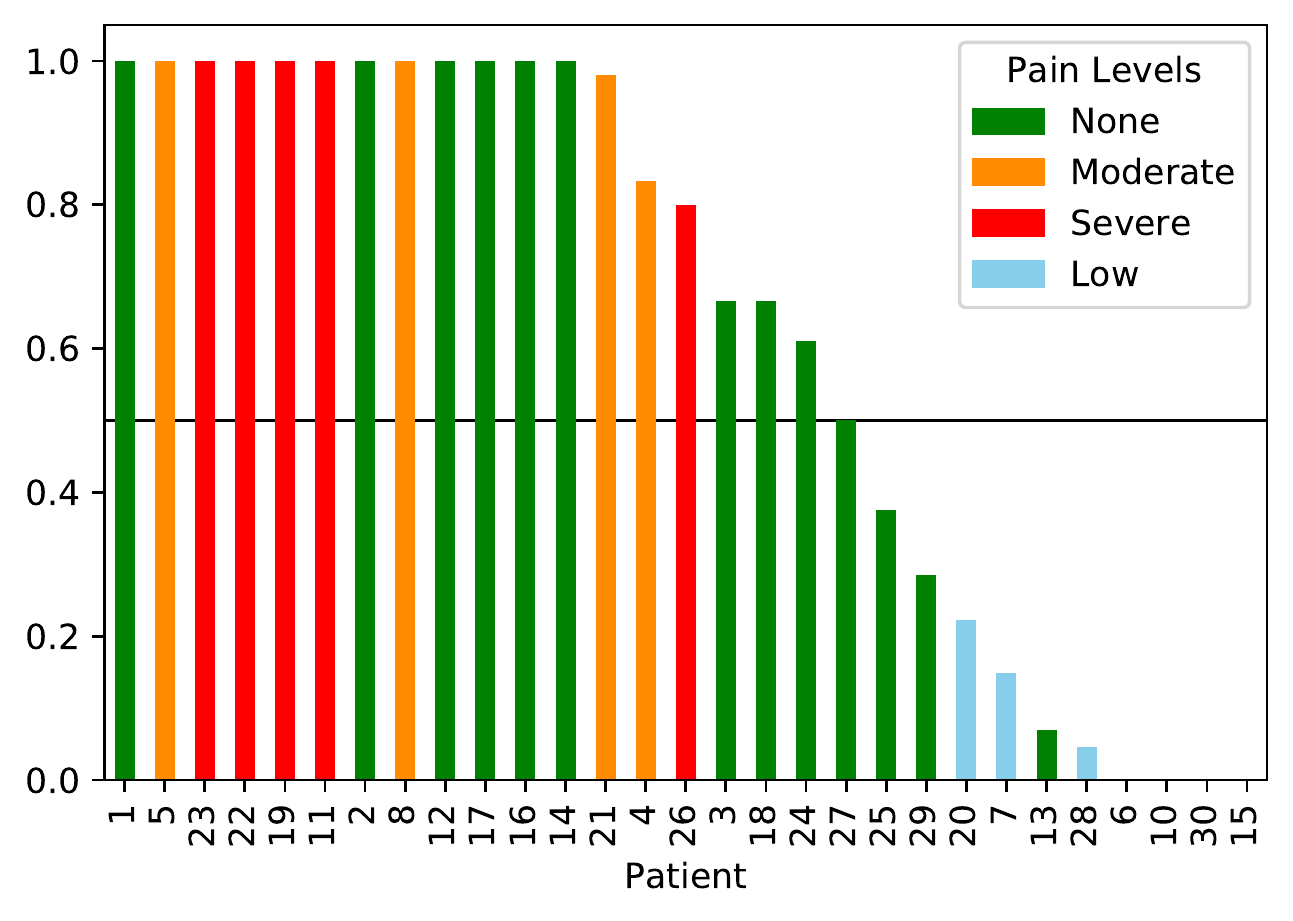}
        \caption{Pain - 4 Labels}
        \label{pain_4}
     \end{subfigure}
    \caption{
    \small{
    \textbf{Accuracy Scores per Test Patient by Model - Fusion 2, 3, and Pain, Visualized with 4 Original Labels.} We show the resulting scores per test patient for the binary pain classifiers. Horizontal bar indicates 50\% accuracy. Percentages in sub-captions indicates the number of patients exceeding 50\% test accuracy. Notation: Pain=Random Forest Pain; F1=Fusion1; F2=Fusion2; F3=Fusion3. Best viewed in color and zoomed in.}}
    \label{faceplot2}
\end{figure}

As a result, in Table \ref{results}, we show the mean accuracy computed for each ``Pain Target" level across all test patients (``No Pain" or ``Pain" for two levels, and ``None", ``Low", ``Moderate", or ``Severe" for four levels of pain). For example, in Experiment 1 ``ResNet50-4-static", the accuracy scores for all patients with ground truth pain labels of ``None", were averaged together to calculate the result of 0.583.  In Figures \ref{faceplot1} and \ref{faceplot2}, the bars are color-coded by the ground truth ``Pain Target" level for each patient. The y-axis is the accuracy predicted for the patient. For example, upon zooming into Figure \ref{a}, Patient 0029's (8 marks from the right of the x-axis) ground truth ``Pain Target" level is ``No Pain". However, the Experiment 1 static binary model only predicts it with 0.309 accuracy. 

\textbf{Experiment Results} The Experiment 6 multimodal network combining multiple features from the facial images, landmarks, pain scores, sex, skin, and timeframe labels performs the best for overall pain classification. Compared to training on a single modality alone (Experiments 1, 2, 3, 7), Experiment 6 (Fusion 3) shows the best overall class accuracy of 0.657 shown in Table \ref{results}. Fusion 3 also shows the highest accuracy for the ``Pain" level at 0.717.  Experiment 6 (Fusion 3) led to 72.4\% of test patients exceeding 50\% accuracy per frame as noted in Figure \ref{B}. However, it ties with Experiment 5 (Fusion2) and Experiment 2 (Random Forest PM) for 51.7\% of test patients achieving over 75\% accuracy per Figure \ref{A} and Figure \ref{d}. While the Random Forest pain model (Figure \ref{pain_4}) shows greater ``No Pain" accuracy, using only the self-reported nine pain scores does not detect the original ``Low" pain levels as well as the multimodal Fusion 3 model visualized in Figure \ref{mmn6_4} shown in blue bars.

Experiment 3 (Random Forest Pain) shows the highest ``No Pain" accuracy scores at 0.706 per Table \ref{results}. Adding the nine self-reported pain scores appears to boost accuracy, compared to training only on faces and landmarks per Experiment 4 (Fusion 1, 0.513) in Table \ref{results}. This is likely due to high correlations between the nine reported pain scores. Analysis shows strong Pearson correlation values exceeding 0.89 among activity, mood, work, enjoyment, and relationship scores. Continued analysis as more patients enroll in the study is required to understand the effect of the nine pain scores across all patients. The facial landmarks perform the worst in Experiment 2 (Random Forest LM) with only 37.9\% of test patients exceeding better than random at over 50\% accuracy per Figure \ref{b}. However, when adding landmarks to facial images in Experiment 4 (Fusion 1), several test patients completely fail to be detected (1, 2, 16, 13, 29, 3, 28, 25) per Figure \ref{d}. This may be consistent with recent research by ~\cite{asgarian2019limitations} who show that landmark detection declines when comparing different populations, such as older patients with dementia, to healthy adults.

\begin{table}[htbp!]
\centering
\caption{\small{\textbf{Experiment Results by Pain Level Accuracy.} ``LM" indicates facial landmarks.}}
\label{results}
\begin{adjustbox}{width=0.99\textwidth}
\begin{tabular}{@{}llllllll@{}}
\toprule
\textbf{Experiment} & \textbf{4-Class Model} & \textbf{Data} & \textbf{All Classes} & \textbf{None} & \textbf{Low} & \textbf{Moderate} & \textbf{Severe} \\ \midrule
Exp. 1 & ResNet50-4-static & Faces , only & 0.378 & 0.583 & 0.168 & 0.252 & 0.213 \\ \toprule
 & \textbf{2-Class Model} & \textbf{} & \textbf{All Classes} & \textbf{No Pain} & \textbf{Pain} &  &  \\ \toprule
Exp. 1 & ResNet50-2-static & Faces, only & 0.568 & 0.513 & 0.612 &  &  \\
Exp. 2 & Random Forest LM & Landmarks, only & 0.373 & 0.479 & 0.287 &  &  \\
Exp. 3 & Random Forest Pain & Pain Scores, only & 0.650 & \textbf{0.706} & 0.602 &  &  \\
Exp. 4 & Fusion 1 & Faces + Landmarks & 0.513 & 0.304 & 0.683 &  &  \\
Exp. 5 & Fusion 2 & Faces + Landmarks +   Pain Scores & 0.631 & 0.563 & 0.687 &  &  \\
Exp. 6 & Fusion 3 & Faces + Landmarks +   Pain Scores & \textbf{0.657} & 0.582 & \textbf{0.717} &  &  \\ \toprule
Exp. 7 & Static Audio & Audio, only & 0.456 & 0.645 & 0.303 &  &  \\ \bottomrule
\end{tabular}
\end{adjustbox}
\end{table}

\section{Discussion and Future Work}
Due to a variety of state-of-the-art techniques, we sought to implement simple models to demonstrate baseline results using fairly minimal preprocessing, transformations, and architectures. The results of our models show the dataset's difficulty. For comparison, acute pain detection studies have shown accuracy scores up to 82.4\% (hit rate) \cite{ashraf2009painful} using the UNBC-McMaster Shoulder Dataset and 95\% for multimodal infant pain detection using a custom dataset by ~\cite{zamzmi2016approach}. Chronic pain detection using psychological inventories have achieved 86.5\% (cross-validated balanced accuracy) using a support vector machine \cite{antonucci2020ensemble}. 

\textbf{Limitations} The first limitation of the dataset is the low number of currently enrolled patients at only 29 patients and the imbalance across pain levels. However, we observe that two new patients enroll into the study every month. As the number of patients grow, we expect a more balanced distribution of pain levels, sex, skin type, and increased volume of data, consistent with the cohort design indicated in Table \ref{recruitment}. However, medical datasets using active patient populations for major diseases such as cancer, are extremely scarce due to the time and review required for medical privacy and ethics. This differs greatly from current pain datasets that have recruited fairly healthy patients, who are not actively undergoing disease treatment. Due to the special sensitivity of the ISS study population, we believe that our current initial results offers important insights currently missing in the medical AI community.  

Next, despite the patient instructions to complete the submission in a quiet, brightly lit room with a white wall or background, many videos submitted varied in quality and resolution. The following examples observed in the dataset present challenges to machine learning: 1) Patient sitting in front of a door with signage in the background showing letters and numbers; 2) Patient occasionally wears a mask in some videos (due to Covid-19); 3) Patient records video in area of intense sunshine and glare causing reflection from various surfaces; 4) Patient records in a dark, shady room, leading to grainy resolution and video quality; 5) Patient speaks very quietly or muffled, making it difficult to hear the patient narrative; 6) Missing data as is the case of Patient 0009 and absent self-reported nine pain scores from Patient 0015. 

\textbf{Ethics} Publicly available acute pain datasets have lacked ethnic diversity. For example, the UNBC-McMaster Database \cite{prkachin2008structure} uses ethnicity as a demographic indicator where out of the original 129 patients (63 Male, 66 Female), a minimum of 13.2\% (17 patients) consisted of non-Caucasian ethnicity (refer to Table 1 of \cite{prkachin2008structure}). It may be less given how studies using the UNBC-McMaster dataset have access to data from only 25 out of 129 patients \cite{zhou2016recurrent, neshov2015pain, khan2013pain, bellantonio2016spatio}. The BioVid and MIntPAIN datasets provide no information about ethnicity and race \cite{walter2013biovid, haque2018deep}. EmoPAIN contains 22 patients (18 Caucasian, 3 African-American, 1 South-Asian) who are majority white \cite{aung2015automatic}. As a result, we sought to increase the diversity of enrolled patients by using cohorts that include sex and skin type specifications. While the Fitzpatrick Skin Type scale was originally developed for dermatological use, it has recently been criticized for its conflation with race and ethnicity \cite{ware2020racial}. It has been found to overestimate the prevalence of Type IV skin classification in African Americans \cite{pichon2010measuring}. The visual grouping of patients into lighter tones (Skin Types I - III) or darker tones (Skin Types IV - VI) may be too restrictive and biased in terms of broadening our diversity of patients.  As a result, the ISS dataset requires careful monitoring and a regular ethics review. 

\textbf{Future Work} The second phase of the study will analyze more diverse modalities. First, we will extract text from the audio files and explore its utility towards multimodal pain models. Next, since patients were unable to conduct in-clinic visits, we were unable to gather thermal imagery captured from a thermal camera stationed at the clinic. Thermal imagery offers insights into physiological states that is unseen on visible images alone \cite{ordun2020use}. Our intent is to generate paired visible-thermal datasets  as collected by the Iris, Eurecom, and Equinox datasets ~\cite{selinger2006appearance, iris, mallat2018benchmark}. Lastly, we estimate that after enrolling 112 patients, the ISS dataset will contain an additional 1,456 videos, 543,733 frames, and 3.8 hours of content. 


\bibliographystyle{IEEEtranN}
\small{
\bibliography{main}

\begin{thebibliography}{65}
\providecommand{\natexlab}[1]{#1}
\providecommand{\url}[1]{#1}
\csname url@samestyle\endcsname
\providecommand{\newblock}{\relax}
\providecommand{\bibinfo}[2]{#2}
\providecommand{\BIBentrySTDinterwordspacing}{\spaceskip=0pt\relax}
\providecommand{\BIBentryALTinterwordstretchfactor}{4}
\providecommand{\BIBentryALTinterwordspacing}{\spaceskip=\fontdimen2\font plus
\BIBentryALTinterwordstretchfactor\fontdimen3\font minus
  \fontdimen4\font\relax}
\providecommand{\BIBforeignlanguage}[2]{{%
\expandafter\ifx\csname l@#1\endcsname\relax
\typeout{** WARNING: IEEEtranN.bst: No hyphenation pattern has been}%
\typeout{** loaded for the language `#1'. Using the pattern for}%
\typeout{** the default language instead.}%
\else
\language=\csname l@#1\endcsname
\fi
#2}}
\providecommand{\BIBdecl}{\relax}
\BIBdecl

\bibitem[Van den Beuken-van Everdingen et~al.(2007)Van den Beuken-van
  Everdingen, De~Rijke, Kessels, Schouten, Van~Kleef, and
  Patijn]{van2007prevalence}
M.~Van den Beuken-van Everdingen, J.~De~Rijke, A.~Kessels, H.~Schouten,
  M.~Van~Kleef, and J.~Patijn, ``Prevalence of pain in patients with cancer: a
  systematic review of the past 40 years,'' \emph{Annals of oncology}, vol.~18,
  no.~9, pp. 1437--1449, 2007.

\bibitem[Sun et~al.(2007)Sun, Borneman, Ferrell, Piper, Koczywas, and
  Choi]{sun2007overcoming}
V.~C.-Y. Sun, T.~Borneman, B.~Ferrell, B.~Piper, M.~Koczywas, and K.~Choi,
  ``Overcoming barriers to cancer pain management: an institutional change
  model,'' \emph{Journal of pain and symptom management}, vol.~34, no.~4, pp.
  359--369, 2007.

\bibitem[Cleary(2000)]{cleary2000cancer}
J.~F. Cleary, ``Cancer pain management,'' \emph{Cancer Control}, vol.~7, no.~2,
  pp. 120--131, 2000.

\bibitem[L{\"o}tsch and Ultsch(2018)]{lotsch2018machine}
J.~L{\"o}tsch and A.~Ultsch, ``Machine learning in pain research,''
  \emph{Pain}, vol. 159, no.~4, p. 623, 2018.

\bibitem[Hu et~al.(2018)Hu, Kim, Ning, and Xu]{hu2018using}
B.~Hu, C.~Kim, X.~Ning, and X.~Xu, ``Using a deep learning network to recognise
  low back pain in static standing,'' \emph{Ergonomics}, vol.~61, no.~10, pp.
  1374--1381, 2018.

\bibitem[Prkachin and Solomon(2008)]{prkachin2008structure}
K.~M. Prkachin and P.~E. Solomon, ``The structure, reliability and validity of
  pain expression: Evidence from patients with shoulder pain,'' \emph{Pain},
  vol. 139, no.~2, pp. 267--274, 2008.

\bibitem[Lucey et~al.(2011)Lucey, Cohn, Prkachin, Solomon, and
  Matthews]{lucey2011painful}
P.~Lucey, J.~F. Cohn, K.~M. Prkachin, P.~E. Solomon, and I.~Matthews, ``Painful
  data: The unbc-mcmaster shoulder pain expression archive database,'' in
  \emph{2011 IEEE International Conference on Automatic Face \& Gesture
  Recognition (FG)}.\hskip 1em plus 0.5em minus 0.4em\relax IEEE, 2011, pp.
  57--64.

\bibitem[Walter et~al.(2013)Walter, Gruss, Ehleiter, Tan, Traue, Werner,
  Al-Hamadi, Crawcour, Andrade, and da~Silva]{walter2013biovid}
S.~Walter, S.~Gruss, H.~Ehleiter, J.~Tan, H.~C. Traue, P.~Werner, A.~Al-Hamadi,
  S.~Crawcour, A.~O. Andrade, and G.~M. da~Silva, ``The biovid heat pain
  database data for the advancement and systematic validation of an automated
  pain recognition system,'' in \emph{2013 IEEE international conference on
  cybernetics (CYBCO)}.\hskip 1em plus 0.5em minus 0.4em\relax IEEE, 2013, pp.
  128--131.

\bibitem[Haque et~al.(2018)Haque, Bautista, Noroozi, Kulkarni, Laursen, Irani,
  Bellantonio, Escalera, Anbarjafari, Nasrollahi, et~al.]{haque2018deep}
M.~A. Haque, R.~B. Bautista, F.~Noroozi, K.~Kulkarni, C.~B. Laursen, R.~Irani,
  M.~Bellantonio, S.~Escalera, G.~Anbarjafari, K.~Nasrollahi \emph{et~al.},
  ``Deep multimodal pain recognition: a database and comparison of
  spatio-temporal visual modalities,'' in \emph{2018 13th IEEE International
  Conference on Automatic Face \& Gesture Recognition (FG 2018)}.\hskip 1em
  plus 0.5em minus 0.4em\relax IEEE, 2018, pp. 250--257.

\bibitem[EkmanW and Friesen(1978)]{ekmanw1978facial}
P.~EkmanW and V.~Friesen, ``Facial action coding system manual,'' 1978.

\bibitem[Wilkie(1995)]{wilkie1995facial}
D.~J. Wilkie, ``Facial expressions of pain in lung cancer,'' \emph{Analgesia},
  vol.~1, no.~2, pp. 91--99, 1995.

\bibitem[(NCI)(2020)]{NIH}
\BIBentryALTinterwordspacing
N.~C.~I. (NCI), ``Machine learning to analyze facial imaging, voice and spoken
  language for the capture and classification of cancer/tumor pain - full text
  view,'' Jun 2020. [Online]. Available:
  \url{https://clinicaltrials.gov/ct2/show/NCT04442425}
\BIBentrySTDinterwordspacing

\bibitem[Pomeraniec et~al.(2020)Pomeraniec, Cha, Rule, Ordun, Hirlinger, and
  Gulley]{pomeraniec2020intelligent}
I.~J. Pomeraniec, A.~Cha, M.~Rule, C.~Ordun, M.~Hirlinger, and J.~Gulley,
  ``Intelligent sight \& sound: Machine learning to analyze facial imaging and
  voice mapping for classification of cancer pain,'' \emph{Neurosurgery},
  vol.~67, no. Supplement\_1, p. nyaa447\_524, 2020.

\bibitem[Cootes et~al.(2001)Cootes, Edwards, and Taylor]{cootes2001active}
T.~F. Cootes, G.~J. Edwards, and C.~J. Taylor, ``Active appearance models,''
  \emph{IEEE Transactions on pattern analysis and machine intelligence},
  vol.~23, no.~6, pp. 681--685, 2001.

\bibitem[Kartynnik et~al.(2019)Kartynnik, Ablavatski, Grishchenko, and
  Grundmann]{kartynnik2019real}
Y.~Kartynnik, A.~Ablavatski, I.~Grishchenko, and M.~Grundmann, ``Real-time
  facial surface geometry from monocular video on mobile gpus,'' \emph{arXiv
  preprint arXiv:1907.06724}, 2019.

\bibitem[Cleeland(1991)]{cleeland1991pain}
C.~S. Cleeland, ``Pain assessment in cancer,'' \emph{Effect of cancer on
  quality of life}, vol. 293, p. 305, 1991.

\bibitem[Cleeland and Ryan(1991)]{cleeland1991brief}
C.~S. Cleeland and K.~Ryan, ``The brief pain inventory,'' \emph{Pain Research
  Group}, pp. 143--147, 1991.

\bibitem[Kunz et~al.(2017)Kunz, Seuss, Hassan, Garbas, Siebers, Schmid,
  Sch{\"o}berl, and Lautenbacher]{kunz2017problems}
M.~Kunz, D.~Seuss, T.~Hassan, J.~U. Garbas, M.~Siebers, U.~Schmid,
  M.~Sch{\"o}berl, and S.~Lautenbacher, ``Problems of video-based pain
  detection in patients with dementia: a road map to an interdisciplinary
  solution,'' \emph{BMC geriatrics}, vol.~17, no.~1, pp. 1--8, 2017.

\bibitem[Lucey et~al.(2010)Lucey, Cohn, Matthews, Lucey, Sridharan, Howlett,
  and Prkachin]{lucey2010automatically}
P.~Lucey, J.~F. Cohn, I.~Matthews, S.~Lucey, S.~Sridharan, J.~Howlett, and
  K.~M. Prkachin, ``Automatically detecting pain in video through facial action
  units,'' \emph{IEEE Transactions on Systems, Man, and Cybernetics, Part B
  (Cybernetics)}, vol.~41, no.~3, pp. 664--674, 2010.

\bibitem[Zhou et~al.(2016)Zhou, Hong, Su, and Zhao]{zhou2016recurrent}
J.~Zhou, X.~Hong, F.~Su, and G.~Zhao, ``Recurrent convolutional neural network
  regression for continuous pain intensity estimation in video,'' in
  \emph{Proceedings of the IEEE conference on computer vision and pattern
  recognition workshops}, 2016, pp. 84--92.

\bibitem[Neshov and Manolova(2015)]{neshov2015pain}
N.~Neshov and A.~Manolova, ``Pain detection from facial characteristics using
  supervised descent method,'' in \emph{2015 IEEE 8th International Conference
  on Intelligent Data Acquisition and Advanced Computing Systems: Technology
  and Applications (IDAACS)}, vol.~1.\hskip 1em plus 0.5em minus 0.4em\relax
  IEEE, 2015, pp. 251--256.

\bibitem[Khan et~al.(2013)Khan, Meyer, Konik, and Bouakaz]{khan2013pain}
R.~A. Khan, A.~Meyer, H.~Konik, and S.~Bouakaz, ``Pain detection through shape
  and appearance features,'' in \emph{2013 IEEE International Conference on
  Multimedia and Expo (ICME)}.\hskip 1em plus 0.5em minus 0.4em\relax IEEE,
  2013, pp. 1--6.

\bibitem[Lo~Presti and La~Cascia(2015)]{lo2015using}
L.~Lo~Presti and M.~La~Cascia, ``Using hankel matrices for dynamics-based
  facial emotion recognition and pain detection,'' in \emph{Proceedings of the
  IEEE conference on computer vision and pattern recognition workshops}, 2015,
  pp. 26--33.

\bibitem[Bellantonio et~al.(2016)Bellantonio, Haque, Rodriguez, Nasrollahi,
  Telve, Escalera, Gonzalez, Moeslund, Rasti, and
  Anbarjafari]{bellantonio2016spatio}
M.~Bellantonio, M.~A. Haque, P.~Rodriguez, K.~Nasrollahi, T.~Telve,
  S.~Escalera, J.~Gonzalez, T.~B. Moeslund, P.~Rasti, and G.~Anbarjafari,
  ``Spatio-temporal pain recognition in cnn-based super-resolved facial
  images,'' in \emph{Video Analytics. Face and Facial Expression Recognition
  and Audience Measurement}.\hskip 1em plus 0.5em minus 0.4em\relax Springer,
  2016, pp. 151--162.

\bibitem[Hassan et~al.(2019)Hassan, Seu{\ss}, Wollenberg, Weitz, Kunz,
  Lautenbacher, Garbas, and Schmid]{hassan2019automatic}
T.~Hassan, D.~Seu{\ss}, J.~Wollenberg, K.~Weitz, M.~Kunz, S.~Lautenbacher,
  J.-U. Garbas, and U.~Schmid, ``Automatic detection of pain from facial
  expressions: a survey,'' \emph{IEEE transactions on pattern analysis and
  machine intelligence}, vol.~43, no.~6, pp. 1815--1831, 2019.

\bibitem[Ashraf et~al.(2009)Ashraf, Lucey, Cohn, Chen, Ambadar, Prkachin, and
  Solomon]{ashraf2009painful}
A.~B. Ashraf, S.~Lucey, J.~F. Cohn, T.~Chen, Z.~Ambadar, K.~M. Prkachin, and
  P.~E. Solomon, ``The painful face--pain expression recognition using active
  appearance models,'' \emph{Image and vision computing}, vol.~27, no.~12, pp.
  1788--1796, 2009.

\bibitem[Bartlett et~al.(2006)Bartlett, Littlewort, Frank, Lainscsek, Fasel,
  Movellan, et~al.]{bartlett2006automatic}
M.~S. Bartlett, G.~Littlewort, M.~G. Frank, C.~Lainscsek, I.~R. Fasel, J.~R.
  Movellan \emph{et~al.}, ``Automatic recognition of facial actions in
  spontaneous expressions.'' \emph{J. Multim.}, vol.~1, no.~6, pp. 22--35,
  2006.

\bibitem[Gruss et~al.(2019)Gruss, Geiger, Werner, Wilhelm, Traue, Al-Hamadi,
  and Walter]{gruss2019multi}
S.~Gruss, M.~Geiger, P.~Werner, O.~Wilhelm, H.~C. Traue, A.~Al-Hamadi, and
  S.~Walter, ``Multi-modal signals for analyzing pain responses to thermal and
  electrical stimuli,'' \emph{JoVE (Journal of Visualized Experiments)}, no.
  146, p. e59057, 2019.

\bibitem[Werner et~al.(2014)Werner, Al-Hamadi, Niese, Walter, Gruss, and
  Traue]{werner2014automatic}
P.~Werner, A.~Al-Hamadi, R.~Niese, S.~Walter, S.~Gruss, and H.~C. Traue,
  ``Automatic pain recognition from video and biomedical signals,'' in
  \emph{2014 22nd International Conference on Pattern Recognition}.\hskip 1em
  plus 0.5em minus 0.4em\relax IEEE, 2014, pp. 4582--4587.

\bibitem[Aung et~al.(2015)Aung, Kaltwang, Romera-Paredes, Martinez, Singh,
  Cella, Valstar, Meng, Kemp, Shafizadeh, et~al.]{aung2015automatic}
M.~S. Aung, S.~Kaltwang, B.~Romera-Paredes, B.~Martinez, A.~Singh, M.~Cella,
  M.~Valstar, H.~Meng, A.~Kemp, M.~Shafizadeh \emph{et~al.}, ``The automatic
  detection of chronic pain-related expression: requirements, challenges and
  the multimodal emopain dataset,'' \emph{IEEE transactions on affective
  computing}, vol.~7, no.~4, pp. 435--451, 2015.

\bibitem[Zamzmi et~al.(2016)Zamzmi, Pai, Goldgof, Kasturi, Ashmeade, and
  Sun]{zamzmi2016approach}
G.~Zamzmi, C.-Y. Pai, D.~Goldgof, R.~Kasturi, T.~Ashmeade, and Y.~Sun, ``An
  approach for automated multimodal analysis of infants' pain,'' in \emph{2016
  23rd International Conference on Pattern Recognition (ICPR)}.\hskip 1em plus
  0.5em minus 0.4em\relax IEEE, 2016, pp. 4148--4153.

\bibitem[Zamzmi et~al.(2018)Zamzmi, Goldgof, Kasturi, and
  Sun]{zamzmi2018neonatal}
G.~Zamzmi, D.~Goldgof, R.~Kasturi, and Y.~Sun, ``Neonatal pain expression
  recognition using transfer learning,'' \emph{arXiv preprint
  arXiv:1807.01631}, 2018.

\bibitem[Rezaei et~al.(2020)Rezaei, Moturu, Zhao, Prkachin, Hadjistavropoulos,
  and Taati]{rezaei2020unobtrusive}
S.~Rezaei, A.~Moturu, S.~Zhao, K.~M. Prkachin, T.~Hadjistavropoulos, and
  B.~Taati, ``Unobtrusive pain monitoring in older adults with dementia using
  pairwise and contrastive training,'' \emph{IEEE Journal of Biomedical and
  Health Informatics}, vol.~25, no.~5, pp. 1450--1462, 2020.

\bibitem[Asgarian et~al.(2019)Asgarian, Zhao, Ashraf, Browne, Prkachin,
  Mihailidis, Hadjistavropoulos, and Taati]{asgarian2019limitations}
A.~Asgarian, S.~Zhao, A.~B. Ashraf, M.~E. Browne, K.~M. Prkachin,
  A.~Mihailidis, T.~Hadjistavropoulos, and B.~Taati, ``Limitations and biases
  in facial landmark detection d an empirical study on older adults with
  dementia.'' in \emph{CVPR Workshops}, 2019, pp. 28--36.

\bibitem[Salekin et~al.(2021)Salekin, Zamzmi, Goldgof, Kasturi, Ho, and
  Sun]{salekin2021multimodal}
M.~S. Salekin, G.~Zamzmi, D.~Goldgof, R.~Kasturi, T.~Ho, and Y.~Sun,
  ``Multimodal spatio-temporal deep learning approach for neonatal
  postoperative pain assessment,'' \emph{Computers in Biology and Medicine},
  vol. 129, p. 104150, 2021.

\bibitem[Salekin et~al.(2019)Salekin, Zamzmi, et~al.]{salekin2019multi}
M.~S. Salekin, G.~Zamzmi \emph{et~al.}, ``Multi-channel neural network for
  assessing neonatal pain from videos,'' in \emph{2019 IEEE International
  Conference on Systems, Man and Cybernetics (SMC)}.\hskip 1em plus 0.5em minus
  0.4em\relax IEEE, 2019, pp. 1551--1556.

\bibitem[Othman et~al.(2021)Othman, Werner, Saxen, Al-Hamadi, Gruss, and
  Walter]{othman2021automatic}
E.~Othman, P.~Werner, F.~Saxen, A.~Al-Hamadi, S.~Gruss, and S.~Walter,
  ``Automatic vs. human recognition of pain intensity from facial expression on
  the x-ite pain database,'' \emph{Sensors}, vol.~21, no.~9, p. 3273, 2021.

\bibitem[Goldsmith et~al.(2012)Goldsmith, Katz, Bilchrest, Paller, Leffel, and
  Wolff]{goldsmith2012fitzpatrick}
L.~Goldsmith, S.~Katz, B.~Bilchrest, A.~Paller, D.~Leffel, and K.~Wolff,
  ``Fitzpatrick’s dermatology in general medicine, ed,'' \emph{McGrawHill
  Medical}, pp. 2421--2429, 2012.

\bibitem[Haefeli and Elfering(2006)]{haefeli2006pain}
M.~Haefeli and A.~Elfering, ``Pain assessment,'' \emph{European Spine Journal},
  vol.~15, no.~1, pp. S17--S24, 2006.

\bibitem[Apolin{\'a}rio-Hagen et~al.(2018)Apolin{\'a}rio-Hagen, Fritsche,
  Bierhals, and Salewski]{apolinario2018improving}
J.~Apolin{\'a}rio-Hagen, L.~Fritsche, C.~Bierhals, and C.~Salewski, ``Improving
  attitudes toward e-mental health services in the general population via
  psychoeducational information material: A randomized controlled trial,''
  \emph{Internet interventions}, vol.~12, pp. 141--149, 2018.

\bibitem[Fink-Lamotte et~al.(2020)Fink-Lamotte, Widmann, Fader, and
  Exner]{fink2020interpretation}
J.~Fink-Lamotte, A.~Widmann, J.~Fader, and C.~Exner, ``Interpretation bias and
  contamination-based obsessive-compulsive symptoms influence emotional
  intensity related to disgust and fear,'' \emph{PloS one}, vol.~15, no.~4, p.
  e0232362, 2020.

\bibitem[Livesay and Porter(1994)]{livesay1994emg}
J.~R. Livesay and T.~Porter, ``Emg and cardiovascular responses to emotionally
  provocative photographs and text,'' \emph{Perceptual and motor skills},
  vol.~79, no.~1, pp. 579--594, 1994.

\bibitem[MacLachlan and Beech(1997)]{maclachlan1997sarah}
P.~MacLachlan and L.~Beech, \emph{Sarah, plain and tall}.\hskip 1em plus 0.5em
  minus 0.4em\relax Scholastic Inc., 1997.

\bibitem[DiCamillo(2009)]{dicamillo2009because}
K.~DiCamillo, \emph{Because of Winn-Dixie}.\hskip 1em plus 0.5em minus
  0.4em\relax Candlewick Press, 2009.

\bibitem[Geisel(1990)]{geisel1990oh}
T.~S. Geisel, \emph{Oh, the places you'll go!}\hskip 1em plus 0.5em minus
  0.4em\relax Random House Books for Young Readers, 1990.

\bibitem[Zhang et~al.(2016)Zhang, Zhang, Li, and Qiao]{zhang2016joint}
K.~Zhang, Z.~Zhang, Z.~Li, and Y.~Qiao, ``Joint face detection and alignment
  using multitask cascaded convolutional networks,'' \emph{IEEE Signal
  Processing Letters}, vol.~23, no.~10, pp. 1499--1503, 2016.

\bibitem[Lugaresi et~al.(2019)Lugaresi, Tang, Nash, McClanahan, Uboweja, Hays,
  Zhang, Chang, Yong, Lee, et~al.]{lugaresi2019mediapipe}
C.~Lugaresi, J.~Tang, H.~Nash, C.~McClanahan, E.~Uboweja, M.~Hays, F.~Zhang,
  C.-L. Chang, M.~G. Yong, J.~Lee \emph{et~al.}, ``Mediapipe: A framework for
  building perception pipelines,'' \emph{arXiv preprint arXiv:1906.08172},
  2019.

\bibitem[McFee et~al.(2021)McFee, Metsai, McVicar,
  et~al.]{brian_mcfee_2021_4792298}
\BIBentryALTinterwordspacing
B.~McFee, A.~Metsai, M.~McVicar \emph{et~al.}, ``librosa/librosa: 0.8.1rc2,''
  May 2021. [Online]. Available: \url{https://doi.org/10.5281/zenodo.4792298}
\BIBentrySTDinterwordspacing

\bibitem[Chen et~al.(2018)Chen, Ansari, and Wilkie]{chen2018automated}
Z.~Chen, R.~Ansari, and D.~Wilkie, ``Automated pain detection from facial
  expressions using facs: A review,'' \emph{arXiv preprint arXiv:1811.07988},
  2018.

\bibitem[Barrett et~al.(2019)Barrett, Adolphs, Marsella, Martinez, and
  Pollak]{barrett2019emotional}
L.~F. Barrett, R.~Adolphs, S.~Marsella, A.~M. Martinez, and S.~D. Pollak,
  ``Emotional expressions reconsidered: Challenges to inferring emotion from
  human facial movements,'' \emph{Psychological science in the public
  interest}, vol.~20, no.~1, pp. 1--68, 2019.

\bibitem[Crawford(2021)]{crawford2021atlas}
K.~Crawford, \emph{The Atlas of AI}.\hskip 1em plus 0.5em minus 0.4em\relax
  Yale University Press, 2021.

\bibitem[Prkachin(1992)]{prkachin1992consistency}
K.~M. Prkachin, ``The consistency of facial expressions of pain: a comparison
  across modalities,'' \emph{Pain}, vol.~51, no.~3, pp. 297--306, 1992.

\bibitem[Rodriguez et~al.(2017)Rodriguez, Cucurull, Gonz{\`a}lez, Gonfaus,
  Nasrollahi, Moeslund, and Roca]{rodriguez2017deep}
P.~Rodriguez, G.~Cucurull, J.~Gonz{\`a}lez, J.~M. Gonfaus, K.~Nasrollahi, T.~B.
  Moeslund, and F.~X. Roca, ``Deep pain: Exploiting long short-term memory
  networks for facial expression classification,'' \emph{IEEE transactions on
  cybernetics}, 2017.

\bibitem[He et~al.(2016)He, Zhang, Ren, and Sun]{he2016deep}
K.~He, X.~Zhang, S.~Ren, and J.~Sun, ``Deep residual learning for image
  recognition,'' in \emph{Proceedings of the IEEE conference on computer vision
  and pattern recognition}, 2016, pp. 770--778.

\bibitem[Deng et~al.(2009)Deng, Dong, Socher, Li, Li, and
  Fei-Fei]{deng2009imagenet}
J.~Deng, W.~Dong, R.~Socher, L.-J. Li, K.~Li, and L.~Fei-Fei, ``Imagenet: A
  large-scale hierarchical image database,'' in \emph{2009 IEEE conference on
  computer vision and pattern recognition}.\hskip 1em plus 0.5em minus
  0.4em\relax Ieee, 2009, pp. 248--255.

\bibitem[Breiman(2001)]{breiman2001random}
L.~Breiman, ``Random forests,'' \emph{Machine learning}, vol.~45, no.~1, pp.
  5--32, 2001.

\bibitem[Huang et~al.(2020)Huang, Pareek, Seyyedi, Banerjee, and
  Lungren]{huang2020fusion}
S.-C. Huang, A.~Pareek, S.~Seyyedi, I.~Banerjee, and M.~P. Lungren, ``Fusion of
  medical imaging and electronic health records using deep learning: a
  systematic review and implementation guidelines,'' \emph{NPJ digital
  medicine}, vol.~3, no.~1, pp. 1--9, 2020.

\bibitem[Antonucci et~al.(2020)Antonucci, Taurino, Laera, Taurisano, Losole,
  Lutricuso, Abbatantuono, Giglio, De~Caro, Varrassi,
  et~al.]{antonucci2020ensemble}
L.~A. Antonucci, A.~Taurino, D.~Laera, P.~Taurisano, J.~Losole, S.~Lutricuso,
  C.~Abbatantuono, M.~Giglio, M.~F. De~Caro, G.~Varrassi \emph{et~al.}, ``An
  ensemble of psychological and physical health indices discriminates between
  individuals with chronic pain and healthy controls with high reliability: A
  machine learning study,'' \emph{Pain and Therapy}, vol.~9, no.~2, pp.
  601--614, 2020.

\bibitem[Ware et~al.(2020)Ware, Dawson, Shinohara, and Taylor]{ware2020racial}
O.~R. Ware, J.~E. Dawson, M.~M. Shinohara, and S.~C. Taylor, ``Racial
  limitations of fitzpatrick skin type,'' \emph{Cutis}, vol. 105, no.~2, pp.
  77--80, 2020.

\bibitem[Pichon et~al.(2010)Pichon, Landrine, Corral, Hao, Mayer, and
  Hoerster]{pichon2010measuring}
L.~C. Pichon, H.~Landrine, I.~Corral, Y.~Hao, J.~A. Mayer, and K.~D. Hoerster,
  ``Measuring skin cancer risk in african americans: is the fitzpatrick skin
  type classification scale culturally sensitive,'' \emph{Ethn Dis}, vol.~20,
  no.~2, pp. 174--179, 2010.

\bibitem[Ordun et~al.(2020)Ordun, Raff, and Purushotham]{ordun2020use}
C.~Ordun, E.~Raff, and S.~Purushotham, ``The use of ai for thermal emotion
  recognition: A review of problems and limitations in standard design and
  data,'' \emph{arXiv preprint arXiv:2009.10589}, 2020.

\bibitem[Selinger et~al.(2006)]{selinger2006appearance}
A.~Selinger \emph{et~al.}, ``Appearance-based facial recognition using visible
  and thermal imagery: a comparative study,'' Equinox Corp., Tech. Rep., 2006.

\bibitem[iri()]{iris}
\BIBentryALTinterwordspacing
Otcbvs benchmark dataset collection. [Online]. Available:
  \url{http://vcipl-okstate.org/pbvs/bench/}
\BIBentrySTDinterwordspacing

\bibitem[Mallat et~al.(2018)]{mallat2018benchmark}
K.~Mallat \emph{et~al.}, ``A benchmark database of visible and thermal paired
  face images across multiple variations,'' in \emph{BIOSIG}.\hskip 1em plus
  0.5em minus 0.4em\relax IEEE, 2018, pp. 1--5.

\bibitem[Buolamwini and Gebru(2018)]{buolamwini2018gender}
J.~Buolamwini and T.~Gebru, ``Gender shades: Intersectional accuracy
  disparities in commercial gender classification,'' in \emph{Conference on
  fairness, accountability and transparency}.\hskip 1em plus 0.5em minus
  0.4em\relax PMLR, 2018, pp. 77--91.

\end{thebibliography}
}

\newpage
\section*{Checklist}

\begin{enumerate}
\item For all authors...
\begin{enumerate}
  \item Do the main claims made in the abstract and introduction accurately reflect the paper's contributions and scope?
    \textcolor{blue}{[Yes]}
  \item Did you describe the limitations of your work?
    \textcolor{blue}{[Yes]}
  \item Did you discuss any potential negative societal impacts of your work?
    \textcolor{blue}{[Yes]}
  \item Have you read the ethics review guidelines and ensured that your paper conforms to them?
    \textcolor{blue}{[Yes]}
\end{enumerate}

\item If you are including theoretical results...
\begin{enumerate}
  \item Did you state the full set of assumptions of all theoretical results?
    \textcolor{gray}{[N/A]}
	\item Did you include complete proofs of all theoretical results?
    \textcolor{gray}{[N/A]}
\end{enumerate}

\item If you ran experiments (e.g. for benchmarks)...
\begin{enumerate}
  \item Did you include the code, data, and instructions needed to reproduce the main experimental results (either in the supplemental material or as a URL)?
    \textcolor{red}{[No]}
  \item Did you specify all the training details (e.g., data splits, hyperparameters, how they were chosen)?
    \textcolor{blue}{[Yes]}
  \item Did you report error bars (e.g., with respect to the random seed after running experiments multiple times)?
    \textcolor{gray}{[N/A]}
  \item Did you include the total amount of compute and the type of resources used (e.g., type of GPUs, internal cluster, or cloud provider)?
    \textcolor{blue}{[Yes]}
\end{enumerate}

\item If you are using existing assets (e.g., code, data, models) or curating/releasing new assets...
\begin{enumerate}
  \item If your work uses existing assets, did you cite the creators?
    \textcolor{gray}{[N/A]}
  \item Did you mention the license of the assets?
    \textcolor{blue}{[Yes]}
  \item Did you include any new assets either in the supplemental material or as a URL?
    \textcolor{gray}{[N/A]}
  \item Did you discuss whether and how consent was obtained from people whose data you're using/curating?
   \textcolor{blue}{[Yes]}
  \item Did you discuss whether the data you are using/curating contains personally identifiable information or offensive content?
    \textcolor{blue}{[Yes]}
\end{enumerate}

\item If you used crowdsourcing or conducted research with human subjects...
\begin{enumerate}
  \item Did you include the full text of instructions given to patients and screenshots, if applicable?
    \textcolor{blue}{[Yes]}
  \item Did you describe any potential patient risks, with links to Institutional Review Board (IRB) approvals, if applicable?
    \textcolor{blue}{[Yes]}
  \item Did you include the estimated hourly wage paid to patients and the total amount spent on patient compensation?
    \textcolor{blue}{[Yes]}
\end{enumerate}

\end{enumerate}

\newpage
\appendix

\section{Author Statement}
Upon future release of the data, NIH bears all responsibility in case of violation of rights, etc., and confirmation of the data license. Wherever possible, all data are de-identified; however, data consisting of facial image frames are inherently classified as Personally Identifying Information (PII) and as such are unable to be de-identified.  All data is stored under the control of an internal NIH database accessible to other NIH researchers for future research. NIH will share access to outside researchers under their discretion. As part of the informed consent process, governed by the NIH Institutional Review Board (IRB), patients have given permission for their data to be shared under these restricted conditions for future research.

\section{Acknowledgements}
We are grateful for the support and advisement received on the development of the ISS Study Design and Protocol. We acknowledge Mason Rule and Susan Wroblewski at the National Institutes of Health, and Lauren Neal, Jeremy Walsh, and John Larson at Booz Allen Hamilton.

\section{Ethical Considerations}
The generation of this dataset has brought multiple ethical concerns to our attention. This section is intended to share our observation of ethical concerns to raise awareness of the issues and stimulate conversation around potential mitigation. We will discuss the following ethical concerns: 
\begin{itemize}
\itemsep -0.2em 
    \item Use of Fitzpatrick Skin Scale
    \item Data collection biases
    \item Video lighting
    \item Camera quality and availability
    \item Facial Recognition Algorithm Bias and Surveillance
\end{itemize}

\textbf{Use of Fitzpatrick Skin Scale}
The Fitzpatrick Skin Scale was originally developed for dermatological use to measure the sensitivity of skin burn during phototherapy and has recently been critiqued for its conflation with race and ethnicity \cite{ware2020racial}. Further, the scale has been found to overestimate the prevalence of Type IV skin classification in African Americans \cite{pichon2010measuring}. In a sample of 2,086 California Black adults, 59\% categorized themselves as not applicable to any of the four skin types and only 26.8\% identified as Type IV, the darkest skin type. The authors assert that the Fitzpatrick Skin Scale is more in line with Caucasian experiences of sun-reactivity and not Asians, Arabs, and African Americans. The visual grouping of patients into ``Dark" (Skin Types IV - VI) or ``Light" (Skin Types I - III) may be too restrictive and biased in terms of broadening our diversity of patients.
 
\textbf{Data Collection Biases} As we collected videos from patient owned devices (e.g. mobile phones, tablets, etc), we immediately noticed wide ranges of video characteristics. Videos were captured using various illuminations (e.g., front lighting vs. backlighting), varying instances of whether a patient is (or is not) wearing glasses, and masks are occasionally worn in videos due to the global Covid-19 pandemic. We also noticed a variety of video quality and resolution which is most attributed to the condition of the physical device hardware used to capture the video. As we continue to use this data to train models, it may be useful to quantify the extent of these many video characteristics and store them as metadata. Having the video characteristics stored as metadata would allow for bias detection techniques to be for identification and (if possible) mitigation of data biases. Although an analysis of socioeconomic backgrounds of each patient has not been conducted, it is fathomable that those in lower socioeconomic groups may not have access to the latest smartphones and devices. This may lead to lower video quality and as a result, model results that are not representative across the broad spectrum of patients. Intrinsic penalization based on video quality conflated with socioeconomic status should be avoided. 
 
\textbf{Facial Recognition Algorithm Bias and Surveillance} Existing bias is always a concern when using pretrained models for either direct inference or transfer learning. Facial recognition algorithms are no different. Researchers ~\cite{buolamwini2018gender} explore issues with facial recognition algorithms explored through the context of the (already concerning) Fitzpatrick Skin Scale. Further, the research towards chronic cancer facial pain detection is not intended for surveillance purposes. A potential misuse would be for monetary purposes of estimating medical risk and liability by analyzing and estimating pain without consent.
 
\section{Preprocessing}
\subsection{Corrupt Data}
Patient 0009's videos were unreadable and could not be processed. This consists of six potential video submissions. Patient 0015 submitted four videos, but never submitted pain scores at the start of the submission. As a result, there are no pain feature responses for Patient 0015. We set all nine self-reported pain scores for Patient 0015 to zero. 

\subsection{Key Processing Steps}
We describe preprocessing and encoding for each of the six datatypes in the ISS Dataset: 1) Self-Reported Pain Scores, 2) Labels for Sex, Skin Tone, and Timeframe, 3) Patient Narrative Video, 4) Video Extracts: Frames, Faces, Landmarks, and 5) Patient Audio, and 6) Audio Extracts: Mel Spectrogram, Audio Features.

\subsubsection{}{Pain Scores, and Sex, Skin Tone, Timeframe Labels} At the start of each video submission, the patient answers a survey that allows them to input self-reported pain scores for nine questions. No preprocessing is done for the pain scores which are reported on an 11-point Likert-scale or for the ``painnow" variable, Boolean.

\begin{itemize}
    \item Patient surveys are converted to .csv files.
    \item Labels are encoded for Sex: {Male: 0, Female: 1} and Skin Tone: {I - III: 1, IV - VI: 0}. 
    \item The ``timeframe" is manually extracted from the video submission timestamp which is reported like \texttt{20201201T003232}. We encode the hour code (the two characters after ``T") as the following:  early\_am = ['05', '06', '07'],  reg\_am = ['08', '09', '10', '11'], early\_pm = ['12', '13', '14', '15', '16'], reg\_pm = ['17', '18', '19', '20'], early\_nite = ['21', '22', '23', '24'], late\_nite = ['00', '01', '02', '03', '04'].
\end{itemize}

\subsubsection{Videos and Extracts}
\begin{itemize}
    \item \textbf{Process video for all basic datatypes} - This will extract all frames from all videos. Do this for one patient at a time per usage instructions below. Running this script will create directories for each patient per video, with corresponding frames, \texttt{.wav}  files, Mel Spectrogram for each \texttt{.wav}  file, and the encoded csv from the survey. Frames are stored like: \texttt{frames/0001/20201201T003232/1.jpg}. Audio is stored as: \texttt{audio/0001/20201201T003232/20201201T003232.wav}. Mel spectograms are stored as \texttt{audio/0001/20201201T003232.png}. Surveys for pain scores are stored as \texttt{surveys/0001/20201201T003232/targets.csv}.
    
    \item \textbf{Detect faces on spliced frames} - We use Facenet-Pytorch to detect faces and crop them directly from the frame to a 160 x 160 image. They are stored in a sub-directory per patient. This saves crops in the form with a prefix associated to each frame number \texttt{crops/0001/0001\_20201201T003232\_1.jpg}.
    
    \item \textbf{Detect and save facial landmarks} - We use the Google Media Pipe API to extract 468 landmarks for each detected face. Landmarks are stored as a list of arrays for each face, in a pickle file. File names for all files traversed, all files where faces were detected, and all files names where faces were not detected, are also stored as pickle files. The following 4 pickled files are stored for each patient: \texttt{landmarks/0001/all\_files.pkl}, \texttt{landmarks/0001/all\_landmarks.pkl}, \texttt{landmarks/0001/face\_files.pkl}, \texttt{landmarks/0001/no\_face\_files.pkl}. The ``all\_landmarks.pkl" holds all the landmarks as a list of arrays. The other files keep track of what files were iterated over. ``face\_files.pkl" is important because this will allow us to keep track of the exact mapping of 1:1 array:image.
    
\end{itemize}

\subsubsection{Audio and Extracts}
Break the large \texttt{.wav}  file into smaller 4 second chunks. For each smaller \texttt{.wav}  file, generate its own spectrogram and CSV file of acoustic features.
\begin{itemize}
    \item \textbf{Extract statistics about the \texttt{.wav}  audio file} - After the \texttt{.wav}  file has been extracted from the patient video, we extract audio features. This will generate a .csv file labeled per sub-directory for each \texttt{.wav}  file. The will store a single .csv file with all audio files labeled by video timestamp. As a result, for each patient, there will be a single ``audio\_features.csv" file that stores 25 audio features for each patient video. For example, if Patient A submits 12 times, there will be a .csv file with 12 rows, one for each audio file extracted per video. 
    \item \textbf{Audio Features} - 25 features are extracted using the Librosa library. These include chromogram, spectral centroid (center of mass), spectral rolloff (signal shape), spectral bandwidth, zero-crossing rate (smoothness), and Mel Frequency Cepstral Coefficients (MFCCs) (n=20 MFCC) statistics. 
    \item \textbf{Audio Chunks} - We selected an arbitrary number of seconds at 4-seconds to evaluate our experiments. However, any number of seconds can be selected to decompose the original audio file extracted from the basic processing into smaller chunks. We use the Python PyDub library and select \texttt{chunk\_length\_ms = 4000} for 4 second chunks, that do not overlap. 
\end{itemize}

\subsubsection{Labels} Three sets of labels are generated for each of the ten splits. After the training data and test data are assigned to each of the ten splits by patient, we encode the pain class target as either binary (``No Pain", ``Pain") or keep the original four labels. Labels are read in via PyTorch dataloader, specific to each of the seven experiments. Labels are:

\textbf{Labels of Images with Landmarks} - Image filename, pain label, sex label, skin tone label, landmarks. 

\begin{figure}[hbt!]
    \centering
    \includegraphics[width=0.5\textwidth]{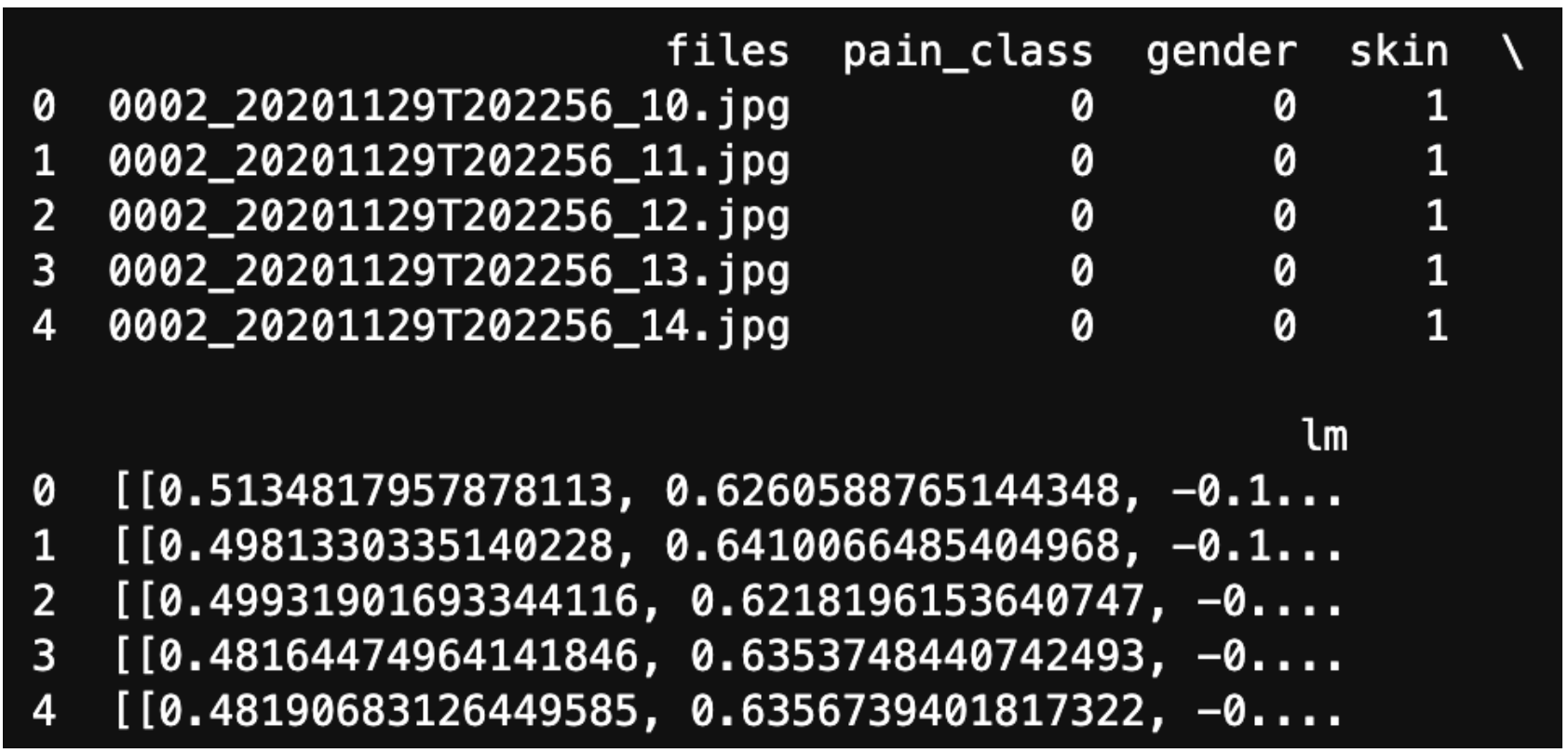}
    \caption{Example of Images with Landmarks Labels}
    \label{images_w_landmarks}
\end{figure}

\textbf{Labels of Images with Landmarks and Pain Scores} - All data from ``Labels of Images with Landmarks" in addition to pain scores, sex, skin tone, and timeframe labels. 

\begin{figure}[hbt!]
    \centering
    \includegraphics[width=0.5\textwidth]{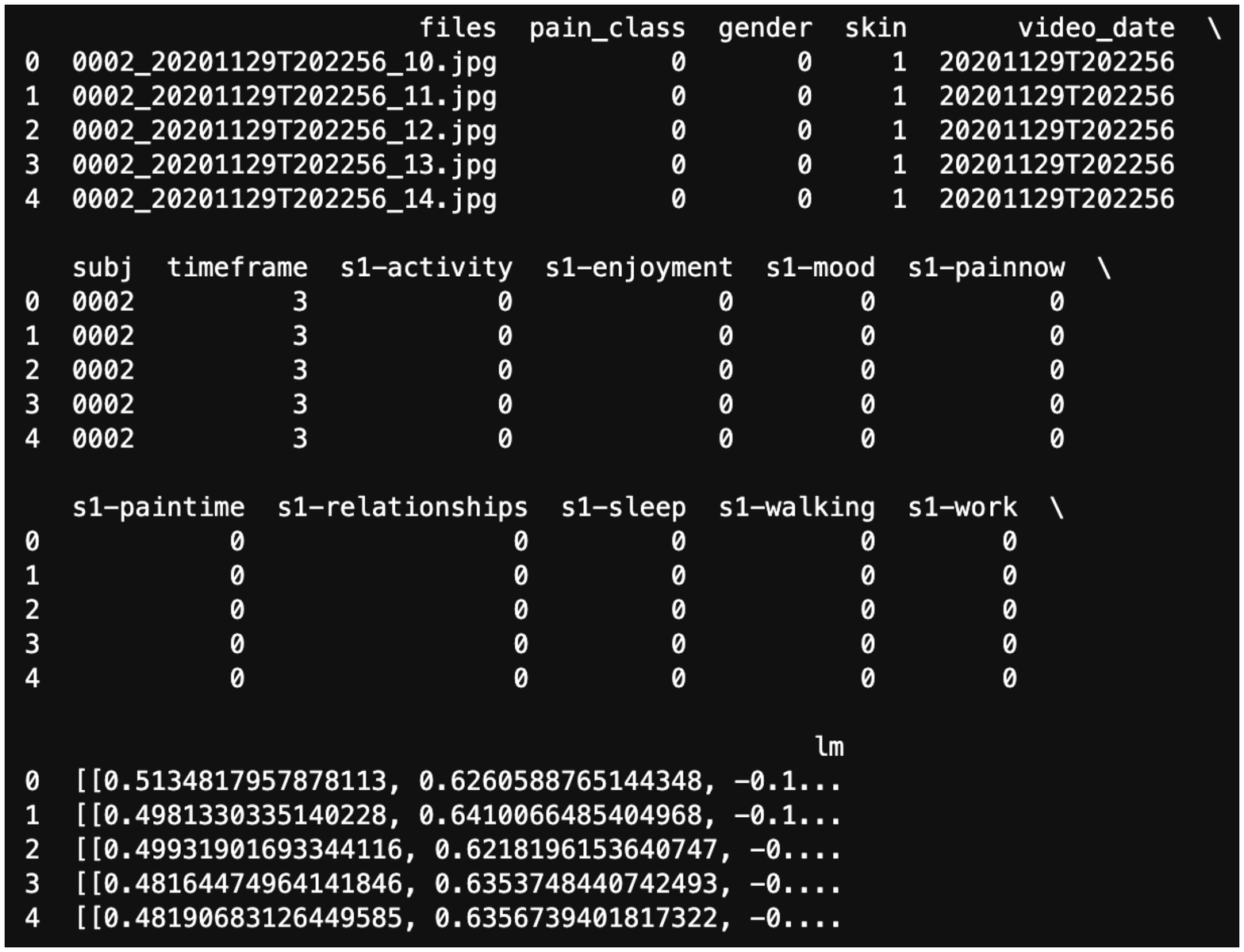}
    \caption{Example of Images, Landmarks, Pain Scores Labels}
    \label{images_wlmpain}
\end{figure}

\textbf{Labels of Audio Spectogram Images with Audio Features} - Contains the filename for the spectrogram and its corresponding set of 25 audio features. Each spectrogram is generated per audio file. Both the original audio file and the 4-second audio chunks are labeled in this manner.
\begin{figure}[hbt!]
    \centering
    \includegraphics[width=0.5\textwidth]{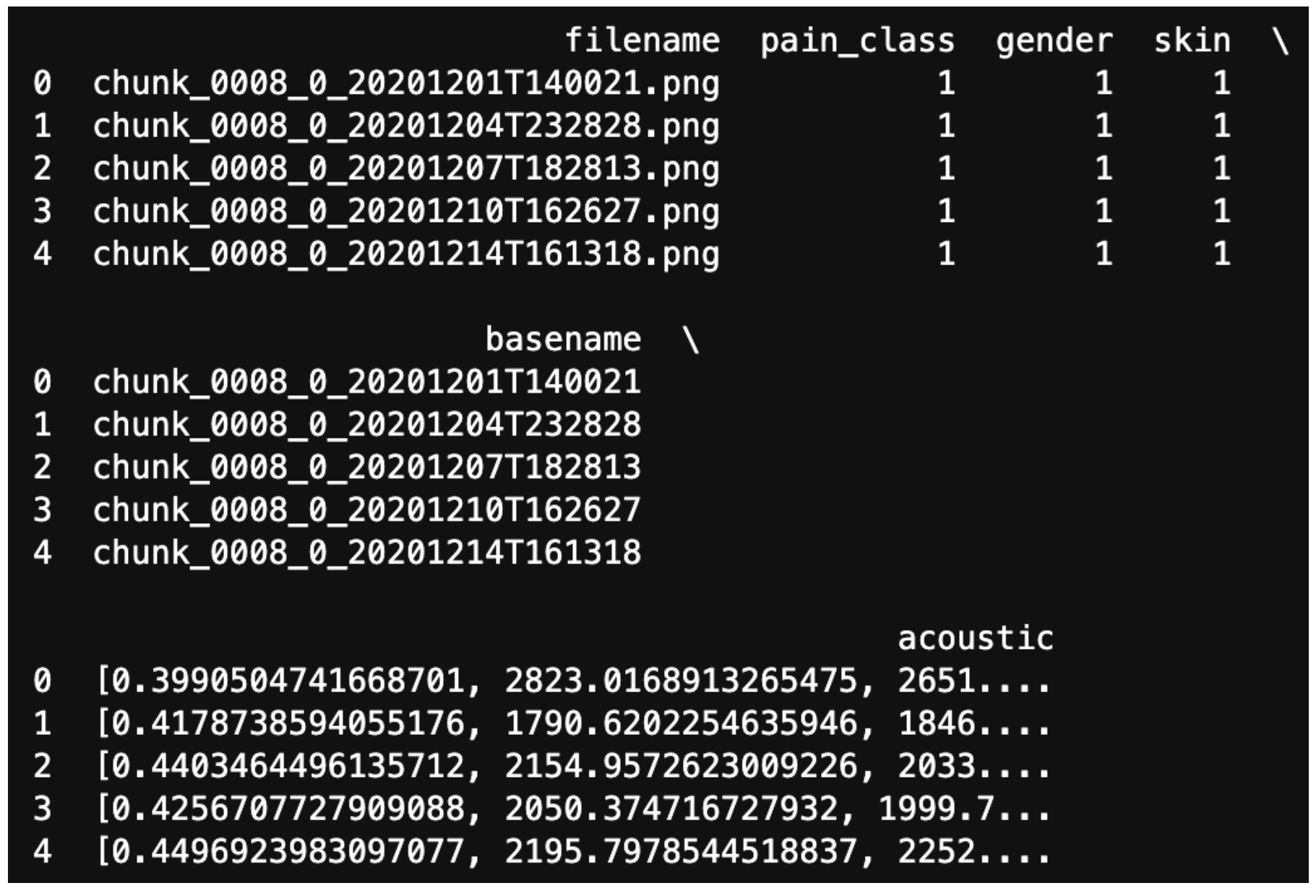}
    \caption{Example of Audio Labels}
    \label{audio_spectr}
\end{figure}

\subsubsection{Model Transforms}
We use PyTorch V1.6.0 for implementing transforms. Limited transforms were conducted to only include \texttt{RandomResizedCrop()} and \texttt{RandomHorizontalFlip}. All images for the facial images and the Mel spectrogram were resized to 224 x 224. Face crops were originally 160 x 160 following MTCNN detection on frames. Images were normalized to ([0.485, 0.456, 0.406], [0.229, 0.224, 0.225]). No image augmentation or synthetic images were used in training.

\section{Additional Data Analysis}
In Table \ref{subjects} we provide counts for the total number of videos, video frames, seconds of video, average seconds per video, and 4-sec. chunk \texttt{.wav}  files for each of the 29 patients. Counts are also provided by the original four levels of pain (``None", ``Low", ``Moderate", ``Severe"), in addition to the two levels (``No Pain", ``Pain") used for the binary classification models. 

\begin{table}[htbp]
\caption{\textbf{Descriptive Statistics for ISS Data Across 29 Enrolled Patients}. The number of Pain Score submissions per patient is equivalent to the number of Videos, since pain scores are submitted at the start of each submission.}
\label{subjects}
\centering
\begin{adjustbox}{width=0.99\textwidth}
\begin{tabular}{@{}llllllllll@{}}
\toprule
\textbf{Patient} & \textbf{Videos} & \textbf{Total Frames} & \textbf{Total Seconds} & \textbf{Avg Sec/Video} & \textbf{4 Sec. Chunks} & \textbf{(4-Pain) Level} & \textbf{(2-Pain) Level} & \textbf{Sex} & \textbf{Skin Tone} \\ \midrule
1 & 23 & 1674 & 167.23 & 7.27 & 53 & None & No Pain & M & IV - VI \\
2 & 23 & 1721 & 172.24 & 7.49 & 54 & None & No Pain & M & I - III \\
3 & 5 & 2558 & 255.91 & 51.18 & 67 & None & No Pain & M & I - III \\
4 & 5 & 1605 & 160.08 & 32.02 & 43 & Moderate & Pain & M & I - III \\
5 & 5 & 2552 & 255.17 & 51.03 & 67 & Moderate & Pain & M & I - III \\
6 & 1 & 193 & 19.39 & 19.39 & 5 & Severe & Pain & F & IV - VI \\
7 & 26 & 5398 & 539.61 & 20.75 & 148 & Low & Pain & M & IV - VI \\
9 & n/a & n/a & n/a & n/a & n/a & Low & Pain & F & I - III \\
8 & 25 & 6938 & 694.21 & 27.77 & 185 & Moderate & Pain & F & I - III \\
10 & 18 & 3523 & 352.43 & 19.58 & 95 & Severe & Pain & F & IV - VI \\
11 & 26 & 35249 & 3525.39 & 135.59 & 896 & Severe & Pain & M & I - III \\
12 & 55 & 41373 & 4136.35 & 75.21 & 1064 & None & No Pain & M & I - III \\
13 & 28 & 6953 & 694.98 & 24.82 & 187 & None & No Pain & M & I - III \\
14 & 43 & 16685 & 1669.08 & 38.82 & 435 & None & No Pain & M & I - III \\
15 & 4 & 1435 & 143.57 & 35.89 & 38 & Severe & Pain & F & IV - VI \\
16 & 16 & 2438 & 243.61 & 15.23 & 66 & None & No Pain & M & I - III \\
17 & 33 & 12556 & 1253.93 & 38 & 328 & None & No Pain & F & I - III \\
18 & 3 & 2779 & 278.08 & 92.69 & 71 & None & No Pain & M & I - III \\
19 & 18 & 635 & 63.38 & 3.52 & 21 & Severe & Pain & F & I - III \\
20 & 8 & 3222 & 322.18 & 40.27 & 84 & Low & Pain & M & I - III \\
21 & 50 & 14904 & 1489 & 29.78 & 397 & Moderate & Pain & F & I - III \\
22 & 11 & 5413 & 541.28 & 49.21 & 141 & Severe & Pain & F & I - III \\
23 & 9 & 3711 & 371.34 & 41.26 & 98 & Severe & Pain & M & IV - VI \\
24 & 17 & 5332 & 533.44 & 31.38 & 142 & None & No Pain & F & I - III \\
25 & 15 & 3504 & 350.6 & 23.37 & 95 & None & No Pain & M & IV - VI \\
26 & 4 & 1073 & 107.43 & 26.86 & 28 & Severe & Pain & M & IV - VI \\
27 & 9 & 1478 & 147.93 & 16.44 & 40 & None & No Pain & M & IV - VI \\
28 & 19 & 1467 & 146.48 & 7.71 & 44 & Low & Pain & M & IV - VI \\
29 & 6 & 1933 & 193.27 & 32.21 & 51 & None & No Pain & M & IV - VI \\
30 & 4 & 1697 & 169.73 & 42.43 & 45 & Low & Pain & M & IV - VI \\ \bottomrule
\end{tabular}
\end{adjustbox}
\end{table}

\begin{figure}[ht!]
     \centering
     \begin{subfigure}[b]{0.89\textwidth}
         \centering
         \includegraphics[width=\textwidth]{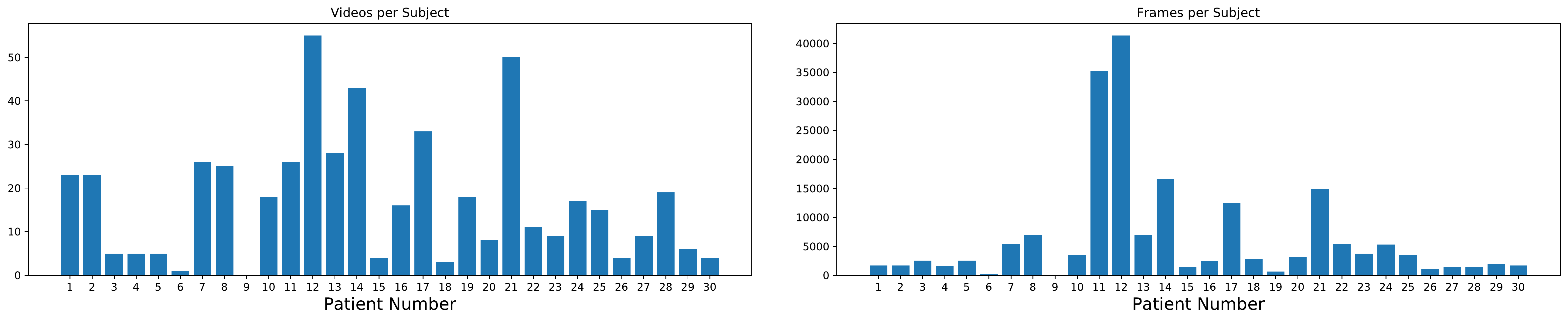}
     \end{subfigure}
     \begin{subfigure}[b]{0.89\textwidth}
         \centering
         \includegraphics[width=\textwidth]{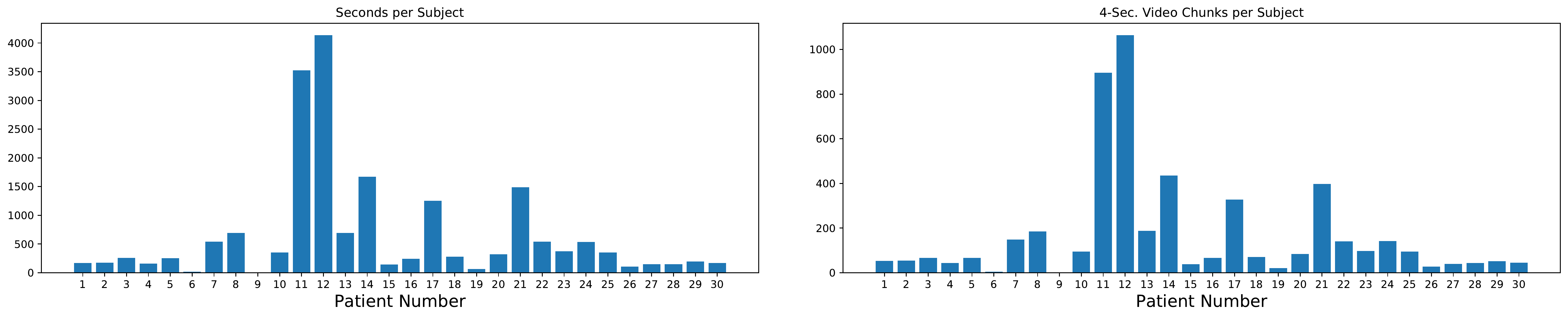}
     \end{subfigure}
    \begin{subfigure}[b]{0.89\textwidth}
        \centering
        \includegraphics[width=\textwidth]{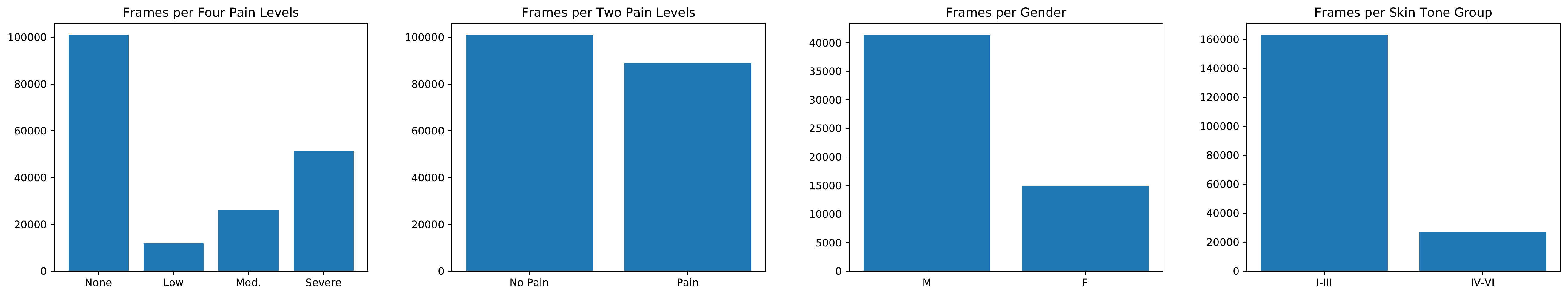}
     \end{subfigure}
     \begin{subfigure}[b]{0.89\textwidth}
        \centering
        \includegraphics[width=\textwidth]{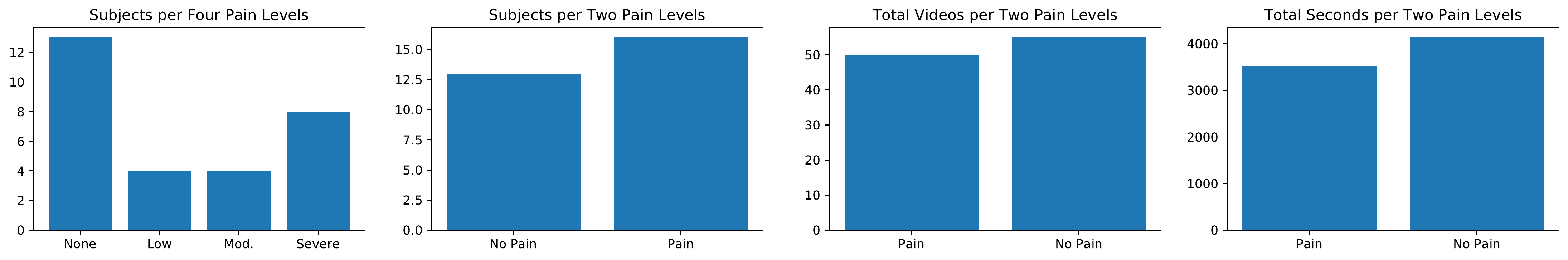}
     \end{subfigure}
    \caption{\textbf{Plots of Descriptive Statistics}. Plots are consistent with Table \ref{subjects} shown above.}
    \label{distribution}
\end{figure}

\begin{figure}[htbp!]
    \centering
    \includegraphics[width=0.75\textwidth]{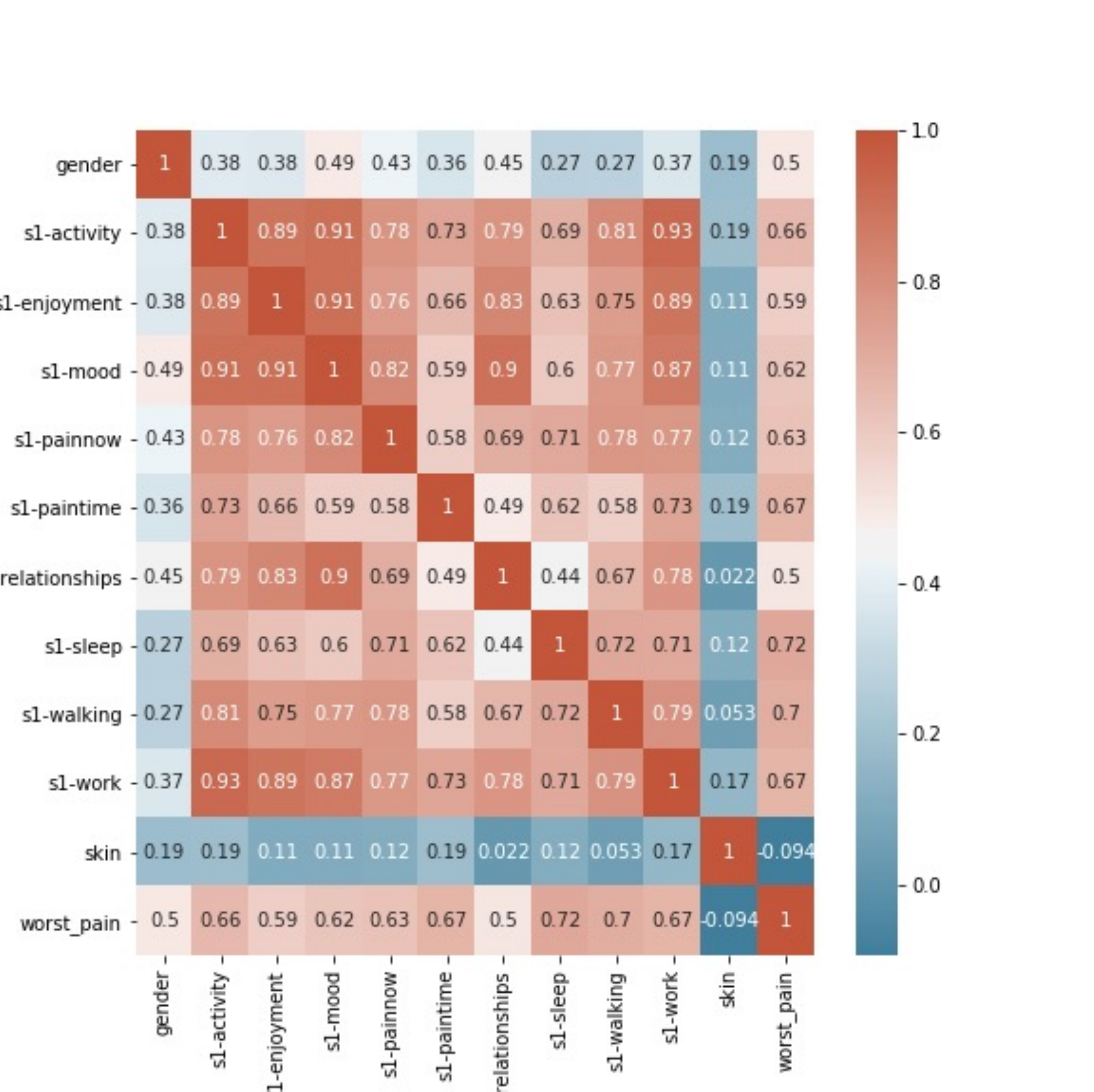}
    \caption{\small\textbf{Correlation Analysis of Self-Reported Pain Scores}}
    \label{corr}
\end{figure}

\begin{table}[htbp]
\caption{\textbf{Source Devices Used to Submit Data by each Patient}}
\label{devices}
\centering
\begin{tabular}{@{}ll@{}}
\toprule
\textbf{Patient} & \textbf{Source Device} \\ \midrule
0001 & Android \\
0002 & iOS \\
0003 & iOS \\
0004 & iOS \\
0005 & Clinic Only \\
0006 & Clinic Only \\
0007 & Android \\
0008 & iOS \\
0009 & Android \\
0010 & iOS \\
0011 & iOS \\
0012 & Android \\
0013 & Android \\
0014 & iOS \\
0015 & iOS \\
0016 & Android \\
0017 & Android \\
0018 & iOS \\
0019 & Android \\
0020 & iOS \\
0021 & Android \\
0022 & iOS \\
0023 & iOS \\
0024 & iOS \\
0025 & iOS \\
0026 & Android \\
0027 & iOS \\
0028 & iOS \\
0029 & iOS \\ \bottomrule
\end{tabular}
\end{table}

\newpage
\section{Details of the Models}
\subsection{Cross Validation Details} 
We use 10-fold-cross-validation leading to 10 different training splits, where 9 of the splits reserve 3 patients for the test set and the tenth split uses 2 patients, since there are 29 patients. 

\begin{table}[ht!]
\caption{\textbf{10-fold-CV details - Facial Image Counts per Split.}}
\label{image10x}
\centering
\begin{adjustbox}{width=0.99\textwidth}
\begin{tabular}{@{}lllllll@{}}
\toprule
 & \multicolumn{6}{c}{\textbf{Face Images}} \\ \midrule
 & \multicolumn{2}{c}{\textbf{Train Set}} & \multicolumn{2}{c}{\textbf{Test Set}} & \multicolumn{2}{c}{\textbf{Total Data}} \\
\textbf{Split} & Face Images & Face Images w/LMs & Face Images & Face Images w/LMs & Face Images & Face Images w/LMs \\
\textbf{1} & 157269 & 153322 & 15742 & 14741 & 173011 & 168063 \\
\textbf{2} & 163128 & 158310 & 9883 & 9753 & 173011 & 168063 \\
\textbf{3} & 163734 & 158809 & 9277 & 9254 & 173011 & 168063 \\
\textbf{4} & 162678 & 157865 & 10333 & 10198 & 173011 & 168063 \\
\textbf{5} & 156622 & 151809 & 16389 & 16254 & 173011 & 168063 \\
\textbf{6} & 168339 & 163578 & 4672 & 4485 & 173011 & 168063 \\
\textbf{7} & 145522 & 141804 & 27489 & 26259 & 173011 & 168063 \\
\textbf{8} & 103418 & 100462 & 69593 & 67601 & 173011 & 168063 \\
\textbf{9} & 166739 & 161816 & 6272 & 6247 & 173011 & 168063 \\
\textbf{10} & 169650 & 164792 & 3361 & 3271 & 173011 & 168063 \\ \bottomrule
\end{tabular}
\end{adjustbox}
\end{table}

\begin{table}[ht!]
\caption{\textbf{10-fold-CV details - 4-Second Chunks of Mel Spectrograms and Audio Features, Counts per Split.}}
\label{audio10x}
\centering
\begin{adjustbox}{width=0.30\textwidth}
\begin{tabular}{@{}llll@{}}
\toprule
\textbf{Split} & \textbf{Train Set} & \textbf{Test Set} & \textbf{Total} \\ \midrule
\textbf{1} & 4486 & 502 & 4988 \\
\textbf{2} & 4719 & 269 & 4988 \\
\textbf{3}& 4742 & 246 & 4988 \\
\textbf{4} & 4710 & 278 & 4988 \\
\textbf{5} & 4561 & 427 & 4988 \\
\textbf{6} & 4858 & 130 & 4988 \\
\textbf{7} & 4264 & 724 & 4988 \\
\textbf{8} & 2841 & 2147 & 4988 \\
\textbf{9} & 4821 & 167 & 4988 \\
\textbf{10} & 4890 & 98 & 4988 \\ \bottomrule
\end{tabular}
\end{adjustbox}
\end{table}

\begin{table}[ht!]
\caption{\textbf{10-fold-CV details - Pain Scores that include Skin, sex, and Timeframe Labels, Counts per Split.}}
\label{pain10x}
\centering
\begin{adjustbox}{width=0.30\textwidth}
\begin{tabular}{@{}llll@{}}
\toprule
\textbf{Split} & \multicolumn{1}{c}{\textbf{Train Set}} & \multicolumn{1}{c}{\textbf{Test Set}} & \textbf{Total} \\ \midrule
\textbf{1} & 455 & 82 & 537 \\
\textbf{2} & 479 & 58 & 537 \\
\textbf{3} & 505 & 32 & 537 \\
\textbf{4} & 511 & 26 & 537 \\
\textbf{5} & 497 & 40 & 537 \\
\textbf{6} & 495 & 42 & 537 \\
\textbf{7} & 449 & 88 & 537 \\
\textbf{8} & 425 & 112 & 537 \\
\textbf{9} & 510 & 27 & 537 \\
\textbf{10} & 507 & 30 & 537 \\ \bottomrule
\end{tabular}
\end{adjustbox}
\end{table}

\begin{table}[ht!]
\caption{\textbf{10-fold-CV details - Test Patients per Split.}  We show each patient held out in the test set for each split. For each test patient, the table shows the original pain level and the binary pain level in order of the patient. For example, for Split 1 Test Patient 0002, the original pain level is ``Low" and the binary pain level is ``Pain".}
\label{pain10x_patients}
\centering
\begin{adjustbox}{width=0.95\textwidth}
\begin{tabular}{@{}llllllllll@{}}
\toprule
\textbf{Split} & \multicolumn{3}{l}{\textbf{Test Patients}} & \multicolumn{3}{l}{\textbf{Respective Original Pain Level}} & \multicolumn{3}{l}{\textbf{Respective Binary Pain Level}} \\ \midrule
\textbf{1} & 0002 & 0029 & 0021 & Low & None & Moderate & Pain & No Pain & Pain \\ \midrule
\textbf{2} & 0028 & 0027 & 0008 & Low & None & Moderate & Pain & No Pain & Pain \\ \midrule
\textbf{3} & 0020 & 0025 & 0005 & Low & None & Moderate & Pain & No Pain & Pain \\ \midrule
\textbf{4} & 0015 & 0024 & 0023 & Severe & None & Severe & Pain & No Pain & Pain \\ \midrule
\textbf{5} & 0018 & 0017 & 0026 & None & None & Severe & No Pain & No Pain & Pain \\ \midrule
\textbf{6} & 0004 & 0016 & 0019 & Moderate & None & Severe & Pain & No Pain & Pain \\ \midrule
\textbf{7} & 0014 & 0022 & 0007 & None & Severe & Low & No Pain & Pain & No Pain \\ \midrule
\textbf{8} & 0013 & 0011 & 0012 & None & Severe & None & No Pain & Pain & No Pain \\ \midrule
\textbf{9} & 0003 & 0010 & 0006 & None & Severe & Severe & No Pain & Pain & Pain \\ \midrule
\textbf{10} & 0030 & 0001 &  & None & None &  & No Pain & No Pain &  \\ \bottomrule
\end{tabular}
\end{adjustbox}
\end{table}

\newpage
\subsection{Model Architectures}
Diagrams provide model architecture details for Experiments 4 - 7. The architecture for Experiment 1 is ResNet50, unmodified, pre-trained on ImageNet and fine-tuned. Models used for Experiments 2 and 3 are Random Forest Classifiers using all default settings in the \texttt{sklearn.ensemble.RandomForestClassifier} library with 100 estimators

\begin{figure}[ht!]
    \centering
    \includegraphics[width=0.9\textwidth]{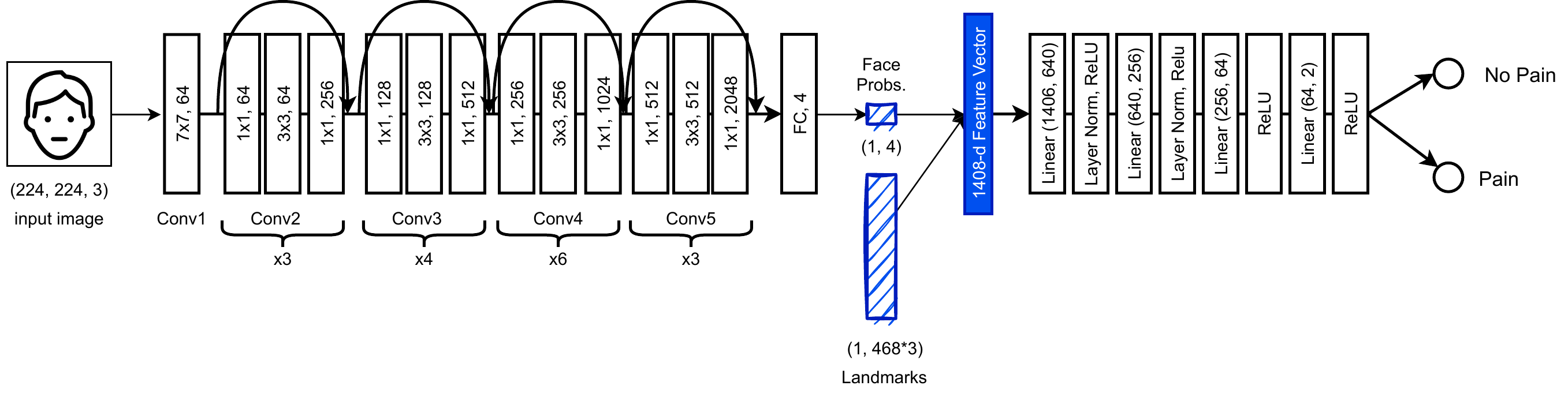}
    \caption{Experiment 4: Fusion 1, Faces + Landmarks.}
    \label{arch_fus1}
\end{figure}

\begin{figure}[ht!]
    \centering
    \includegraphics[width=0.9\textwidth]{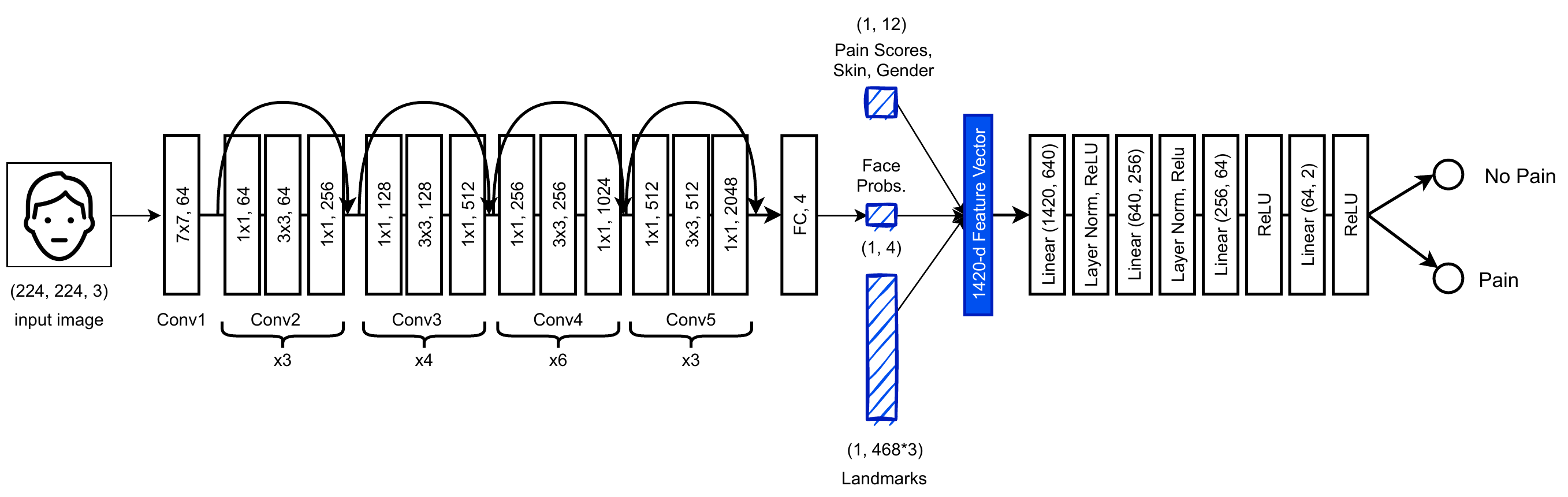}
    \caption{Experiment 5: Fusion 2, Faces + Landmarks + Pain Scores.}
    \label{arch_fus2}
\end{figure}

\begin{figure}[ht!]
    \centering
    \includegraphics[width=0.9\textwidth]{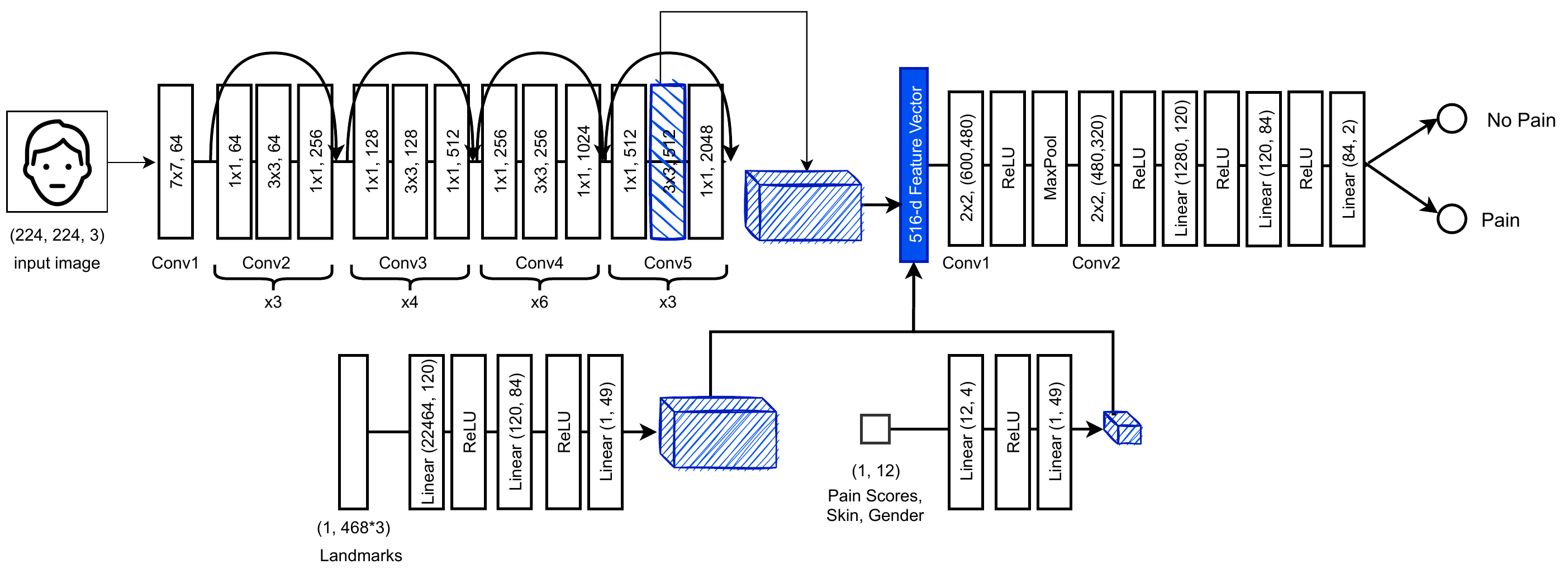}
    \caption{Experiment 6: Fusion 3, Faces + Landmarks + Pain Scores.}
    \label{arch_fus3}
\end{figure}

\begin{figure}[ht!]
    \centering
    \includegraphics[width=0.9\textwidth]{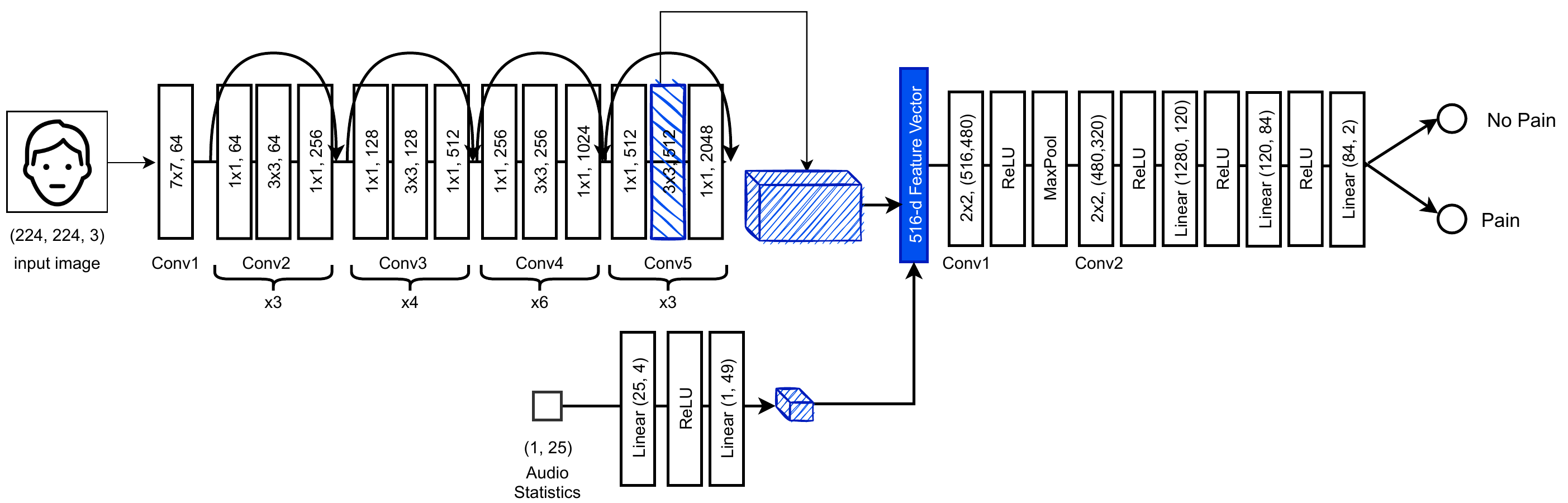}
    \caption{Experiment 7: Static Audio, Audio only.}
    \label{arch_audioonly}
\end{figure}

\newpage
\section{Additional Model Results}
\begin{figure}[ht!]
     \centering
     \begin{subfigure}[b]{0.30\textwidth}
         \centering
         \includegraphics[width=\textwidth]{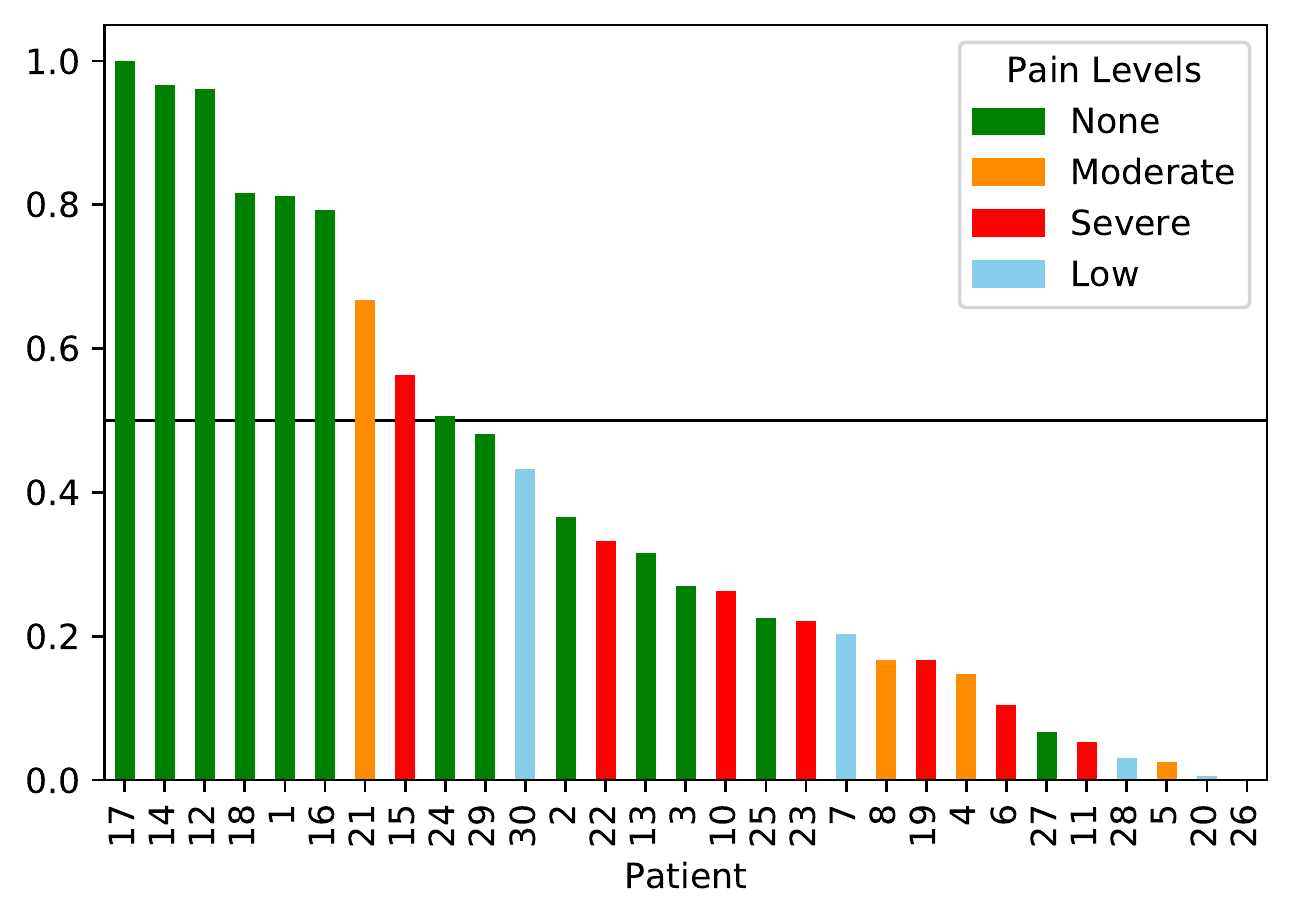}
         \caption{ResNet50-4-static}
     \end{subfigure}
    \begin{subfigure}[b]{0.30\textwidth}
        \centering
        \includegraphics[width=\textwidth]{images/static2_2.pdf}
        \caption{ResNet50-2-static - Binary}
     \end{subfigure}
    \begin{subfigure}[b]{0.30\textwidth}
        \centering
        \includegraphics[width=\textwidth]{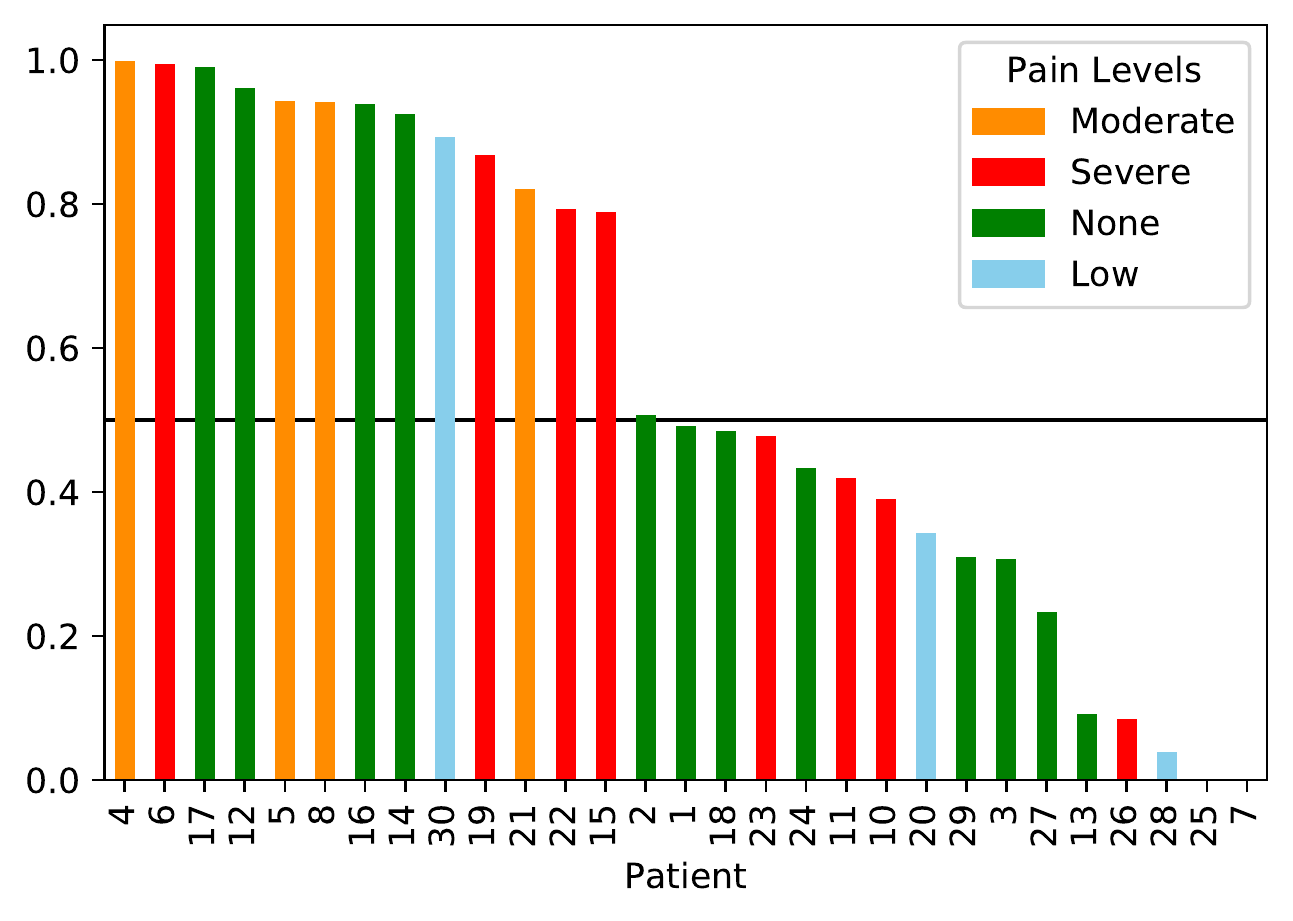}
        \caption{ResNet50-2-static - 4 Labels}
     \end{subfigure}
    \caption{
    \small{
    \textbf{Accuracy per Patient for Static Face Models.} Patients are ranked in descending order of test accuracy and color coded by pain labels. (a) shows the results of Experiment 1 (ResNet50-4-static). (b) Experiment 1 (ResNet50-2-static) binary pain classification results. (c) Experiment 1 (ResNet50-4-static) results but visualized with the original four pain labels.}}
    \label{exp1_plots}
\end{figure}

\begin{figure}[ht!]
     \centering
    \begin{subfigure}[b]{0.43\textwidth}
         \centering
         \includegraphics[width=\textwidth]{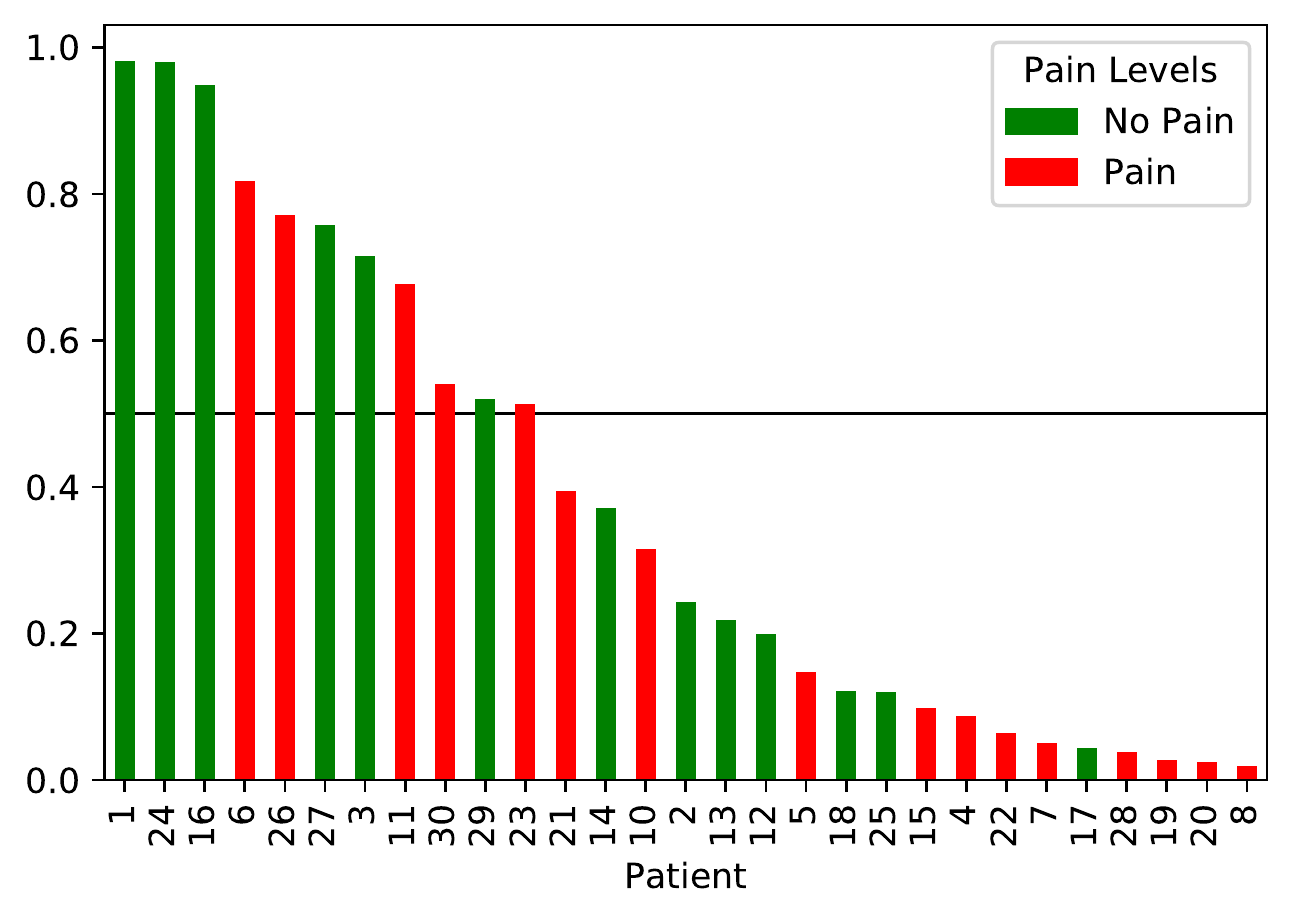}
         \caption{Random Forest LM - Binary}
     \end{subfigure}
    \begin{subfigure}[b]{0.43\textwidth}
        \centering
        \includegraphics[width=\textwidth]{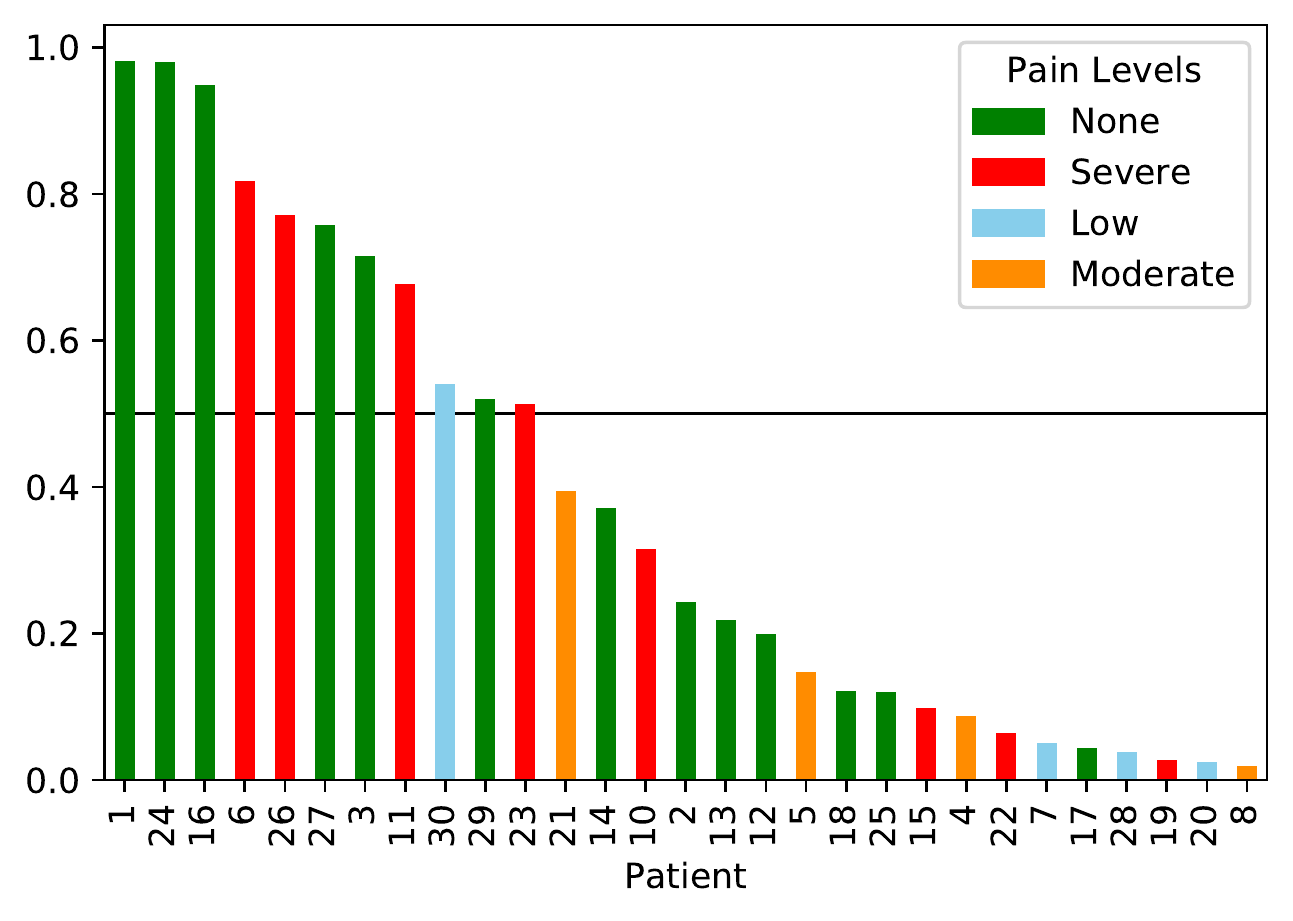}
        \caption{Random Forest LM - 4 Labels}
     \end{subfigure}
    \caption{
    \small{
    \textbf{Accuracy per Patient for Landmark Models.}Patients are ranked in descending order of test accuracy and color coded by pain labels. (a) shows the results of Experiment 2 (Random Forest LM) binary pain classification results. (b) shows Experiment 2 (Random Forest LM) results but visualized with the original four pain labels.}}
    \label{lm_plots}
\end{figure}

\begin{figure}[ht!]
     \centering
     \begin{subfigure}[b]{0.43\textwidth}
         \centering
         \includegraphics[width=\textwidth]{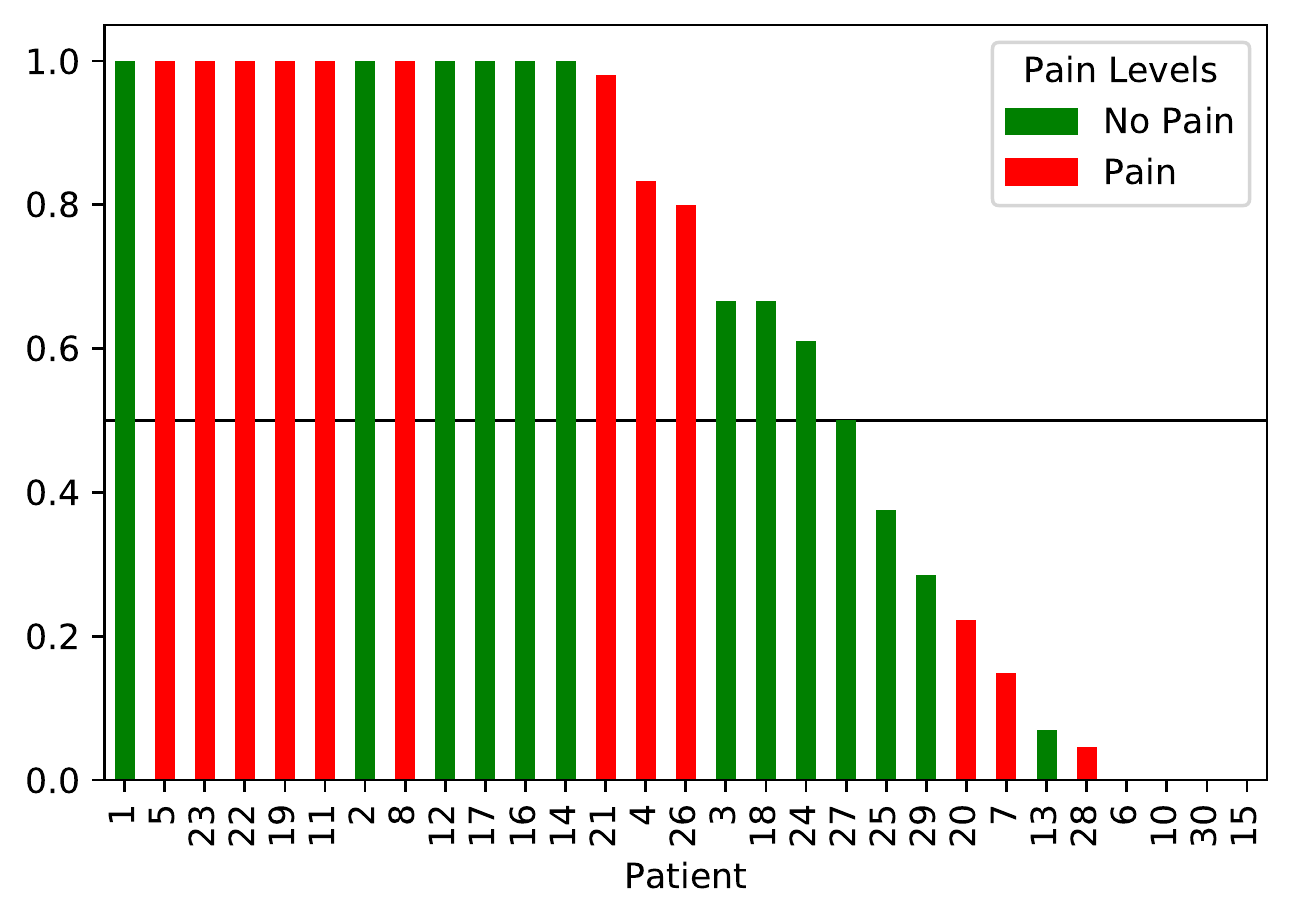}
         \caption{Random Forest Pain - Binary}
     \end{subfigure}
    \begin{subfigure}[b]{0.43\textwidth}
        \centering
        \includegraphics[width=\textwidth]{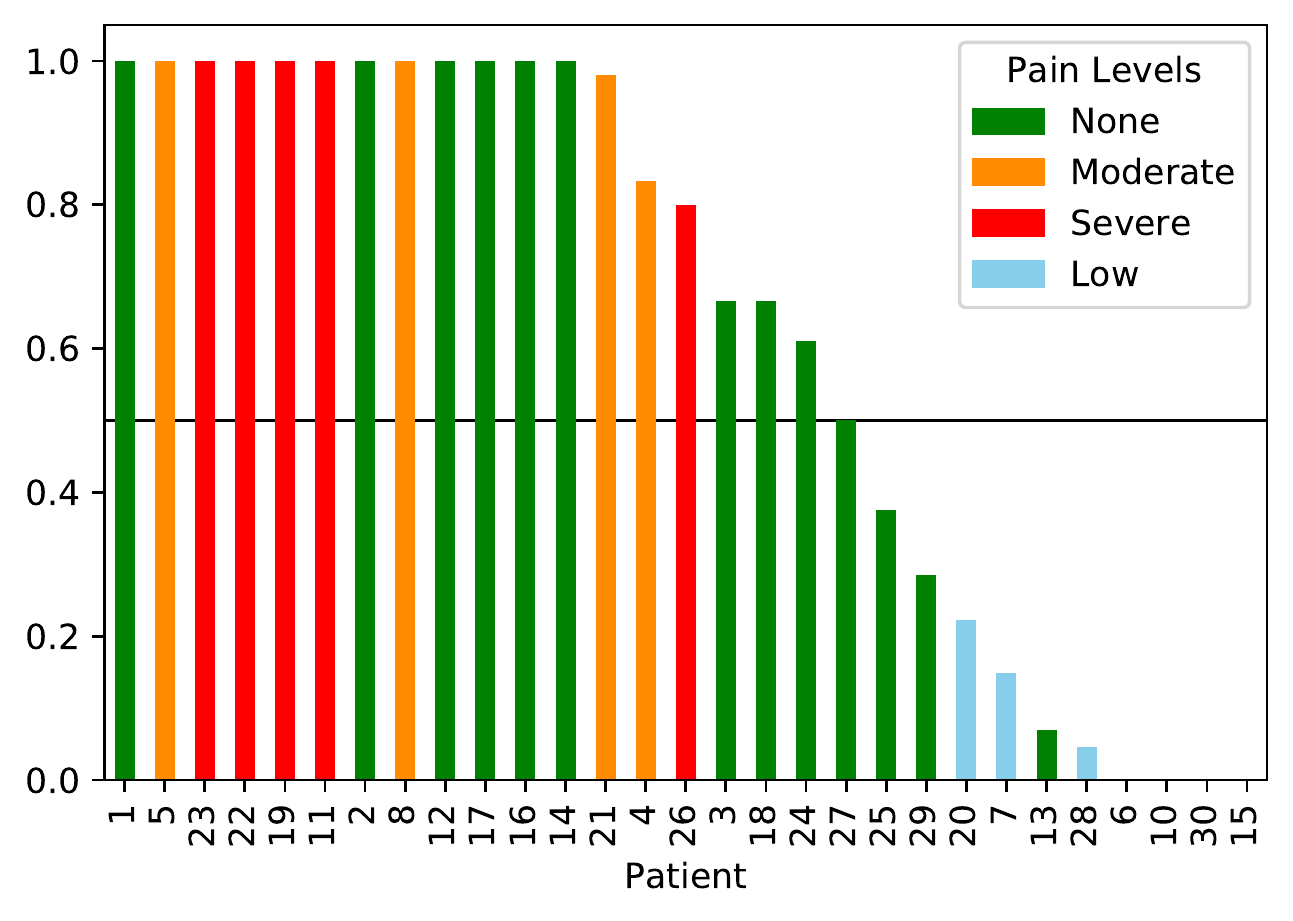}
        \caption{Random Forest Pain - 4 Labels}
     \end{subfigure}
    \caption{
    \small{
    \textbf{Accuracy per Patient for Pain Models.} Patients are ranked in descending order of test accuracy and color coded by pain labels. (a) shows the results of Experiment 3 (Random Forest Pain) binary pain classification results. (b) shows Experiment 3 (Random Forest Pain) results but visualized with the original four pain labels.}}
    \label{pain_plots}
\end{figure}

We provide additional results from baseline experiments. In Table \ref{ranges}, the number and percentage of patients with accuracy scores in the ranges of 0 - 25\%, 25\% - 50\%, 50\% - 75\%, and 75\% is shown.  In Figure \ref{exp1_plots}, we show the test patient accuracy scores by binary labels as well as the original four pain labels. There is an imbalance of pain levels in the ISS dataset, where ``None" frames outnumber ``Low" and ``Moderate" classes at a ratio of 8.50:1 and 3.88:1, respectively.  When inspecting the binary classifier according to the original four pain labels, the ``Moderate" test patients show greater accuracy. In Figure \ref{lm_plots} we see a decrease in the accuracy of the ``Moderate" and ``Low" test patients. In Figure \ref{pain_plots}, while the ``Moderate" classes exceed 50\% accuracy, the model struggles with ``Low" pain test patients. However, 13 test patients reach test accuracies close to 1.0. 

\begin{table}[ht!]
\caption{\textbf{Model Results by Patient Scores by Accuracy Range.}}
\label{ranges}
\centering
\begin{adjustbox}{width=0.99\textwidth}
\begin{tabular}{@{}llllllllllll@{}}
\toprule
 &  & \multicolumn{8}{c}{\textbf{Test Accuracy Ranges}} &  &  \\ \midrule
\textbf{} & \textbf{} & \multicolumn{2}{c}{\textbf{{[}0. 0.25)}} & \multicolumn{2}{c}{\textbf{{[}0.25, 0.50)}} & \multicolumn{2}{c}{\textbf{{[}0.50, 0.75)}} & \multicolumn{2}{c}{\textbf{{[}0.75, 1.0)}} & \multicolumn{2}{c}{\textbf{Over 50\%}} \\
\textbf{Experiments} & \textbf{Model} & \multicolumn{1}{c}{\textbf{No. Patients}} & \multicolumn{1}{c}{\textbf{Perc. Patients}} & \multicolumn{1}{c}{\textbf{No. Patients}} & \multicolumn{1}{c}{\textbf{Perc. Patients}} & \multicolumn{1}{c}{\textbf{No. Patients}} & \multicolumn{1}{c}{\textbf{Perc. Patients}} & \multicolumn{1}{c}{\textbf{No. Patients}} & \multicolumn{1}{c}{\textbf{Perc. Patients}} & \multicolumn{1}{c}{\textbf{No. Patients}} & \multicolumn{1}{c}{\textbf{Perc. Patients}} \\
Exp. 1 & ResNet50-4-static & 13 & 44.8\% & 7 & 24.1\% & 3 & 10.3\% & 6 & 20.7\% & 9 & 0.31034483 \\
Exp. 1 & ResNet50-2-static & 6 & 20.7\% & 9 & 31.0\% & 1 & 3.4\% & 13 & 44.8\% & 14 & 0.48275862 \\
Exp. 2 & Random Forest LM & 15 & 51.7\% & 3 & 10.3\% & 5 & 17.2\% & 6 & 20.7\% & 11 & 0.37931034 \\
Exp. 3 & Random Forest Pain & 7 & 24.1\% & 2 & 6.9\% & 4 & 13.8\% & 15 & \textbf{51.7\%} & 19 & 0.65517241 \\
Exp. 4 & Fusion 1 & 11 & 37.9\% & 3 & 10.3\% & 4 & 13.8\% & 11 & 37.9\% & 15 & 0.51724138 \\
Exp. 5 & Fusion 2 & 8 & 27.6\% & 3 & 10.3\% & 3 & 10.3\% & 15 & \textbf{51.7\%} & 18 & 0.62068966 \\
Exp. 6 & Fusion 3 & 7 & 24.1\% & 1 & 3.4\% & 6 & 20.7\% & 15 & \textbf{51.7\%} & 21 & \textbf{0.72413793} \\
Exp. 7 & Static Audio & 9 & 31.0\% & 7 & 24.1\% & 7 & 24.1\% & 6 & 20.7\% & 13 & 0.44827586 \\ \bottomrule
\end{tabular}
\end{adjustbox}
\end{table}

\begin{figure}[ht!]
     \centering
     \begin{subfigure}[b]{0.30\textwidth}
         \centering
         \includegraphics[width=\textwidth]{images/mmn4_2.pdf}
         \caption{Fusion 1 - Binary}
         \label{mmna}
     \end{subfigure}
    \begin{subfigure}[b]{0.30\textwidth}
        \centering
        \includegraphics[width=\textwidth]{images/mmn5_2.pdf}
        \caption{Fusion 2 - Binary}
        \label{mmnb}
     \end{subfigure}
    \begin{subfigure}[b]{0.30\textwidth}
         \centering
         \includegraphics[width=\textwidth]{images/mmn6_2.pdf}
         \caption{Fusion 3 - Binary}
         \label{mmnc}
     \end{subfigure}
    \begin{subfigure}[b]{0.30\textwidth}
        \centering
        \includegraphics[width=\textwidth]{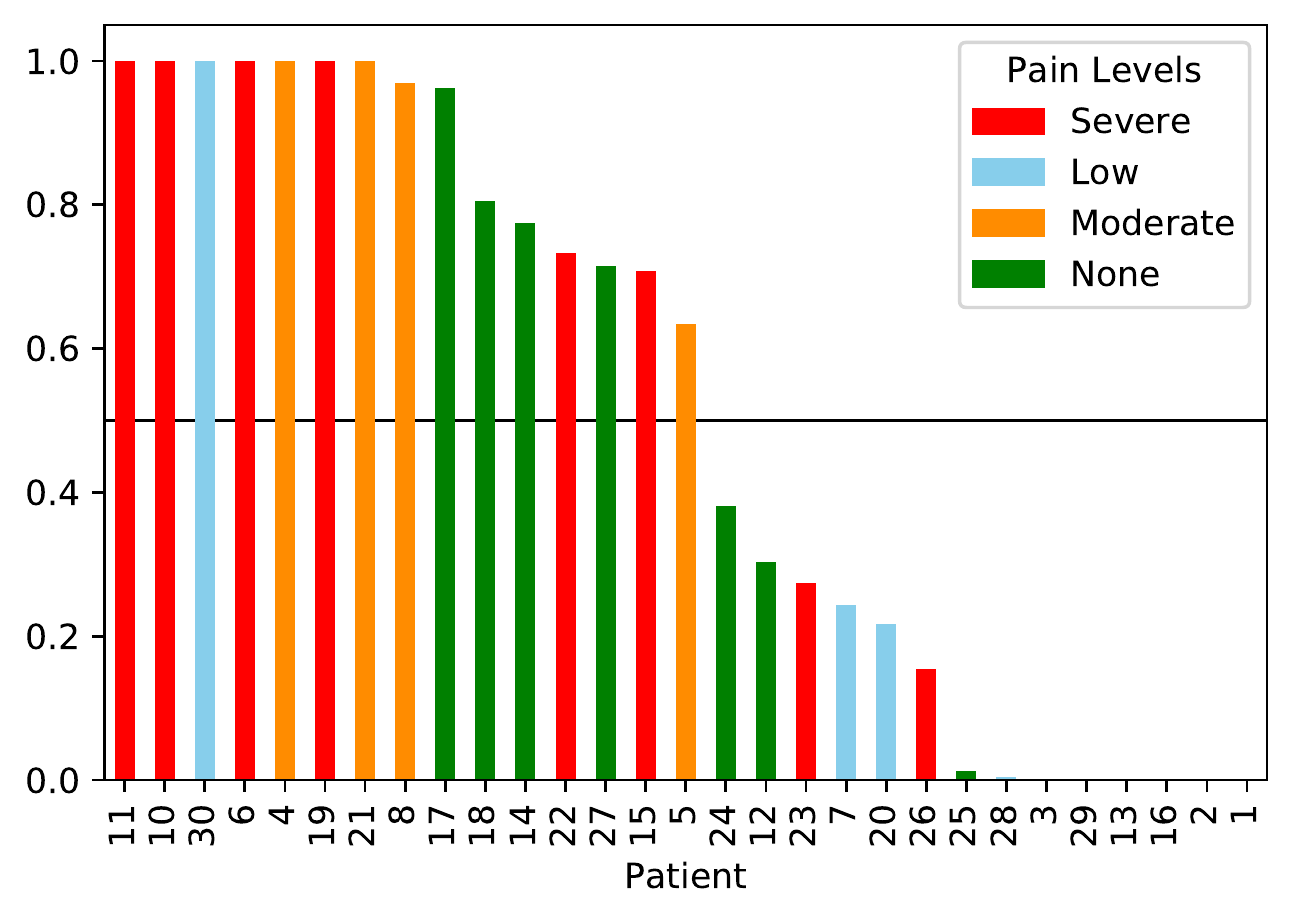}
        \caption{Fusion 1 - 4 Labels}
        \label{mmnd}
     \end{subfigure}
    \begin{subfigure}[b]{0.30\textwidth}
        \centering
        \includegraphics[width=\textwidth]{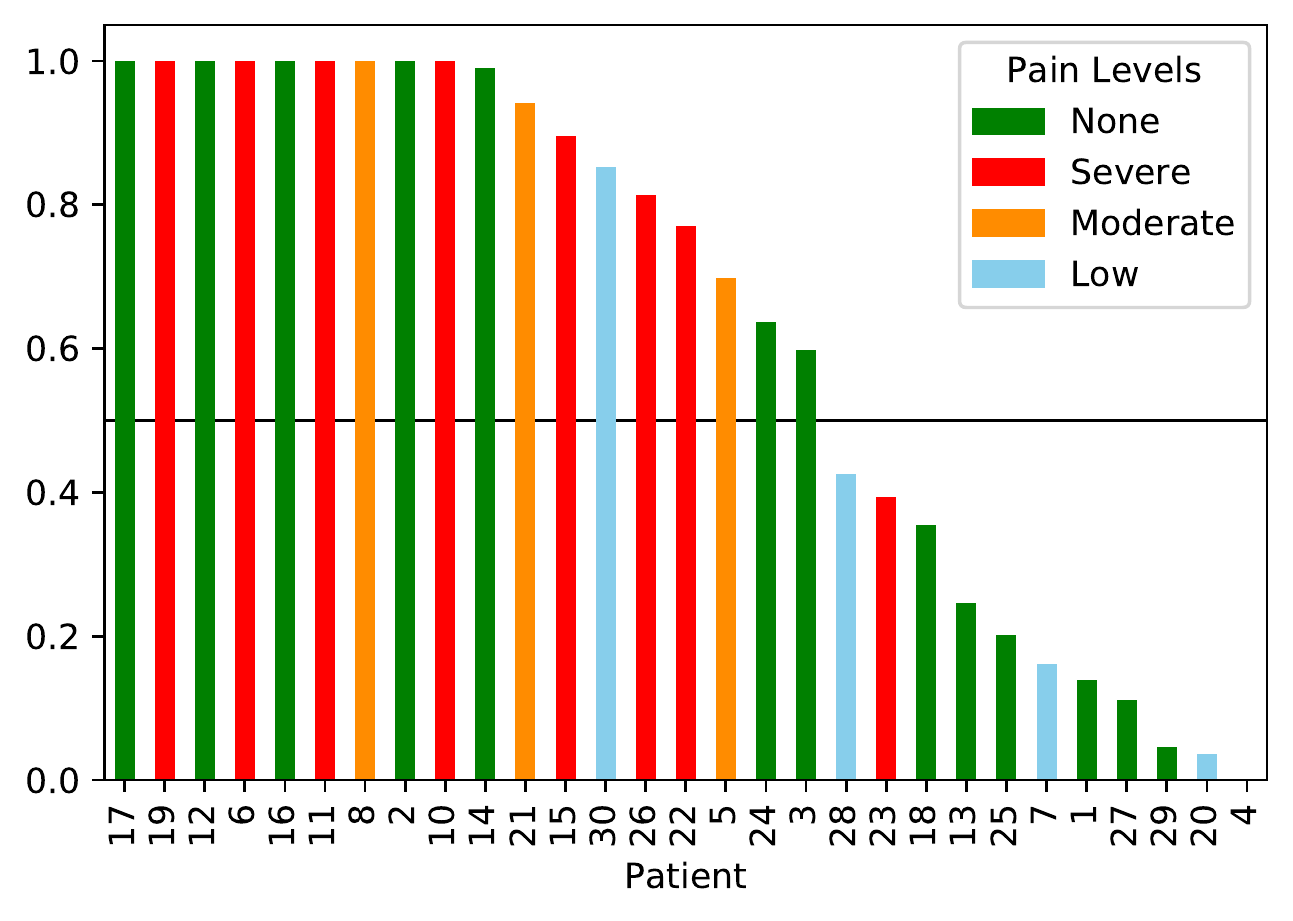}
        \caption{Fusion 2 - 4 Labels}
        \label{mmne}
     \end{subfigure}
    \begin{subfigure}[b]{0.30\textwidth}
        \centering
        \includegraphics[width=\textwidth]{images/mmn6_4.pdf}
        \caption{Fusion 3 - 4 Labels}
        \label{mmnf}
     \end{subfigure}
    \caption{
    \small{
    \textbf{Accuracy per Patient for Multimodal Face, Landmark and Pain Score Models.} Each subplot shows the results of Experiments 4, 5, and 6, with the binary predictions as well as visualized with the original four pain levels to provide greater granularity. Plots (a) and (d) provide accuracy scores for each test patient for Experiment 4 that combines facial images and landmarks. Plots (b) and (e) provide accuracy scores for Experiment 5 that combines facial images, raw landmarks, and raw pain scores, with sex, skin tone, and timeframe labels by simple concatenations.  Plots (c) and (f) provide results of Experiment 6 that combines learned features from separate networks for each modality of the facial images, landmarks, and pain scores with sex, skin tone, and timeframe labels, into a single multimodal network. }}
    \label{mmnplot}
\end{figure}

\begin{figure}[ht!]
     \centering
    \begin{subfigure}[b]{0.45\textwidth}
         \centering
         \includegraphics[width=\textwidth]{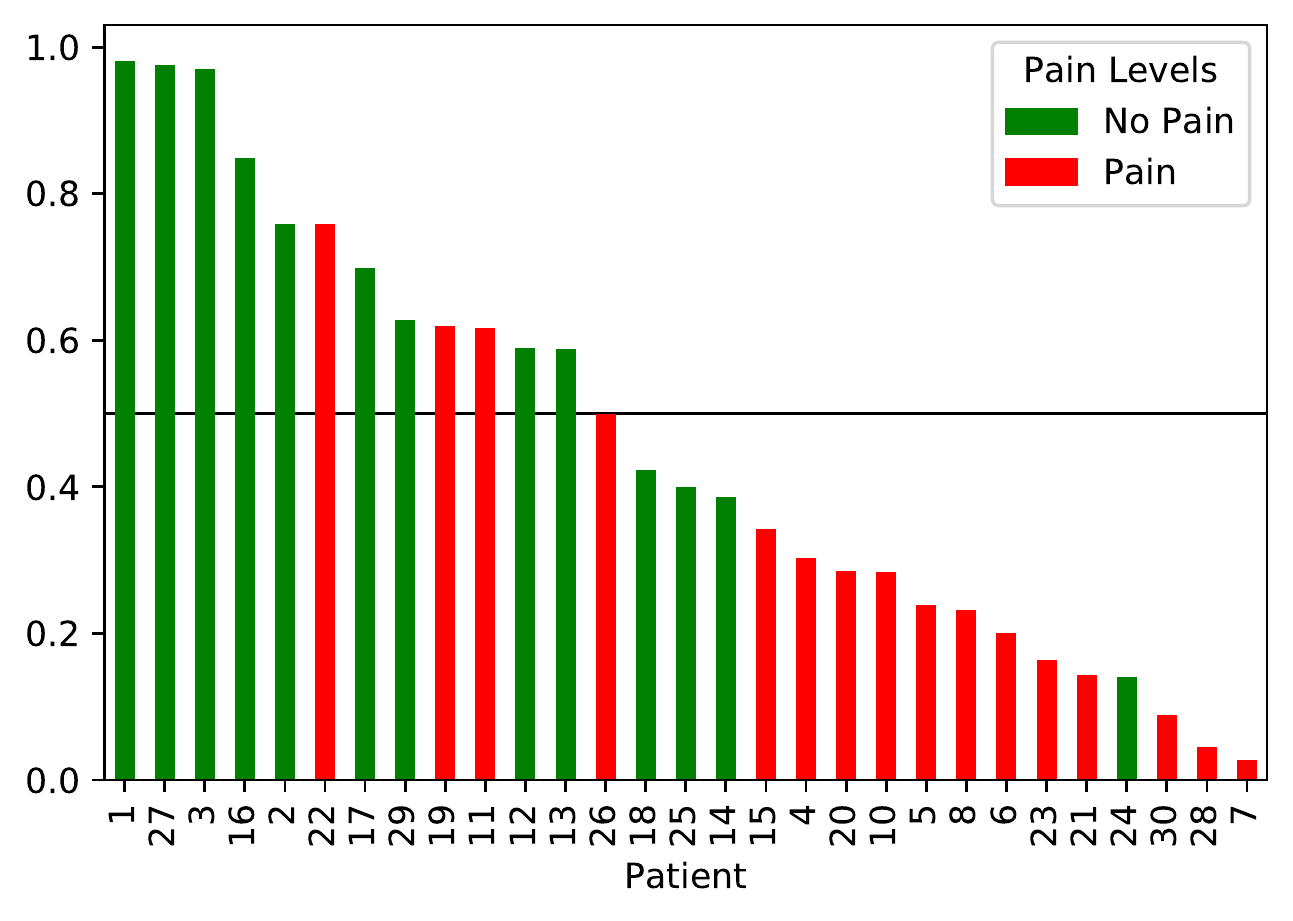}
         \caption{Audio, 2 Labels}
     \end{subfigure}
    \begin{subfigure}[b]{0.45\textwidth}
        \centering
        \includegraphics[width=\textwidth]{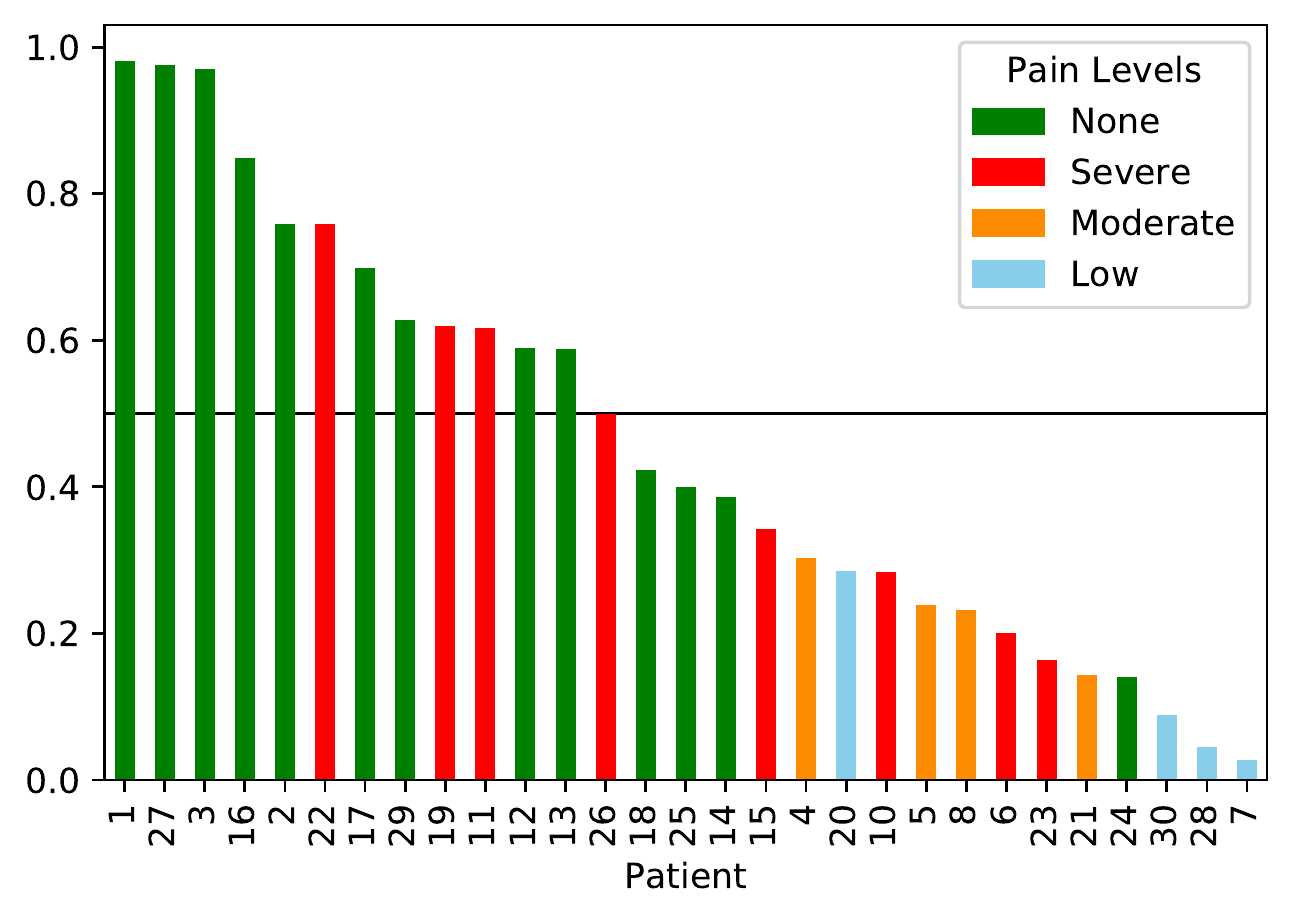}
        \caption{Audio, 4 Labels}
     \end{subfigure}
    \caption{
    \small{
    \textbf{Accuracy per Patient for Audio Models.}}}
    \label{audio}
\end{figure}

\begin{table}[ht!]
\centering
\label{failures}
\caption{Bottom Five Test Patients for Each Model, with Near Zero Accuracy.}
\begin{tabular}{@{}lll@{}}
\toprule
\textbf{Experiments} & \textbf{Model} & \textbf{Bottom 5 Test Patients} \\ \midrule
Exp. 1 & Resnet50-2-static & 7, 25, 28, 26, 13 \\
Exp. 2 & Random Forest LM & 8, 20, 19, 28, 17 \\
Exp. 3 & Random Forest Pain & 15, 30, 10, 6, 28 \\
Exp. 4 & Fusion 1 & 1, 2, 16, 13, 29, 25* \\
Exp. 5 & Fusion 2 & 4, 20, 29, 17, 1 \\
Exp. 6 & Fusion 3 & 4, 27, 29, 25, 20 \\ \bottomrule
\end{tabular}
\end{table}

\clearpage
\newpage
\section{Datasheets for Datasets}

\subsection{Motivation}
\textbf{For what purpose was the dataset created?} The ISS dataset was created to explore whether facial video, images, audio, text, and pain scores, could be used to predict chronic cancer self-reported pain across a diverse patient group. [Paper section reference: Introduction]

\textbf{Who created the dataset?} The National Institutes of Health (NIH) National Cancer Institute (NCI). [Paper section reference: Introduction]

\textbf{Who funded the creation of the dataset?} The National Institutes of Health (NIH) National Cancer Institute (NCI). [Paper section reference: Introduction]

\textbf{Any other comments?} [N/A]

\subsection{Composition}
\textbf{What do the instances that comprise the dataset represent (e.g., documents, photos, people,
countries)? Are there multiple types of instances (e.g., movies, users, and ratings; people and
interactions between them; nodes and edges)? Please provide a description.} The ISS dataset consists of 29 cancer patients currently undergoing active treatment at the NIH, scoring their self-reported pain levels and narrating their feelings about pain. [Paper section reference: ISS Dataset/Sample and Study Design]

\textbf{How many instances are there in total (of each type, if appropriate)?} 29 patients are currently enrolled in the study. [Paper section reference: ISS Dataset/Sample and Study Design]

\textbf{Does the dataset contain all possible instances or is it a sample (not necessarily random) of instances from a larger set?} It is the current number of enrolled patients to date. The larger set will encompass 112 patients once the study has concluded. [Paper section reference: ISS Dataset/Sample and Study Design]

\textbf{What data does each instance consist of? “Raw” data (e.g., unprocessed text or images) or features?} Each patient submits a video through their smartphone or home computer. The video is extracted into video frames, facial images, facial landmarks, audio files, spectrograms, and audio features. [Paper section reference: ISS Dataset/Data Description]

\textbf{Is there a label or target associated with each instance?} Yes, the label is ``worst pain" which represents the patient's pain level reported in the past 30 days upon enrollment in the study. This label does not change through the patient's time in the study. [Paper section reference: ISS Dataset/Sample and Study Design; Baselines]

\textbf{Is any information missing from individual instances?} Yes, Patient 0015 lacks self-reported pain scores, and Patient 0009 video data was unusable. [Paper section reference: ISS Dataset/Sample and Study Design]

\textbf{Are relationships between individual instances made explicit (e.g., users' movie ratings, social network links)?} N/A.

\textbf{Are there recommended data splits (e.g., training, development/validation, testing)?} Yes. We use 10-fold-cross-validation that withholds three patients from nine of the splits, and two patients from the tenth split. [Paper section reference: Appendix/Details of the Models/Cross Validation Details]

\textbf{Are there any errors, sources of noise, or redundancies in the dataset?} Patients are inconsistent in their video recording, leading to noisy backgrounds such as low illumination, muffled or quiet speaking, letters and signage in the background, shiny bright sunlight and glare, and occasional mask wearing due to Covid-19. [Paper section reference: Discussion and Future Work]

\textbf{Is the dataset self-contained, or does it link to or otherwise rely on external resources (e.g.,websites, tweets, other datasets)?} The ISS dataset will be self-contained and hosted under the supervision of the NIH. [Paper section reference: ISS Dataset/Data Storage and Access]

\textbf{Does the dataset contain data that might be considered confidential (e.g., data that is protected by legal privilege or by doctor–patient confidentiality, data that includes the
content of individuals’ non-public communications)?} Yes. The dataset consists of PII in the form of people's video, faces, and medical discussions. [Paper section reference: ISS Dataset/Data Storage and Access]

\textbf{Does the dataset contain data that, if viewed directly, might be offensive, insulting, threatening, or might otherwise cause anxiety?} Unknown.

\textbf{Does the dataset relate to people?} Yes, it consists of 29 cancer patients undergoing active treatment. [Paper section reference: ISS Dataset/Sample and Study Design]

\textbf{Does the dataset identify any subpopulations (e.g., by age, sex)?} The dataset is grouped into cohorts assigned by age and sex. [Paper section reference: ISS Dataset/Sample and Study Design]

\textbf{Is it possible to identify individuals (i.e., one or more natural persons), either directly or indirectly (i.e., in combination with other data) from the dataset?} No. An anonymous identifier has been used to label each patient. The dataset has been deidentified of patient name, date of birth, address, or medical record number. [Paper section reference: ISS Dataset/Data Storage and Access; Appendix/Author Statement]

\textbf{Does the dataset contain data that might be considered sensitive in any way (e.g., data that reveals racial or ethnic origins, sexual orientations, religious beliefs, political opinions or union memberships, or locations; financial or health data; biometric or genetic data; forms of government identification, such as social security numbers; criminal history)?} No.

\textbf{Any other comments?} N/A

\subsection{Collection Process}
\textbf{How was the data associated with each instance acquired?} Each patient submits a video through a mobile or web-based application. [Paper section reference: ISS Dataset/Patient Protocol]

\textbf{What mechanisms or procedures were used to collect the data (e.g., hardware apparatus or sensor, manual human curation, software program, software API)?} A mobile/web-based application is the main interface used to collect the self-reported pain scores and video recording. Each patient uses their own smartphone or computer in their personal home setting to submit the video. [Paper section reference: ISS Dataset/Patient Protocol]

\textbf{If the dataset is a sample from a larger set, what was the sampling strategy (e.g., deterministic, probabilistic with specific sampling probabilities)?} N/A.

\textbf{Who was involved in the data collection process (e.g., students, crowdworkers, contractors) and how were they compensated (e.g., how much were crowdworkers paid)?} Compensation for patients enrolled in the study is explained in the Consent Form as follows ([Paper section reference: Appendix/Consent Form]): 

\begin{itemize}
    \item \$15 per week in which you successfully submit three (3) or more completed questionnaires, including self-video recordings. A maximum of one (1) questionnaire may be completed per day.
    \item \$15 per in-clinic questionnaire and video recording session successfully completed. A minimum of one in-clinic recording must be completed within three months.
    \item No matter how many questionnaires or in-clinic sessions you complete, the total maximum allowed compensation for this study is \$225.
\end{itemize} 

\textbf{Over what timeframe was the data collected?} Patients reported in this dataset were enrolled in the study between December 2020 and July 2021. [Paper section reference: ISS Dataset/Sample and Study Design; Related Works Table 1]

\textbf{Were any ethical review processes conducted (e.g., by an institutional review board)?} Yes, NCI protocol 20C0130 entitled, “A Feasibility Study Investigating the Use of Machine Learning to Analyze Facial Imaging, Voice and Spoken Language for the Capture and Classification of Cancer/Tumor Pain.” \url{https://clinicaltrials.gov/ct2/show/NCT04442425} [Paper section reference: Introduction]

\textbf{Does the dataset relate to people?} Yes.

\textbf{Did you collect the data from the individuals in question directly, or obtain it via third parties or other sources (e.g., websites)?} From the individuals directly. [Paper section reference: ISS Dataset/Patient Protocol]

\textbf{Were the individuals in question notified about the data collection?} Yes. ([Paper section reference: Appendix/Consent Form])

\textbf{Did the individuals in question consent to the collection and use of their data?} Yes. [Paper section reference: Appendix/Consent Form]

\textbf{If consent was obtained, were the consenting individuals provided with a mechanism to revoke their consent in the future or for certain uses?} Yes. Each patient signed an NIH Assent to Participate in a Clinical Research Study, reviewed alongside their physician which provides information about the study such as its purpose, requirements, benefits, and monetary incentives. The form indicates that the patient may change their mind at any point in time and drop out of the study. [Paper section reference: ISS Dataset/Sample and Study Design; Appendix/Consent Form]

\textbf{Has an analysis of the potential impact of the dataset and its use on data subjects (e.g., a data protection impact analysis) been conducted?} Yes, during the IRB approval process.

\textbf{Any other comments?} N/A

\subsection{Preprocessing/cleaning/labeling}

\textbf{ Was any preprocessing/cleaning/labeling of the data done (e.g., discretization or bucketing, tokenization, part-of-speech tagging, SIFT feature extraction, removal of instances, processing of missing values)?} Yes. Video data was extracted for video frames at 10 frames per second, faces cropped using a facial detection algorithm, and landmarks (AAMs) detected. Video data was extracted form audio, and the resulting audio file was used to generate a Mel spectrogram and audio features. Minimal transformations were applied to the extracted data. [Paper section reference: ISS Dataset/Data Description; Appendix/Preprocessing]

\textbf{Was the “raw” data saved in addition to the preprocessed/cleaned/labeled data (e.g., to support unanticipated future uses)?} Yes. 

\textbf{Is the software used to preprocess/clean/label the instances available?} Yes, we use open source Python libraries to include OpenCV for image processing, ffmpeg for video extraction, PyDub for audio file chunking, Librosa for audio feature extraction and Mel spectrogram generation, Scikit-Learn for machine learning models, and PyTorch for neural networks. [Paper section reference: ISS Dataset/Data Description; Appendix/Preprocessing]

\textbf{Any other comments?} N/A

\subsection{Uses}

\textbf{Has the dataset been used for any tasks already?} No.

\textbf{Is there a repository that links to any or all papers or systems that use the dataset?} No.

\textbf{What (other) tasks could the dataset be used for?}
Our baseline experiments only addressed one task of static frame-by-frame pain classification. However, there are a variety of tasks specific to improving self-reported chronic pain prediction, that can be explored with the ISS dataset. They are:

\begin{itemize}
\itemsep -0.2em 
    \item Temporal-based, video classification model, integrating image, audio, and text.
    \item Disentangling the model's dependency on patients and their identity.
    \item Predicting multiple features of pain such as the average self-reported affective and activity scores. 
    \item Development and automation of chronic cancer pain-specific facial action units.
\end{itemize}

\textbf{Is there anything about the composition of the dataset or the way it was collected and
preprocessed/cleaned/labeled that might impact future uses?} No.

\textbf{Are there tasks for which the dataset should not be used?} General emotion detection, facial recognition, surveillance applications.

\textbf{Any other comments?} N/A

\subsection{Distribution}

\textbf{Will the dataset be distributed to third parties outside of the entity (e.g., company, institution, organization) on behalf of which the dataset was created? If so, please provide a description.} No. NIH will maintain guardianship of the data.

\textbf{How will the dataset will be distributed (e.g., tarball on website, API, GitHub)? Does the dataset have a digital object identifier (DOI)?} To be determined.

\textbf{When will the dataset be distributed?} To be determined, under the guidance of the NIH.

\textbf{Will the dataset be distributed under a copyright or other intellectual property (IP) license, and/or under applicable terms of use (ToU)? If so, please describe this license and/or ToU, and provide a link or other access point to, or otherwise reproduce, any relevant licensing terms or ToU, as well as any fees associated with these restrictions.} To be determined, under the guidance of the NIH. [Paper section reference: Appendix/Author Statement]

\textbf{Have any third parties imposed IP-based or other restrictions on the data associated with the instances? If so, please describe these restrictions, and provide a link or other access point to, or otherwise reproduce, any relevant licensing terms, as well as any fees associated with these restrictions.} No.

\textbf{Do any export controls or other regulatory restrictions apply to the dataset or to individual instances? If so, please describe these restrictions, and provide a link or other access point to, or otherwise reproduce, any supporting documentation.} No.

\textbf{Any other comments?} Prior to accessing the data, it is likely that users will be required to provide proof of successful completion of some or all of the following types of training: Protection of Human Subjects in Research, Global Privacy and Data Protection, a HIPAA and Health Privacy, General Information Security, Social and Behavioral Research, and/or Conflicts of Interest.

\subsection{Maintenance}
\textbf{Who will be supporting/hosting/maintaining the dataset?} The NIH. [Paper section reference: ISS Dataset/Data Storage and Access]

\textbf{How can the owner/curator/manager of the dataset be contacted (e.g., email address)?} \url{gulleyj@mail.nih.gov}

\textbf{Is there an erratum? If so, please provide a link or other access point.} N/A.

\textbf{Will the dataset be updated (e.g., to correct labeling errors, add new instances, delete instances)? If so, please describe how often, by whom, and how updates will be communicated to users (e.g., mailing list, GitHub)?} To be determined, under the guidance of the NIH.

\textbf{If the dataset relates to people, are there applicable limits on the retention of the data associated with the instances (e.g., were individuals in question told that their data would be retained for a fixed period of time and then deleted)? If so, please describe these limits and explain how they will be enforced.} Patients were informed through the Consent Form that their data may be held indefinitely. [Paper section reference: Appendix/Consent Form]

\textbf{Will older versions of the dataset continue to be supported/hosted/maintained? If so, please describe how. If not, please describe how its obsolescence will be communicated to users.} To be determined, under the guidance of the NIH.

\textbf{If others want to extend/augment/build on/contribute to the dataset, is there a mechanism for them to do so? If so, please provide a description. Will these contributions be validated/verified? If so, please describe how. If not, why not? Is there a process for communicating/distributing these contributions to other users? If so, please provide a description.} To be determined, under the purview of the NIH.

\textbf{Any other comments?} N/A.

\section{Risk Categorization}
Our study was classified as “Minimal Risk under 45 CFR 46 / 21 CFR 56” after review and approval by the National Institutes of Health’s Federalwide Assurance (FWA00005897), in accordance with Federal regulations 45 CFR 46 and 21 CFR 56, including Good Clinical Practice.  The study also was determined to fall under three “Expedited Review Categories” due to three applicable characteristics including (1b) clinical studies of medical devices for which an investigational device exemption application (21 CFR Part 812) is not required, (6) collection of data from voice, video, digital, or image recording made for research purposes, and (7) research on individual or group characteristics or behavior.  The data collection which involved the use of a mobile or web app to input answers to questions and record videos do not pose risk to patients.  The tasks also do not involve deception.

\end{document}